\documentclass[12pt]{article}
\usepackage{amsfonts,amsmath,amssymb,epsf,framed}
\usepackage{amsmath,braket}
\usepackage{graphicx,hyperref,color,comment}
\usepackage{cite}

\usepackage{subfigure}

\hypersetup{
    bookmarks=true,         % show bookmarks bar?
    unicode=false,          % non-Latin characters in Acrobat’s bookmarks
    pdftoolbar=true,        % show Acrobat’s toolbar?
    pdfmenubar=true,        % show Acrobat’s menu?
    linktocpage=true,       % hyperlink by page number in contents
    pdffitwindow=false,     % window fit to page when opened
    pdfstartview={FitH},    % fits the width of the page to the window
    pdfnewwindow=true,      % links in new PDF window
    colorlinks=true,       % false: boxed links; true: colored links
    linkcolor=blue,          % color of internal links (change box color with linkbordercolor)
    citecolor=blue,        % color of links to bibliography
    filecolor=blue,      % color of file links
    urlcolor=blue           % color of external links
}
\numberwithin{equation}{section}									% equation numbering by section

\def\d#1{\,{\rm d}#1}

\newcommand{\de}{\partial}
\newcommand{\be}{\begin{equation}}
\newcommand{\ba}{\begin{eqnarray}}
\newcommand{\ea}{\end{eqnarray}}
\newcommand{\ee}{\end{equation}}

\newcommand{\s}{\sqrt}

\newcommand{\ti}{\tilde}
\newcommand{\ap}{\alpha}

\newcommand{\ddd}{\cdots}
\newcommand{\no}{\nonumber \\}
\newcommand{\la}{\langle}
\newcommand{\lb}{\rangle}
\newcommand{\bea}{\begin{eqnarray}}
\newcommand{\eea}{\end{eqnarray}}
\newcommand{\bes}{\begin{equation*}}
\newcommand{\beas}{\begin{eqnarray*}}
\newcommand{\eeas}{\end{eqnarray*}}
\newcommand{\bas}{\begin{array*}}
\newcommand{\eas}{\end{array*}}
\newcommand{\ees}{\end{equation*}}
\newcommand{\nn}{\nonumber}
\newcommand{\p}{\partial}
\newcommand{\ep}{\epsilon}

\def\pd{\partial}
\def\bs{\boldsymbol}

\newcommand{\D}{{\cal D}}

\newcommand{\vp}{\varphi}

\newcommand{\na}{\nabla}
\newcommand{\tn}{\tilde\nabla}

\newcommand{\tB}{\tilde\Box}

\newcommand{\cL}{{\cal L}}

\newcommand{\Ga}[1]{{\Gamma \left({#1} \right)}}

\newcommand{\Z}{\mathbb{Z}}
\newcommand{\N}{\mathbb{N}}
\newcommand{\eq}[1]{(\ref{#1})}

\newcommand{\ra}{\rangle}

\let\a=\alpha  \let\c=\chi \let\d=\delta \let\e=\epsilon  \let\g=\gamma \let\h=\eta  \let\l=\lambda \let\m=\mu \let\n=\nu
 \let\p=\phi \let\r=\rho %\let\s=\sigma
\let\t=\tau \let\th=\theta  \let\vp=\varphi   \let\z=\zeta
 \let\D=\Delta \let\G=\Gamma \let\L=\Lambda \let\O=\Omega      
\def\nn{\nonumber}
\def\inf{\infty}
\def\na{\nabla}
\def\pa{\partial}

\begin{document}

\begin{titlepage}
\thispagestyle{empty}

\vspace*{-2cm}
\begin{flushright}
YITP-22-56
\\
IPMU22-0032
\\
\end{flushright}

\bigskip

\begin{center}
\noindent{{\Large \textbf{Brane Dynamics of Holographic BCFTs}}}\\
\vspace{1cm}

\quad 
Keisuke Izumi$^{a,b}$, Tetsuya Shiromizu$^{b,a}$, 
Kenta Suzuki$^c$, \\
Tadashi Takayanagi$^{c,d,e}$ and Norihiro Tanahashi$^f$
\vspace{1cm}\\

{\it $^a$ Kobayashi-Maskawa Institute, Nagoya University, Nagoya 464-8602, Japan}\\
\vspace{1mm}
{\it $^b$ Department of Mathematics, Nagoya University, Nagoya 464-8602, Japan}\\
\vspace{1mm}
{\it $^c$Center for Gravitational Physics and Quantum Information (CGPQI),\\
Yukawa Institute for Theoretical Physics,
Kyoto University, \\
Kitashirakawa Oiwakecho, Sakyo-ku, Kyoto 606-8502, Japan}\\
\vspace{1mm}
{\it $^d$Inamori Research Institute for Science,\\
620 Suiginya-cho, Shimogyo-ku,
Kyoto 600-8411 Japan}\\
\vspace{1mm}
{\it $^{e}$Kavli Institute for the Physics and Mathematics
 of the Universe (WPI),\\
University of Tokyo, Kashiwa, Chiba 277-8582, Japan}\\
\vspace{1mm}
{\it $^{f}$Department of Physics, Chuo University, Kasuga, Bunkyo-ku, Tokyo 112-8551, Japan}

\bigskip \bigskip
\vskip 2em
\end{center}

\begin{abstract}
In this paper we study various dynamical aspects of the AdS/BCFT correspondence in higher dimensions. We study properties of holographic stress energy tensor by analyzing the metric perturbation in the gravity dual. We also calculate the stress energy tensor for a locally excited state on a half plane in a free scalar CFT. Both of them satisfy a reflective boundary condition that is expected for any BCFTs. We also study the behavior of the scalar field perturbation in the AdS/BCFT setup and show that they also show complete reflections.
Moreover, we find that the entanglement entropy of a BCFT computed from the AdS/BCFT matched with that calculated from the Island formula, which supports the Island/BCFT correspondence in higher dimensions. Finally we show how we can calculate one point functions in a BCFT in our gravity dual.
\end{abstract}

\end{titlepage}

\newpage

\tableofcontents

%%%%%%%%%%%%%%%%%%%%%%%%%%%%%%%%%%%%%%%%%%%%%%%%%%%%%%%%
%%%%%%%%%%%%%%%%%%%%%%%%%%%%%%%%%%%%%%%%%%%%%%%%%%%%%%%%
\section{Introduction}
\label{sec:introduction}
%%%%%%%%%%%%%%%%%%%%%%%%%%%%%%%%%%%%%%%%%%%%%%%%%%%%%%%%
%%%%%%%%%%%%%%%%%%%%%%%%%%%%%%%%%%%%%%%%%%%%%%%%%%%%%%%%

A boundary conformal field theory (BCFT) is a conformal field theory (CFT) defined on a manifold with boundaries such that a part of conformal symmetry is preserved by the boundaries \cite{Cardy:1984bb,Cardy:2004hm,McAvity:1993ue,McAvity:1995zd}.
Recently, a class of gravity duals of BCFTs, called the AdS/BCFT \cite{Takayanagi:2011zk,Fujita:2011fp,Nozaki:2012qd} (for an earlier model refer to \cite{Karch:2000gx}), have been actively studied. One reason for this is that the AdS/BCFT makes calculations of the entanglement entropy much more tractable. Moreover it provides useful setups which model black hole evaporation processes from which we can derive the Page curve \cite{Almheiri:2019hni}, where the Island prescription \cite{Penington:2019npb,Almheiri:2019psf} is expected to be realized via the brane-world holography \cite{Randall:1999ee,Randall:1999vf,Gubser:1999vj,Karch:2000ct,Giddings:2000mu,Shiromizu:2001jm,Shiromizu:2001ve,Nojiri:2000eb,Nojiri:2000gb,Hawking:2000kj,Koyama:2001rf,Kanno:2002iaa}.
Refer to e.g.\cite{Almheiri:2019psy,Rozali:2019day,Chen:2019uhq,Balasubramanian:2020hfs,Geng:2020qvw,Chen:2020uac,Chen:2020hmv,Bousso:2020kmy,Chen:2020jvn,Chen:2020tes,Akal:2020twv,Miyaji:2021lcq,Akal:2021foz,Geng:2021mic,Bhattacharya:2021nqj,Hu:2022ymx,Suzuki:2022xwv,Anous:2022wqh,Kawamoto:2022etl,Bianchi:2022ulu,Akal:2021dqt,Akal:2022qei} for applications of AdS/BCFT to black hole information problem.

The AdS/BCFT has also been successfully applied to many other problems. For example, this includes condensed matter and field theoretic aspects of chaotic BCFTs \cite{Fujita:2012fp,Ugajin:2013xxa,Erdmenger:2015xpq,Seminara:2017hhh,Seminara:2018pmr,Hikida:2018khg,Shimaji:2018czt,Caputa:2019avh,Mezei:2019zyt,Reeves:2021sab,Kusuki:2021gpt, Numasawa:2022cni}, studies of boundary renormalization group flow and boundary entropy, \cite{Gutperle:2012hy,Estes:2014hka,Kobayashi:2018lil,Sato:2020upl}, holographic calculations of the computational complexity \cite{Chapman:2018bqj,Sato:2019kik,Braccia:2019xxi,Sato:2021ftf,Hernandez:2020nem,Collier:2021ngi,Chalabi:2021jud,Belin:2021nck} and cosmological models \cite{Cooper:2018cmb,Antonini:2019qkt,VanRaamsdonk:2020tlr,VanRaamsdonk:2021qgv,Waddell:2022fbn}.
String theory embeddings have been studied in \cite{Chiodaroli:2011nr,Chiodaroli:2012vc,Karch:2020iit,Bachas:2020yxv,Simidzija:2020ukv,Ooguri:2020sua,Raamsdonk:2020tin,Uhlemann:2021nhu,Coccia:2021lpp}. Moreover, the AdS/BCFT correspondence was generalized in  \cite{Akal:2020wfl,Miao:2020oey} to construct gravity duals of higher codimension holography. Refer also to \cite{Suzuki:2021pyw} for an analysis of AdS/BCFT with one loop quantum corrections and to \cite{Omiya:2021olc} for an appearance of non-locality in the brane-world description of AdS/BCFT.

\begin{figure}
  \centering
   \includegraphics[width=7cm]{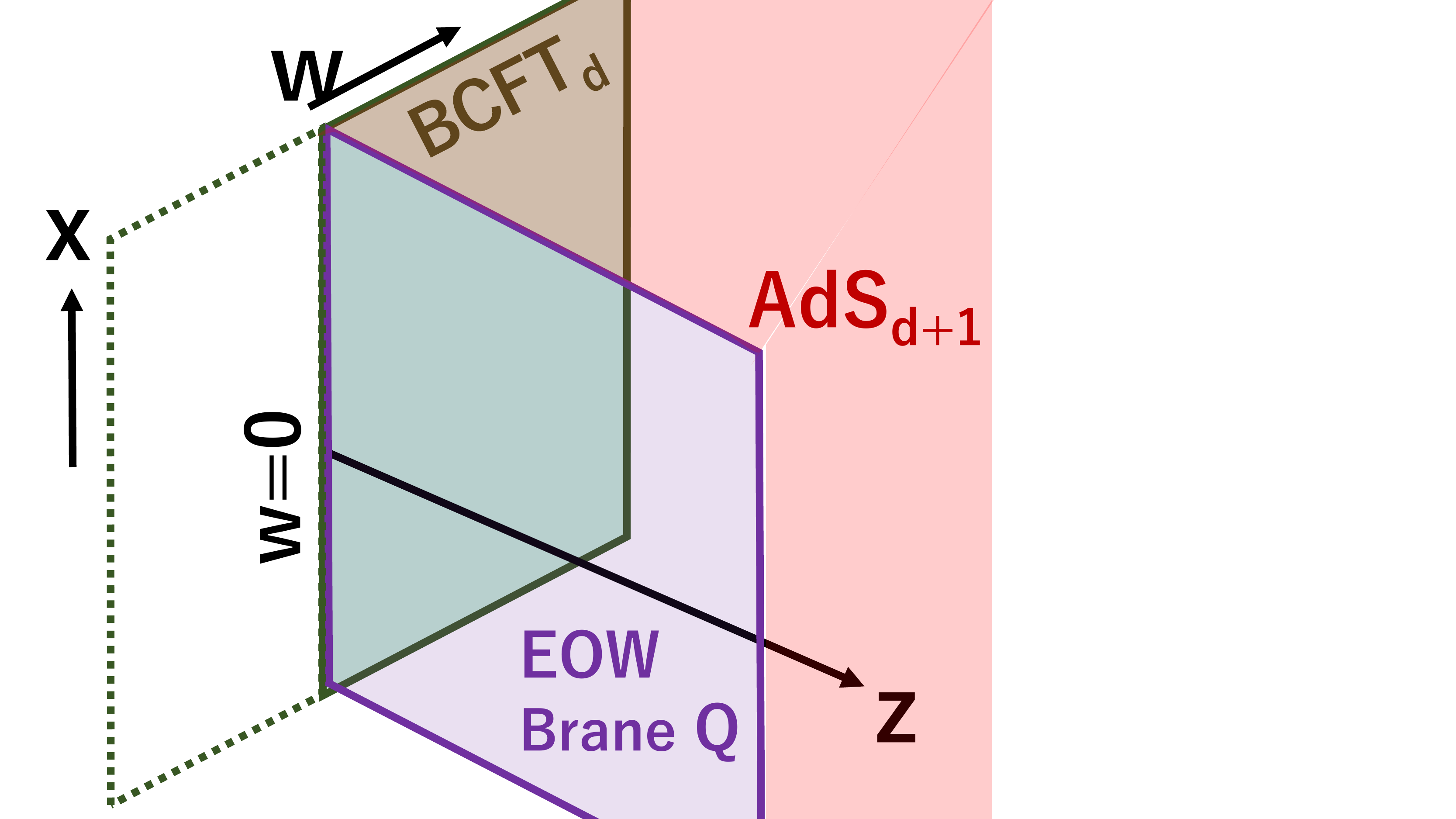}
    \caption{A sketch of AdS/BCFT construction. The gravity dual of a BCFT is given by the red colored region which is surrounded by the boundary where the BCFT is situated (blown colored) and the the end of the world-brane (EOW brane, purple colored). }
    \label{fig:setupsk}
\end{figure}

The basic idea of AdS/BCFT is to simply extend the BCFT towards the bulk AdS by introducing the end of the world-brane (EOW brane) such that the EOW brane at the AdS boundary coincides with the boundary of the BCFT. The AdS$_{d+1}/$BCFT$_d$ argues that a $d$ dimensional BCFT (BCFT$_d$) is dual to the $d+1$ dimensional gravity with a negative cosmological constant on the region surrounded by the boundary where the BCFT is situated and the the EOW brane, as sketched in Fig.~\ref{fig:setupsk}.

The EOW brane is defined by a Neumann boundary condition and has a tension $\sigma$. Therefore in general, its gravitational  backreaction deforms the geometry. This problem of backreaction can be solved analytically in the lowest dimensional AdS/BCFT for $d=2$, i.e.\ 
AdS$_{3}/$BCFT$_2$. This is because any solutions in the three dimensional pure gravity locally coincide with the AdS$_3$, reflecting the infinite dimensional Virasoro symmetry in the dual BCFT. Accordingly, various dynamical aspects of AdS$_{3}/$BCFT$_2$ have been studied quite well. On the other hand, in higher dimensions $d\geq 3$, there have been little results known for the dynamics of higher dimensional AdS/BCFT. The purpose of this paper is to extensively study dynamical aspects of the higher dimensional AdS/BCFT from several perspectives. We will study dynamical properties of stress energy tensor in BCFTs by analyzing metric perturbations in the AdS/BCFT and also by directly calculating a simple free scalar BCFT. This is also interesting from the brane-world interpretation. In the AdS/BCFT, gravity is localized on the EOW  brane and this mode is dual to the boundary degrees of freedom of the BCFT, while the bulk gravity is dual to the bulk degrees of freedom of the BCFT. One basic question is whether the AdS/BCFT correctly explains the complete reflection of the energy and momentum flux in BCFTs, which is not so obvious from the profile of  the gravity dual geometry. A similar question will also be considered in a scalar field excitation in the bulk. We will also study how one 
point functions of scalar operators in BCFTs can be reproduced from the gravity dual calculations, generalizing the previous result in \cite{Suzuki:2022xwv} for $d=2$.

The triality between (i) a $d$ dimensional BCFT, (ii) its gravity dual in $d+1$ dimension via the AdS$_{d+1}/$BCFT$_d$ and (iii) its brane-world interpretation, leads to  a purely $d$ dimensional duality relation that a $d$ dimensional BCFT is equivalent to a $d$ dimensional CFT on a half plane, coupled to $d$ dimensional gravity on AdS$_d$, which is called Island/BCFT correspondence \cite{Suzuki:2022xwv}. As we mentioned, this duality can be derived from known conjectures. However, we are still far from a complete justification of the Island/BCFT correspondence. 
The recent paper \cite{Suzuki:2022xwv} examined this problem and presented quantitative evidences in the lowest dimension $d=2$. In this paper, we would like to examine the Island/BCFT correspondence by calculating entanglement entropy in higher dimensions $d\geq 3$. 

The rest of the paper is organized as follows.
In section \ref{sec:stress tensors}, we present field theory results of stress energy tensor in BCFTs. We also give gravity dual results for two dimensional BCFTs via AdS$_3/$BCFT$_2$.
In section \ref{sec:GWa}, we analyze metric perturbations in AdS$_{d+1}/$BCFT$_d$ for higher dimensions $d\geq 3$. We work out the behavior of holographic stress energy tensor.
In section \ref{sec:SF}, we analyze scalar field excitations in AdS/BCFT setup and show that the scalar wave is completely reflected.
In section \ref{sec:island}, we give evidences of the Island/BCFT correspondence by comparing the calculation of entanglement entropy in a holographic BCFT and in a CFT coupled to gravity.
In section \ref{sec:one-point}, we give a prescription of computing one point functions in higher dimensional AdS/BCFT.
In section \ref{sec:conclusions}, we summarize our conclusions.
In appendix \ref{app:d=3}, we present a class of explicit solutions of metric perturbations for $d=3$.
In appendix \ref{app:minimal surface}, we show our detailed analysis of minimal surfaces in AdS$_{d+1}$.

{\it Note added:} When we were completing this draft, we became aware of the independent work \cite{Suzuki:2022yru},
which has a partial overlap with ours in that
the metric perturbations in a higher dimensional AdS/BCFT were analyzed.

%%%%%%%%%%%%%%%%%%%%%%%%%%%%%%%%%%%%%%%%%%%%%%%%%%%%%%%%
%%%%%%%%%%%%%%%%%%%%%%%%%%%%%%%%%%%%%%%%%%%%%%%%%%%%%%%%
\section{Stress energy tensors in BCFTs}
\label{sec:stress tensors}
%%%%%%%%%%%%%%%%%%%%%%%%%%%%%%%%%%%%%%%%%%%%%%%%%%%%%%%%
%%%%%%%%%%%%%%%%%%%%%%%%%%%%%%%%%%%%%%%%%%%%%%%%%%%%%%%%

We consider a $d$ dimensional conformal field theory (CFT$_d$) on a manifold with $d-1$ dimensional boundaries. In particular, we impose those classes of boundary conditions which preserve a part of conformal invariance. This is called a boundary conformal field theory (BCFT$_d$). The CFT$_d$ originally has the conformal symmetry $SO(d,2)$ in Lorentzian signature (or $SO(d+1,1)$ in Euclidean signature) and the symmetry preserved by BCFT$_d$ is its subgroup $SO(d-1,2)$ (or $SO(d,1)$). The main purpose of this paper is to explore various dynamical aspects of gravity duals of BCFTs. In this section, as a preparation of our later arguments, we would like to explain what we expect as the behavior of the stress energy tenor in BCFTs. We will also describe the gravity results for BCFT$_2$, leaving the gravity analysis in higher dimensions left as the main problems discussed in later section.

%%%%%%%%%%%%%%%%%%%%%%%%%%%%%%%%%%%%%%%%%%%%%%%%%%%%%%%%
\subsection{Free scalar BCFT$_d$}
%%%%%%%%%%%%%%%%%%%%%%%%%%%%%%%%%%%%%%%%%%%%%%%%%%%%%%%%

In this section, we consider the simplest example of BCFT$_d$ given by the conformally coupled real free scalar theory
	\begin{align}
		I \, = \, \frac{1}{2} \int d^{d}x \left[ \pa_a \p \pa^a \p \, + \, \frac{d-2}{4(d-1)} \, R_0 \p^2 \right] \, ,
	\end{align}
where $R_0$ is the background scalar curvature and we assume an Euclidean signature.
We consider this theory on a flat $d$ dimensional Euclidean space\footnote{
Since we consider a flat space, $R_0=0$. However, this terms is needed to derive the conformal stress energy tensor.
}
$\xi^a = (x^i,w)$ where $i=0,1, \cdots, d-2$, and place a boundary at $w=0$, so that the BCFT lives on the half plane $w>0$.
The stress energy tensor of this theory is given by
	\begin{align}
    	{\cal T}_{ab} \, = \, \frac{d}{2(d-1)}\pa_a \p \pa_b \p - 
    \frac{\delta_{ab}}{2(d-1)}(\de\phi)^2-\frac{d-2}{2(d-1)} (\pa_a \pa_b\phi)\phi+\frac{(d-2)\delta_{ab}}{2d(d-1)}(\de^2\phi)\phi.
    \label{constitutive}
	\end{align}
With this expression of the stress energy tensor, it is traceless ${\cal T}^a_a = 0$ and, provided the equation of motion $\pa^2 \p=0$, it is conserved $\pa^a {\cal T}_{ab} = 0$.

Now, we would like to compute the one-point function of this stress energy tensor.
This can be done by using the two-point function of the scalar fields, which is given by \cite{McAvity:1993ue}
	\begin{align}
    	&\qquad \big\la \p(w_1, \vec{x}_1) \p(w_2, \vec{x}_2) \big\ra \\
    	&= \, \frac{1}{(d-2) S_d}
    	\Bigg( \frac{1}{\big((w_1-w_2)^2 + |\vec{x}_1-\vec{x}_2|^2 \big)^\frac{d-2}{2}} + \frac{(-1)^\h}{\big((w_1+w_2)^2 + |\vec{x}_1-\vec{x}_2|^2 \big)^\frac{d-2}{2}} \Bigg) \, ,\nn
	\label{2pt}
	\end{align}
where $S_d=2\pi^{d/2}/\G(d/2)$.
The boundary coalition parameter $\h$ is $\h=0$ for Neumann boundary condition at $w=0$: $\pa_w \p(0, \vec{x}) = 0$,
or $\h=1$ for Dirichlet boundary condition at $w=0$: $ \p(0, \vec{x}) = 0$.
First using the expression (\ref{constitutive}) and this two-point function (\ref{2pt}), one can explicitly check that the one-point function of the stress energy tensor in the vacuum state indeed vanishes everywhere
	\begin{align}
    	\big\la {\cal T}_{ab}(w, \vec{x}) \big\ra \, = \, 0 \, .
	\end{align}

Next, we consider an excited state $|\Psi_\phi\lb$ which is created from the vacuum by inserting a single operator located at $w=l$ and $\vec{x}=0$. However, since the local operator excitation is singular, we make a regularization by introducing an infinitesimally small imaginary time evolution by $\ep$ \cite{Nozaki:2014hna} 
(refer to \cite{Horowitz:1999gf,Nozaki:2013wia} for gravity duals) as follows
\ba
|\Psi_\phi\lb=e^{-\ep H}\phi(w=l,\vec{x}=0)|0\lb,
\ea
which is equivalent to shifting the value of $x_0$ into $x_0=-\ep$.
This regularization leads to a finite inner product
\ba
\la \Psi_\phi|\Psi_\phi\lb= \frac{1}{(d-2) S_d}(2\ep)^{-(d-2)}.
\ea
However, note that it is also meaningful to consider a fine value of 
$\ep$ which defines a class of excitation states. Thus below we allow $\ep$ to take any positive values.

The one-point function of the stress energy tensor in this state is expressed as
	\begin{align}
    	\big\la {\cal T}_{ab}(w, \vec{x}) \big\ra_{\p}
    	\, &\equiv \, \frac{\big\la \Psi_\phi | {\cal T}_{ab}(w, \vec{x}) |\Psi_\phi \big\ra}{\la \Psi_\phi|\Psi_\phi\lb}
    	\\
    	&= \, (d-2) S_d (2\ep)^{d-2} \, \big\la \p(l, \ep,0) | {\cal T}_{ab}(w,\vec{x}) | \p(l,-\ep, 0) \big\ra \, , \nn
	\end{align}
and this one-point function can be computed from the four-point function of the scalar fields (we ignore the Wick contraction of 
$\p(w_1, \vec{x}_1)$ and  $\p(w_2, \vec{x}_2)$ as this leads to the vacuum expectation value of stress energy tensor)
%	\begin{align}
%    	&\quad \ \big\la \p(l,\ep,0) \p(w_1,\tau_1,x_1) \p(w_2,\tau_2, x_2) \p(l,-\ep, 0) \big\ra \nn\\
%    	&= \, \frac{1}{(d-2)^2 S_d^2} 
%    	\Bigg( \frac{1}{\big((w_1-l)^2 +(\tau_1-\ep)^2+(x_1)^2 \big)^\frac{d-2}{2}} + \frac{1}{\big((w_1+l)^2 +(\tau_1-\ep)^2+(x_1)^2\big)^\frac{d-2}{2}} \Bigg) \nn\\
%    	&\hspace{60pt} \times 
%    	\Bigg( \frac{1}{\big((w_2-l)^2 + (\tau_2+\ep)^2+(x_2)^2 \big)^\frac{d-2}{2}} + \frac{1}{\big((w_2+l)^2 + (\tau_2+\ep)^2+(x_2)^2\big)^\frac{d-2}{2}} \Bigg) \nn\\
%    	&\qquad + \, (\e \leftrightarrow - \e) \, .
%	\end{align}
%	\begin{align}
%    	&\quad \ \big\la \p(l,\ep,0) \p(w_1,\tau_1,x_1) \p(w_2,\tau_2, x_2) \p(l,-\ep, 0) \big\ra \nn\\
%    	&= \, \frac{1}{(d-2)^2 S_d^2} \, \sum_{\pm, \pm, \pm} 
%     	\Bigg( \frac{1}{\big((w_1 \pm l)^2 + (\tau_1 \pm \ep)^2+(x_1)^2 \big)^\frac{d-2}{2}} \Bigg)\nn\\
%     	&\hspace{110pt}\times \Bigg( \frac{1}{\big((w_2 \pm l)^2 + (\tau_2 \mp \ep)^2+(x_2)^2 \big)^\frac{d-2}{2}} \Bigg) \, ,
%    \end{align}
	\begin{align}
    	&\quad \ \big\la \p(l,\ep,0) \p(w_1,\tau_1,x_1) \p(w_2,\tau_2,x_2) \p(l,-\ep, 0) \big\ra \\
    	&= \, \frac{1}{(d-2)^2 S_d^2} \, \sum_{\pm, \pm, \pm} 
     	\frac{(\pm1)^\h}{\big((w_1 \pm l)^2 + (\tau_1 \pm \ep)^2+(x_1)^2 \big)^\frac{d-2}{2}\big((w_2 \pm l)^2 + (\tau_2 \mp \ep)^2+(x_2)^2 \big)^\frac{d-2}{2}} \, , \nn
    \end{align}
where we introduced the expression $\vec{x}_{1,2}=(\tau_{1,2},x_{1,2})$ and the summation contains eight terms for all possible combinations of the $\pm$'s in the denominator.
However, the $\mp$ sign in $(\tau_2 \mp \ep)^2$ is not independent but the sign must be opposite from the sign in $(\tau_1 \pm \ep)^2$.
The numerator is $(+1)^\h$ when the two signs in $(w_1 \pm l)^2$ and $(w_2 \pm l)^2$ are same, and 
it is $(-1)^\h$ when the two signs in $(w_1 \pm l)^2$ and $(w_2 \pm l)^2$ are opposite.
Then by taking the derivative with respect to $(w_{1,2},\tau_{1,2},x_{1,2})$ following (\ref{constitutive}), we can find the expectation value of stress energy tensors.
Since the full expression of this one-point function of the stress energy tensor is quite lengthy, we do not write down all components explicitly. Instead we will present the expression of ${\cal T}_{ab}$ at the boundary $w=0$ by choosing the dimension $d=3$ and Neumann boundary condition $\h=0$.
They are given as follows:
%
%	\begin{align}
%& {\cal T}_{\tau\tau}|^{d=3}_{w=0}\no
%&=\frac{4L^2\!(\!L^2\!+\!\tau^2\!+\!x^2)^2\ep\!-\!8\left(\!L^4+L^2(\!-7\tau^2\!+\!3x^2\!)\!+\!2\!(\!\tau^4\!-\!4\tau^2x^2\!+\!x^4\!)\right)\ep^3\!-4(\!7L^2\!
%-\!8\tau^2\!+\!8x^2\!)\ep^5\!-\!16\ep^7\!}{16\pi^2(L^2+x^2+(\tau-\ep)^2)^{\frac{5}{2}}(L^2+x^2+(\tau+\ep)^2)^{\frac{5}{2}}},\no
%& {\cal T}_{xx}|^{d=3}_{w=0}\no
%&=\frac{4L^2\!(\!L^2\!+\!\tau^2\!+\!x^2)^2\ep\!+\!8\left(2L^4\!+\!\tau^4\!-\!10\tau^2x^2+x^4+L^2(\tau^2\!+\!3x^2\!)\right)\ep^3
%+4(5L^2\!-\!4\tau^2\!+\!4x^2\!)\ep^5\!+\!8\ep^7}{16\pi^2(L^2+x^2+(\tau-\ep)^2)^{\frac{5}{2}}(L^2+x^2+(\tau+\ep)^2)^{\frac{5}{2}}},\no
%&{\cal T}_{ww}|^{d=3}_{w=0}\no
%&=\frac{-8L^2\!(\!L^2\!+\!\tau^2\!+\!x^2)^2\ep\!+\!8\left(-L^4\!-8L^2\tau^2+(\tau^2+x^2)^2\right)\ep^3+8(L^2\!-\!2\tau^2\!+\!2x^2\!)\ep^5\!+\!8\ep^7}{16\pi^2(L^2+x^2+(\tau-\ep)^2)^{\frac{5}{2}}(L^2+x^2+(\tau+\ep)^2)^{\frac{5}{2}}},\no
%&{\cal T}_{\tau x}|^{d=3}_{w=0}=\frac{48\tau x (L^2-\tau^2+x^2+\ep^2)\ep^3}{16\pi^2(L^2+x^2+(\tau-\ep)^2)^{\frac{5}{2}}(L^2+x^2+(\tau+\ep)^2)^{\frac{5}{2}}},\no
%& {\cal T}_{\tau w}|^{d=3}_{w=0}={\cal T}_{xw}|^{d=3}_{w=0}=0.
%    \end{align}
%    
	\begin{align}
& {\cal T}_{\tau\tau}|^{d=3}_{w=0}\no
&=\frac{4L^2\!(\!L^2\!+\!\tau^2\!+\!x^2)^2\ep\!-\!8\left(\!L^4+L^2(\!-7\tau^2\!+\!3x^2\!)\!+\!2\!(\!\tau^4\!-\!4\tau^2x^2\!+\!x^4\!)\right)\ep^3\!-4(\!7L^2\!
-\!8\tau^2\!+\!8x^2\!)\ep^5\!-\!16\ep^7\!}{16\pi^2(L^2+x^2+(\tau-\ep)^2)^{\frac{5}{2}}(L^2+x^2+(\tau+\ep)^2)^{\frac{5}{2}}},\no
& {\cal T}_{xx}|^{d=3}_{w=0}\no
&=\frac{4L^2\!(\!L^2\!+\!\tau^2\!+\!x^2)^2\ep\!+\!8\left(2L^4\!+\!\tau^4\!-\!10\tau^2x^2+x^4+L^2(\tau^2\!+\!3x^2\!)\right)\ep^3
+4(5L^2\!-\!4\tau^2\!+\!4x^2\!)\ep^5\!+\!8\ep^7}{16\pi^2(L^2+x^2+(\tau-\ep)^2)^{\frac{5}{2}}(L^2+x^2+(\tau+\ep)^2)^{\frac{5}{2}}},\no
&{\cal T}_{ww}|^{d=3}_{w=0}\no
&
=\frac{-8L^2\!(\!L^2\!+\!\tau^2\!+\!x^2)^2\ep\!+\!8\left(-L^4\!-8L^2\tau^2+(\tau^2+x^2)^2\right)\ep^3+8(L^2\!-\!2\tau^2\!+\!2x^2\!)\ep^5\!+\!8\ep^7}{16\pi^2(L^2+x^2+(\tau-\ep)^2)^{\frac{5}{2}}(L^2+x^2+(\tau+\ep)^2)^{\frac{5}{2}}},
    \end{align}
and
	\begin{gather}
        {\cal T}_{\tau x}|^{d=3}_{w=0}
        \, = \, \frac{48\tau x (L^2-\tau^2+x^2+\ep^2)\ep^3}{16\pi^2(L^2+x^2+(\tau-\ep)^2)^{\frac{5}{2}}(L^2+x^2+(\tau+\ep)^2)^{\frac{5}{2}}} \, , \nn\\[6pt]
        {\cal T}_{\tau w}|^{d=3}_{w=0} \, = \, {\cal T}_{xw}|^{d=3}_{w=0} \, = \, 0 \, .
    \end{gather}

In general, the stress energy tensor in BCFT satisfies the boundary condition \cite{Herzog:2017xha}:
\ba
{\cal T}_{wi}|_{w=0}=0,\ \ (i=0,1,2,\ddd,d-2), \label{bcbcft}
\ea
while other components of stress tensor are non-vanishing at the boundary in general. In the bulk, it satisfies as usual the traceless and conservation condition 
\ba
&& {\cal T}^a_a=0,  \no
&&\de^a {\cal T}_{ab}=0. \label{bcbfta}
\ea

Notice that this boundary condition (\ref{bcbcft}) means that the energy flux and momentum flux are completely reflected at the boundary. Indeed the total energy and 
momentum are conserved 
\ba
\frac{d}{dt}\int dx^{d-2}dw {\cal T}_{0i}=0,\ \ \ (i=0,1,\ddd,d-2)
\ea
owing to the conservation law (\ref{bcbfta}) and the boundary condition (\ref{bcbcft}).

\subsection{Holographic stress energy tensor}

In the AdS/CFT, the stress energy tensor can be computed from the behavior of the metric near the AdS boundary, so called holographic stress energy tensor \cite{Balasubramanian:1999re,deHaro:2000vlm}. When the CFT is defined on a flat space, which we always assume in this paper, its takes a simple form as we will explain below.
We can write the metric of $d+1$ dimensional asymptotically Poincare AdS background in the Fefferman-Graham expansion form:
\ba
&& ds^2=\frac{dz^2+g_{ab}(\xi,z)d\xi^a d\xi^b}{z^2}, \no
&& g_{ab}\simeq \eta_{ab}+\frac{16\pi G_N}{d}z^d~ {\cal T}_{ab}(\xi)+O(z^{d+1})\ \ (z\to 0),
\label{holEM}
\ea
where $\xi^a\ \ (a=0,1,2,\ddd,d-1)$ is the coordinate of the flat space on which the CFT is defined. In the above expansion, ${\cal T}_{ab}(\xi)$ is the holographic stress energy tensor, normalized such that it is computed from the variation of the CFT action $S_{CFT}$ with respect to the background metric $g_{ab}$
\ba
{\cal T}_{ab}(\xi)=-\frac{2}{\s{-g}}\frac{\delta S_{CFT}}{\delta g^{ab}}.
\ea
The bulk Einstein equation leads to the traceless condition and the conservation
law (\ref{bcbfta}).

In this paper we will consider the BCFT defined on the $d$ dimensional half plane $\xi_{d-1}(\equiv w)>0$. Then from the standard BCFT result\cite{McAvity:1993ue,McAvity:1995zd,Herzog:2017xha}, we expect the boundary condition for the stress tensor at $w=0$ (\ref{bcbcft}). Note that we can still apply the calculation of holographic stress energy tensor (\ref{holEM}) even for BCFTs.
Refer to \cite{Nozaki:2012qd,Miao:2017gyt,Chu:2017aab,Chu:2021mvq} for earlier calculations.

\subsection{Holographic stress energy tensor in AdS$_3/$BCFT$_2$}
\label{sec:ads3}

The gravity dual for BCFT$_2$ based on AdS/BCFT \cite{Takayanagi:2011zk,Fujita:2011fp} is exceptionally simple and the computations of stress energy tensor is straightforward as opposed to those in the higher dimensional BCFTs. Therefore we would like to briefly present this lower dimensional analysis below.

Consider the pure Einstein gravity with negative cosmological constant.
Solutions to the Einstein equation are all locally equivalent to the pure AdS$_3$.
Therefore we start with the Poincar\'{e} metric in AdS$_3$ (we set $U=T-W$ and $V=T+W$)
\ba
ds^2=\frac{d\eta^2-dUdV}{\eta^2}, 
\label{pol}
\ea
which is dual to the vacuum state in a CFT$_2$. In this paper, we set the AdS radius to unity for simplicity. In AdS/BCFT, the gravity dual of BCFT is given by considering the end of the world-brane (EOW brane) $Q$ such that it gives an extension of the boundary of BCFT toward the bulk AdS and such that it satisfied the Neumann boundary condition 
\ba
K_{\mu\nu}-Kh_{\mu\nu}+\frac{\sigma}{2} h_{\mu\nu}=0,  \label{beina}
\ea
where $h_{\mu\nu}$ and $K_{\mu\nu}$ are the induced metric and the extrinsic curvature (with the out-going normal vector) of the EOW brane $Q$. The parameter $\sigma$ is the tension of the EOW brane. A canonical solution is the hyperplane given by 
\ba
W+\lambda \eta=0,
\ea
where the parameter $\lambda$ is related to the tension of the EOW brane via 
\ba
\lambda=\frac{\sigma}{\s{4-\sigma^2}}.
\ea
The AdS/BCFT argues that the gravity dual of BCFT vacuum on the right half plane $W\geq 0$ is given by the region $W+\lambda\eta\geq 0$ in the AdS$_3$ (\ref{pol}). 

To construct the gravity dual for generic excited states, we consider a conformal transformation of the form (we set $u=t-w$ and $v=t+w$):
\ba
\ti{u}=p(u),\ \ \ \ti{v}=q(v).  \label{cmapcft}
\ea
Via the AdS/CFT, this is dual to the following coordinate transformation in AdS$_3$
\ba
\begin{split}
U&=p(u)+\frac{2z^2(p')^2 q''}{4p'q'-z^2p''q''},\\
V&=q(v)+\frac{2z^2(q')^2p''}{4p'q'-z^2p''q''},\\
\eta&=\frac{4z(p'q')^{3/2}}{4p'q'-z^2p''q''},
\label{corads}
\end{split}
\ea
which is known as the so-called Ba$\tilde{\text{n}}$ados map 
\cite{Banados:1998gg,Roberts:2012aq,Shimaji:2018czt}.
The metric expressed in terms of the new coordinates $(u,v,z)$ looks like
\ba
ds^2=\frac{dz^2}{z^2}+{\cal T}_{++}(u)(du)^2+{\cal T}_{--}(v)(dv)^2-\left(\frac{1}{z^2}
+z^2{\cal T}_{++}(u) {\cal T}_{--}(v)\right)dudv, 
\label{metads}
\ea
where
\ba
{\cal T}_{++}(u)=\frac{3(p'')^2-2p'p'''}{4p'^2},\qquad {\cal T}_{--}(v)=\frac{3(q'')^2-2q'q'''}{4q'^2}
\label{emgt}
\ea
are proportional to the chiral and anti-chiral part of the holographic stress energy tensor, respectively.

We consider a BCFT defined on the right half $w\geq 0$ of a two dimensional plane. 
If we require that after the conformal map (\ref{cmapcft}) of this right half plane, we have the same geometry (i.e.\ right half plane), we need to require that $p$ and $q$ are the same function: $p(t)=q(t)$. In this case it is easy to find\footnote{This relation can be violated if $p\neq q$ and this is realized when the boundary has a non-trivial time dependent profile.
Such a BCFT is known as a moving mirror (see e.g.\cite{Akal:2020twv,Akal:2021foz}). In this paper, we always assume that the boundary of the background spacetime is time-independent and is just given by a straight line or plane.} 
\ba
{\cal T}_{++}(u)={\cal T}_{--}(v),
\ea
on the boundary $w=0$. Therefore the energy flux is conserved. In this way, the perfect reflection of energy flux is clear in the gravity dual also.

%%%%%%%%%%%%%%%%%%%%%%%%%%%%%%%%%%%%%%%%%%%%%%%%%%%%%%%%%%%%%%%%%
\section{Gravitational dynamics in higher dimensional AdS/BCFT}
\label{sec:GWa}
%%%%%%%%%%%%%%%%%%%%%%%%%%%%%%%%%%%%%%%%%%%%%%%%%%%%%%%%%%%%%%%%%

\begin{figure}
  \centering
   \includegraphics[width=7cm]{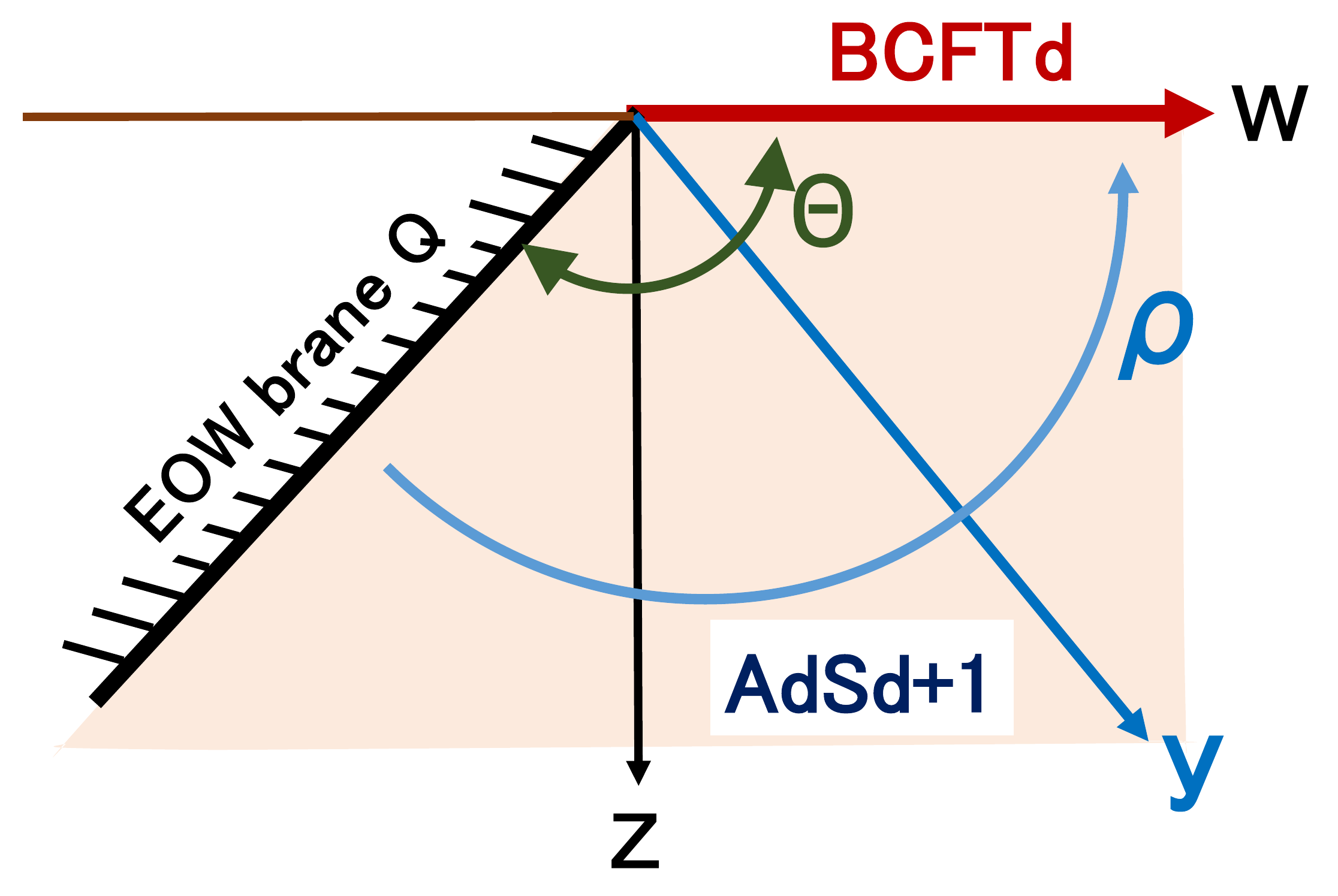}
    \caption{A sketch of AdS/BCFT setup.}
    \label{setuphigher}
\end{figure}

Now we would like to move on to our main problem, higher dimensional AdS/BCFT.
Here we investigate the metric perturbations for the standard AdS/BCFT setup in $d+1$ dimensions, depicted in Fig.~\ref{setuphigher}. This allows up to calculate the holographic stress energy tensor in the dual BCFT$_d$.

First, let us present the original background for which we perform the perturbation. Consider the ($d+1$) dimensional Poincare AdS
\ba
ds^2=\frac{dz^2+dw^2+\sum_{i=0}^{d-2}dx_i^2}{z^2}.
\ea
Via the coordinate transformation 
\ba
z=\frac{y}{\cosh\rho},\qquad  w=y\tanh\rho,   \label{cortra}
\ea
we obtain the following metric in terms of the hyperbolic slice of AdS$_d$:
\ba
ds^2=d\rho^2+\cosh^2\rho\left(\frac{dy^2+\sum_{i=0}^{d-2}dx_i^2}{y^2}\right).
\label{metric ansatz}
\ea

The ground state of a BCFT defined on the half plane $w>0$ in the coordinate $(\tau,x,w)$ \cite{Takayanagi:2011zk,Fujita:2011fp} is dual to the part of AdS$_{d+1}$ defined by (refer to Fig.~\ref{setuphigher})
\ba
w+\lambda z>0,  \label{xlam}
\ea
where $\lambda$ is related to the brane tension $\sigma$  of the end of the world-brane via 
\ba
\sigma=\frac{2(d-1)\lambda}{\s{1+\lambda^2}}.  \label{tensiona}
\ea
It is straightforward to confirm that this satisfies the required boundary condition (\ref{beina}).
In the hyperbolic slicing coordinates, the brane is located at $\r=\r_*$ (with $\r_* <0$), where the brane location $\r_*$ is related to $\l$ by
\ba
\l \, = \, - \sinh \r_* \, .
\ea

Below we study the metric perturbations around this background solution, assuming a pure gravity theory with a negative cosmological constant. Our convention of coordinate labeling is as follows. $(M,N.\cdots)$ and $(\mu,\nu,\cdots)$ are indices for $(d+1)$-dimensional and $d$-dimensional directions without $\r$ coordinate. 
Variables and operators with tilde are those with respect to the $d$-dimensional background metric $\gamma_{\mu\nu}$ 
(see Eq.~\eq{metAdS}).
Prime $'$ denotes a derivative w.r.t.\ $\rho$. 
Equality with hat ($\hat =$) means that the equation holds only on the brane.

%%%%%%%%%%%%%%%%%%%%%%%%%%%%%%%%%%%%%%%%%%%%%%%%%%%%%%%%%%%%%%%%%
\subsection{Setup}
%%%%%%%%%%%%%%%%%%%%%%%%%%%%%%%%%%%%%%%%%%%%%%%%%%%%%%%%%%%%%%%%%

The metric of $(d+1)$-dimensional bulk is 
\begin{eqnarray}
ds^2_{AdS_{d+1}}= d \rho^2 + \left(\mathfrak{a}^2(\rho)  \gamma_{\mu\nu} + h_{\mu\nu}\right) dx^\mu dx^\nu,
\label{metAdS}
\end{eqnarray}
where
\begin{eqnarray}
\mathfrak{a}(\rho) := \ell \cosh \left( \frac{\rho}{\ell}\right)
\label{defa}
\end{eqnarray}
and $\gamma_{\mu\nu}$ is the $d$-dimensional AdS metric with the unit curvature.
The Riemann curvature of the background metric (i.e.\ for the case with $h_{\mu\nu}=0$) becomes
\begin{eqnarray}
R^{(0)}_{MNAB}= -\frac{1}{\ell^2} \left( g_{MA}g_{NB}-g_{MB}g_{NA} \right).
\end{eqnarray}
When the Einstein equation with the cosmological constant $\Lambda$ gives this solution, 
$\Lambda$ should be
\begin{eqnarray}
\Lambda= - \frac{d(d-1)}{2\ell^2}.
\end{eqnarray}
%{\color{blue}
%KS: I think this should be
%\begin{eqnarray}
%\Lambda= - \frac{d(d-1)}{2l^2}.
%\end{eqnarray}
%}

For the perturbation $h_{\mu\nu}$, we use the following gauge condition:
\begin{eqnarray}
\tn^\mu h_{\mu\nu} =0, \qquad h^\mu{}_\mu=0.  \label{TTG}
\end{eqnarray}
Note that it gives
\begin{eqnarray}
\na_M h^M{}_N = 0.
\end{eqnarray}
This gauge condition is called transverse-traceless (TT) gauge,
which is often used for the analysis of gravitational waves in the vacuum 
and of gravitational perturbations in 
braneworld model \cite{Karch:2000ct,Karch:2001jb,Garriga:1999yh}.

%%%%%%%%%%%%%%%%%%%%%%%%%%%%%%%%%%%%%%%%%%%%%%%%%%%%%
\subsection{Perturbed equation in the bulk}
%%%%%%%%%%%%%%%%%%%%%%%%%%%%%%%%%%%%%%%%%%%%%%%%%%%%%

The perturbation of the bulk Ricci tensor is calculated as
\begin{eqnarray}
R_{MN} &=& R^{(0)}_{MN} + \frac{1}{2} \left( \na_A \na_M h^A{}_N + \na_A \na_N h^A{}_M -\Box h_{MN}- \na_M \na_N h \right)+ {\cal O} \left( h_{MN}^2\right)
 \nonumber \\
 &=& -\frac{d}{\ell^2} \left(g^{(0)}_{MN}+h_{MN} \right) - \frac12  \Box h_{MN} -\frac{1}{\ell^2} h_{MN} + {\cal O} \left( h_{MN}^2\right).
\end{eqnarray}
Then vacuum equation gives the perturbation equation in the bulk:
\begin{eqnarray}
0&=&R_{MN} - \frac12 R g_{MN} + \Lambda g_{MN}  \qquad
\left(=:  G_{MN} + \Lambda g_{MN} \right)
 \nonumber \\
&=&
\left( -\frac{d}{\ell^2} +\frac12 \frac{d(d+1)}{\ell^2}-\frac12 \frac{d(d-1)}{\ell^2} \right) \left(g^{(0)}_{MN}+h_{MN} \right) 
- \frac12\left(  \Box  +\frac{2}{\ell^2} \right) h_{MN} + {\cal O} \left( h_{MN}^2\right)
 \nonumber \\
&=&- \frac12\left(  \Box  +\frac{2}{\ell^2} \right) h_{MN} + {\cal O} \left( h_{MN}^2\right),
\end{eqnarray}
where $\delta\left(\cdots\right)$ means the first order terms of $\left(\cdots\right)$ with respect to $h_{MN}$.
Therefore, the bulk equation of the metric perturbation becomes
\begin{eqnarray}
0&=&\left(  \Box  +\frac{2}{\ell^2} \right) h_{\mu\nu}
 \nonumber \\
&=&
h_{\mu\nu}''+(d-4) \frac{\mathfrak{a}'}{\mathfrak{a}} h_{\mu\nu}' -2 (d-2) \left( \frac{\mathfrak{a}'}{\mathfrak{a}}\right)^2 h_{\mu\nu} 
-2 \left(\frac{\mathfrak{a}'}{\mathfrak{a}}\right)' h_{\mu\nu} 
+ \mathfrak{a}^{-2}\left(\tB +2 \right) h_{\mu\nu}
 \nonumber \\
&=&
\mathfrak{a}^2\left( H_{\mu\nu}'' + d \frac{\mathfrak{a}'}{\mathfrak{a}} H_{\mu\nu}' \right)+ \left(\tB +2\right) H_{\mu\nu}  \nonumber \\
&=&
\mathfrak{a}^2\left( H_{\mu\nu}'' + d l \tanh^2 \left(\frac{\rho}{\ell}\right) H_{\mu\nu}' + \cosh^{-2} \left(\frac{\rho}{\ell}\right)  \left(\tB +2\right) H_{\mu\nu} \right),
\label{EOMH}
\end{eqnarray}
where
\begin{eqnarray}
h_{MN}=: \mathfrak{a}^2 H_{MN} 
\end{eqnarray}
and
\begin{eqnarray}
\Box:= g^{(0)MN} \nabla_M \nabla_N , \qquad \tilde \Box:= \gamma^{\mu\nu} \tilde \nabla_\mu\tilde \nabla_\nu.
\end{eqnarray}
Note that $\Box$ includes covariant derivatives, and thus, the operation to $h_{\mu\nu}$ is different from that to scalar field. 
Note that ``$2$'' in $\left(\tB +2\right)$ is required for spin-2 fields, which is not mass. 
Due to the curvature of AdS, this ``$2$'' is generated from the kinetic term of spin-2 action.
Equation \eq{EOMH} can be solved with the separation of the variables.

Hereinafter, we set $\ell=1$.
Solutions could be written in separable form $H_{\mu\nu}=\sum_i {\mathcal R}^{(i)}(\rho) Y^{(i)}_{\mu\nu}(x^\mu)$. 
With this separable form of $H_{\mu\nu}$, Eq.~\eq{EOMH} becomes the following two different 
equations with eigenvalues $\lambda_\rho^{(i)}$
\begin{eqnarray}
&&{\mathcal R}^{(i)}_{,\rho\rho}+d\, \tanh \rho {\mathcal R}^{(i)}_{,\rho} 
= -\frac{\lambda^{(i)}_\rho}{\cosh^2\rho} {\mathcal R}^{(i)},
\label{eqR}
\\ 
&& \left(\tilde \Box +2 - \lambda^{(i)}_\rho \right) Y^{(i)}_{\mu\nu} =0.
\label{eqy}
\end{eqnarray}
This decomposition is interpreted as a Kaluza-Klein reduction; 
the equation for extra-dimension Eq.~\eq{eqR} fixes the Kaluza-Klein (KK) mass as $(m^{(i)}_{KK})^2= \lambda^{(i)}_\rho$, while Eq.~\eq{eqy} is $d$-dimensional equation for spin-2 field with (KK) mass $(m^{(i)}_{KK})^2$ on the $d$-dimensional AdS spacetime.
%We show later that normalizable modes should give non-negative KK mass squared. 
Hereinafter, we omit the superscript $(i)$ for the simple notation, when the difference of $(i)$ is not important.

Let us solve Eq.~\eq{eqR} first.
Equation \eq{eqR} can be written as
\begin{eqnarray}
&&\cL_{\mathcal R} {\mathcal R} = \left(1-\zeta^2 \right)^{-\frac{d}2} \lambda_\rho {\mathcal R}, \\
&&\cL_{\mathcal R} := - \frac{d}{d\zeta} \left(1-\zeta^2 \right)^{-\frac{d}2+1}\frac{d}{d\zeta}  , \\
&&\zeta := \tanh \rho.
\label{zetadef}
\end{eqnarray}
Solutions of Eq.~\eq{eqR} are expressed by Ferrers' associated Legendre functions% 
\footnote{Note that it is different from Hobson's associated Legendre functions.}
$P^\mu_l (\zeta)$ and $Q^\mu_l (\zeta)$;
\begin{eqnarray}
{\mathcal R} =\left[ c_1 P^\mu_l (\zeta) + c_2 Q^\mu_l (\zeta) \right] (1-\zeta^2)^{\frac{d}{4}} ,
\label{P+Q}
\end{eqnarray}
where
\begin{eqnarray}
l= \nu -\frac12 , \qquad \mu = \frac{d}{2}, \qquad \nu= \frac12 \sqrt{(d-1)^2+4\lambda_\rho} .
\label{lmunu}
\end{eqnarray}
Since, near the AdS boundary $\zeta\to1$ $(\rho\to\infty)$, $P^\mu_l (\zeta)$ and $Q^\mu_l (\zeta)$ behave as
\begin{eqnarray}
&&P^\mu_l (\zeta) \, \simeq \, (1-\zeta)^{-\frac{\mu}{2}} \frac{2^{\frac{\mu}2} e^{i\pi\mu}}{\Gamma(1-\mu)},
\\
&&2 \sin (\pi \mu) Q^\mu_l (\zeta) \, \simeq \, \pi\cos (\pi \mu) P^\mu_l (\zeta),
\end{eqnarray}
the Dirichlet boundary condition on the AdS boundary gives the relation between $c_1$ and $c_2$;
\begin{eqnarray}
2 \sin (\pi \mu)c_1= -\pi\cos\left(\pi \mu\right)c_2.
\label{c1c2}
\end{eqnarray}
Note that, because of $\mu = d/2$ (See Eq.~\eq{lmunu}), 
Eq.~\eq{c1c2} gives 
\begin{eqnarray}
{\mathcal R} = 
\begin{cases}
P^\mu_l (\zeta) (1-\zeta^2)^{\frac{d}{4}} & \mbox{(if $d$ is even)}\\
Q^\mu_l (\zeta) (1-\zeta^2)^{\frac{d}{4}} & \mbox{(if $d$ is odd)}
\end{cases} .
\label{PorQ}
\end{eqnarray}
However, with the form of Eq.~\eq{P+Q}, 
both cases can be analysed simultaneously 
and if the bulk field has mass, which we will face on in Sec.~\ref{sec:SF}, 
the mode function ${\mathcal R}$ has a generic form of Eq.~\eq{P+Q}.
Therefore, we use Eq.~\eq{P+Q}, although it 
can be expressed with the simple form \eq{PorQ}.
For $l = \mu-n$ where $n$ is any positive integer, Eq.~\eq{PorQ} becomes zero. 
This is because, in the definition of associated Legendre functions, 
there exists a constant factor which goes to zero for $l \to \mu-n$. 
To avoid this situation, we use $P^{-\mu}_l$. 
Mode function ${\mathcal R}$ expressed in Eq.~\eq{P+Q} with the boundary condition \eq{c1c2} is proportional to $P^{-\mu}_l$ for general $\mu$, not only for $\mu=d/2$. 
For $l = \mu-n$, its proportional constant becomes zero, but $P^{-\mu}_l$ is non-zero and
a bulk solution satisfying  the Dirichlet boundary condition on AdS boundary. 
Therefore, $P^{-\mu}_l$ is a unified expression which can be used for any $d$ and any $l$. 
Hence, we write the bulk mode function as 
\begin{eqnarray}
\qquad \qquad {\mathcal R} \, = \, B_0 P^{-\mu}_l (\zeta)  (1-\zeta^2)^{\frac{d}{4}} \, , \qquad  \mbox{(for\, any\, $d$)}
\label{P-}
\end{eqnarray}
where $B_0$ is a normalization constant.
%We can  regularize them with multiplying $P^\mu_l$ by $\Gamma(1-\mu)$ and 
%$Q^\mu_l$ by $\Gamma(l-\mu+1)$.
%One can check that, for normalized ones, the Dirichlet boundary condition is satisfied.

Another boundary, namely junction condition on the brane, will be discussed in the next section,
which specifies the value of $\lambda_\rho$.

%%%%%%%%%%%%%%%%%%%%%%%%%%%%%%%%%%%%%%%%%%%%%%%%%%%%%%%%%%%%%%%%%
\subsection{Junction condition}
\label{sec:JC}
%%%%%%%%%%%%%%%%%%%%%%%%%%%%%%%%%%%%%%%%%%%%%%%%%%%%%%%%%%%%%%%%%

The boundary condition on the brane, that is, the junction condition (\ref{beina}) or equally
\begin{eqnarray}
K^\mu{}_\nu-K \delta^\mu{}_\nu \, \hat{=} \, - \frac12 \sigma \delta^\mu{}_\nu 
\label{junctionfull}
%- \frac{\kappa}{2}T^\mu{}_\nu
\end{eqnarray}
is required locally. 
In the derivation of the junction condition for the metric perturbations, 
%therefore, 
it is better to use the local coordinate based on the brane called Gaussian normal coordinate. 
Note that the brane position can be perturbed, that is, it does not generically exist on $\rho=\rho_*$ surface in the TT gauge, where $\rho_*$ is a constant.
By the coordinate transformation $\bar\rho=\rho+\varphi(x^\mu)$ from the TT gauge, 
we can set the brane to be on $\bar \rho=$ const. surface. 
This coordinate transformation generates the non-zero $(\bar \rho,x^\mu)$-components of metric perturbations.
Considering the coordinate transformation 
\begin{eqnarray}
\bar x^\nu = x^\nu + \xi^\nu,
\end{eqnarray}
we can remove the cross components.
Then, the Gaussian normal coordinate $(\bar \rho,\bar x^\mu)$ is achieved \cite{Garriga:1999yh}.
The relation between the metric perturbations $\bar h_{MN}$ in the Gaussian normal coordinate  and $h_{MN}$ in TT gauge 
is 
\begin{eqnarray}
\bar h_{MN}= h_{MN} +\na_M \xi_N +\na_M \xi_N,
\end{eqnarray}
where $\xi^\rho = \varphi$.
The Gaussian normal coordinate is chosen such that the $(\rho\rho)$ and $(\rho\mu)$ components of the metric perturbation are zero, as with the TT gauge condition. 
The former 
\begin{eqnarray}
\bar h_{\rho\rho}= h_{\rho\rho} (=0)
\end{eqnarray}
gives
\begin{eqnarray}
\partial_\rho \xi_\rho =0.
\end{eqnarray}
Therefore, $\xi_\rho$ depends only on $x^\mu$, which is consistent with 
\begin{eqnarray}
\xi^\rho=\xi_\rho= \varphi (x^\mu).\label{vpmode}
\end{eqnarray}
The latter one, namely vanishing of $(\rho\mu)$ component, means
\begin{eqnarray}
\bar h_{\rho\mu}= h_{\rho\mu}(=0). 
\end{eqnarray}
This gives
\begin{eqnarray}
0= \na_\rho \xi_\mu+\na_\mu \xi_\rho = \partial_\rho \xi_\mu + \partial_\mu \xi_\rho - 2\frac{\mathfrak{a}'}{\mathfrak{a}} \xi_\mu.
\end{eqnarray}
The solution for $\xi^\mu(=g^{(0)\mu M}\xi_M)$ is  
\begin{eqnarray}
\xi^\mu = -\frac{\mathfrak{a}'}{\mathfrak{a}} \gamma^{\mu\nu} \partial_\nu \varphi + \chi^\mu (x^\alpha)
\end{eqnarray}
where $\chi^\mu$ is an integration constant.%
%\footnote{
%$\chi^\mu$ corresponds the gauge degrees of freedom for the $d$-dimensional massless spin-2 mode.
%}
Then, the relation in the $(\mu\nu)$ component is 
\begin{eqnarray}
\bar h_{\mu\nu} &=& h_{\mu\nu} + \na_\mu \xi_\nu +\na_\nu \xi_\mu
\nonumber \\
&=&
h_{\mu\nu}+2 \mathfrak{a}\mathfrak{a}'\left(\gamma_{\mu\nu}\varphi -\tn_\mu\tn_\nu \varphi\right)+ 2\mathfrak{a}^2 \tn_{(\mu} \left(\gamma_{\nu)\alpha}\chi^\alpha\right). 
\label{GtoTT}
\end{eqnarray}

The junction condition \eq{junctionfull} for the leading  order gives
\begin{eqnarray}
\frac{\mathfrak{a}'}{\mathfrak{a}} \, \hat{=} \, - \frac{\sigma}{2(d-1)},
\label{backJ}
\end{eqnarray}
where $\sigma$ is the brane tension.
This fixes the background position of brane, which is consistent with Eq.~\eq{tensiona}.

The perturbed junction condition in the Gaussian normal coordinate is written as 
\begin{eqnarray}
\frac12 \bar h_{\mu\nu}' \, \hat{=} \, - \frac{\sigma}{2(d-1)} \bar h_{\mu\nu} .
%-  \frac{ \kappa^2}{2} \left( T_{\mu\nu}- \frac{1}{d-1}g_{\mu\nu} T \right),
\end{eqnarray}
%where $T_{\mu\nu}$ is the energy momentum tensor on brane and $T$ is its trace. 
Substituting the background equation \eq{backJ} into the above, we have
\begin{eqnarray}
\bar h_{\mu\nu}' -2\frac{\mathfrak{a}'}{\mathfrak{a}} \bar h_{\mu\nu} \, \hat{=} \, 0. 
% -\kappa^2 \left( T_{\mu\nu}- \frac{1}{d-1}g_{\mu\nu} T \right).
\end{eqnarray}
From the relation between the metric perturbations in the Gaussian normal coordinate and TT gauge \eq{GtoTT}, 
the perturbed junction condition for the TT gauge is obtained as
\begin{eqnarray}
h_{\mu\nu}' -2\frac{\mathfrak{a}'}{\mathfrak{a}}  h_{\mu\nu} \, \hat{=} \,  %-\kappa^2 \left( T_{\mu\nu}- \frac{1}{d-1}g_{\mu\nu} T \right)+
\frac{2}{l^2}\left(\tn_\mu \tn_\nu \varphi -  \gamma_{\mu\nu}\varphi \right).
\label{JCwS}
\end{eqnarray}
The trace of this equation gives
\begin{eqnarray}
\left( \tB - d \right) \varphi \, \hat{=} \, 0
%- \kappa^2 \frac{l^2 a^2}{2(d-1)} T
\label{varphieq}
\end{eqnarray}
and the traceless part becomes
\begin{eqnarray}
h_{\mu\nu}' - 2\frac{\mathfrak{a}'}{\mathfrak{a}} h_{\mu\nu} \, \hat{=} \, %- \kappa^2 \left( T_{\mu\nu} - \frac{1}{d} T g_{\mu\nu} \right) + 
2 \left(\tn_\mu \tn_\nu - \frac{1}{d} \tB \gamma_{\mu\nu} \right) \varphi .
\label{TLJ}
\end{eqnarray}
This boundary condition says that the brane bending $\varphi$ excites a metric perturbation in the bulk. 
This mode is called a brane bending mode.

Let us analyze the brane bending mode first.
Note that divergence of the right hand side in Eq.~\eq{TLJ} becomes
\begin{eqnarray}
\tn^\mu \left(\tn_\mu \tn_\nu - \frac{1}{d} \tB \gamma_{\mu\nu} \right) \varphi
&=&
\left(\frac{d-1}{d} \tn_\nu \tB+ \tilde R_\nu{}^\alpha \tn_\alpha\right) \varphi
\nonumber \\
&=&
\frac{d-1}{d}  \tn_\nu \left( \tB -d \right)\varphi.
\label{divvarphi}
\end{eqnarray}
With Eq.~\eq{varphieq}, Eq.~\eq{divvarphi} shows that the divergence of the right hand side of Eq.~\eq{TLJ} becomes zero.
Moreover, the right hand side of Eq.~\eq{TLJ} is traceless. 
Hence, it is a transverse-traceless tensor: 
\begin{eqnarray}
Y^{(\varphi)}_{\mu\nu}= \left(\tn_\mu \tn_\nu - \frac{1}{d} \tB \gamma_{\mu\nu} \right) \varphi
\end{eqnarray}
satisfies
\begin{eqnarray}
\tn^\mu Y^{(\varphi)}_{\mu\nu} =0, \qquad Y^{(\varphi)}_{\mu}{}^{\mu}=0.
\label{S}
\end{eqnarray}
This means that a scalar mode with $m^2=d$ can generate spin-2
(transverse-traceless 2-tensor)%
\footnote{
Then, helicity-0 mode of spin-2 can be expressed by scalar as Eq.~\eq{S}. 
This property is used in partially massless gravity discussed in cosmology~\cite{Deser:2001us}.
The corresponding mass of spin-2 in $4$-dimensional de Sitter is ``2'', which coincides with Higuchi mass.
}~\cite{Izumi:2006ca}.
The corresponding mass in spin-2 field can be understood from the following calculation;
\begin{eqnarray}
\left( \tilde \nabla_\mu \tilde \nabla_\nu - \frac{1}{d} \tilde \Box \gamma_{\mu\nu}\right) \left( \tilde \Box -d \right)\varphi
&=&  \left( \tilde \Box +2+ (d-2) \right)\left( \tilde \nabla_\mu \tilde \nabla_\nu - \frac{1}{d} \tilde \Box \gamma_{\mu\nu}\right) \varphi \nonumber \\
&=&  \left( \tilde \Box +2+ (d-2) \right)Y^{(\varphi)}_{\mu\nu},
\end{eqnarray}
that is, the corresponding mass for spin-2 is given by 
% where the decomposion of a 2-tensor $t_{\mu\nu}$ in to scalar and spin-2 does not work 
%satisfies%
%\footnote{
%In the braneworld context, this mode is not normalizable and thus it is considered as an unphysical excitation.
%However, in the analysis of AdS/BCFT, this mode would be took into consideration. We will see it later.}
\begin{eqnarray}
m_2^2 = \lambda_\rho = -(d-2). \label{BBmodemass}
\end{eqnarray}
Therefore, for the brane bending mode, we have $l=\m-2$.
%This corresponding mass for spin-2 is negative. 
%Therefore, this mode is not normalizable and not exited with finite energy.
%Since the value of $\lambda_\rho$ for the brane bending mode is fixed, 
%eqs.~\eq{P+Q} and \eq{lmunu} gives its mode function $R$, which is consistent with the perturbative junction condition \eq{JCwS}.
Since the $d$-dimensional eigenfunction $Y_{\mu\nu}$ of the brane bending mode is fixed by $Y_{\mu\nu}^{(\varphi)}$, 
the mode decomposition corresponding to the
brane bending is written as
\begin{eqnarray}
h^{\varphi}_{\mu\nu}= \mathfrak{a}^2 {\mathcal R}^{(\varphi)}Y_{\mu\nu}^{(\varphi)} = \mathfrak{a}^2 {\mathcal R}^{(\varphi)}\left(\tn_\mu \tn_\nu - \frac{1}{d} \tB \gamma_{\mu\nu} \right) \varphi. 
\end{eqnarray}
Then, the junction condition \eq{TLJ} gives
\begin{eqnarray}
{{\mathcal R}^{(\varphi)}}' \, \hat{=} \, 2.
\label{BrBendJ}
\end{eqnarray}
We will see that there always exists a mode satisfying this junction condition.

\begin{comment}
Since for the brane bending mode $\lambda_\rho$ satisfies Eq.~\eq{BBmodemass}, 
from eq.~\eq{lmunu} we know that the corresponding $l$ is $l=\mu-2$. 
As discussed below Eq.~\eq{PorQ}, then the associated Legendre function \eq{PorQ} becomes zero, 
and we should use the normalized one.
\end{comment}

Since the perturbative analysis here is linear, any solution is written with superposition of homogeneous solutions and non-homogeneous solutions. 
The source generating non-homogeneous solutions is only the brane bending. 
Let us next derive the homogeneous solutions.
With the brane position fixed  ($\varphi=0$), Eq.~\eq{JCwS} becomes
\begin{eqnarray}
h_{\mu\nu}' -2\frac{\mathfrak{a}'}{\mathfrak{a}}  h_{\mu\nu} \, \hat{=} \,  0. 
%-\left( T_{\mu\nu}- \frac{1}{d-1}g_{\mu\nu} T \right). 
\label{JCwoS}
\end{eqnarray}
%Hereinafter, we consider the case with $T_{\mu\nu}=0$, that is, no matter fields exist on brane. 
%(It would be interesting to consider the case where $T_{\mu\nu}\neq 0$. We would consider such cases later.)
The above equation \eq{JCwoS} means 
\begin{eqnarray}
{\mathcal R}' \, \hat{=} \, 0, \label{JCR}
\end{eqnarray}
that is, the Neumann boundary condition is imposed for ${\mathcal R}$ on the brane. 
With the form of Eq.~\eq{P-} and an identity for the associated Legendre function, Eq.~\eq{JCR} is written as 
\begin{eqnarray}
0&\hat=&\frac{d}{d\zeta} \left(1-\zeta^2\right)^{\frac{d}{4}}  P^{-\mu}_l (\zeta)
\nonumber \\ 
&=&
 -\frac{d}{2}\left(1-\zeta^2\right)^{\frac{d}{4}-1} \zeta P^{-\mu}_l (\zeta)  +\left(1-\zeta^2\right)^{\frac{d}{4}}  \frac{d}{d\zeta} P^{-\mu}_l (\zeta)
\nonumber \\ 
&=&
-\left(1-\zeta^2\right)^{\frac{d-2}{4}}  P^{-\mu+1}_l (\zeta).
\label{ddzPQ}
\end{eqnarray}
%
\begin{comment}
\begin{eqnarray}
0&\hat=&\frac{d}{d\zeta} \left[\left(1-\zeta^2\right)^{\frac{d}{4}} \left( c_1 P^\mu_l (\zeta)+c_2 Q^\mu_l (\zeta)\right) \right]
\nonumber \\ 
&=&
 -\frac{d}{2}\left(1-\zeta^2\right)^{\frac{n}{4}-1} \zeta \left( c_1 P^\mu_l (\zeta)+c_2 Q^\mu_l (\zeta)\right)  +\left(1-\zeta^2\right)^{\frac{d}{4}} \left( c_1 \frac{d}{d\zeta} P^\mu_l (\zeta)+c_2 \frac{d}{d\zeta} Q^\mu_l (\zeta)\right)
\nonumber \\ 
&=&
\left(1-\zeta^2\right)^{\frac{d-2}{4}} (l+\mu)(l-\mu+1) \left( c_1 P^{\mu-1}_l (\zeta)+c_2 Q^{\mu-1}_l (\zeta)\right),
\label{ddzPQ}
\end{eqnarray}
where we use an identity
\begin{eqnarray}
(1-\zeta^2)\frac{d}{d\zeta} P^\mu_l (\zeta) = \mu \zeta P^\mu_l(\zeta) +(l+\mu)(l-\mu+1) \sqrt{1-\zeta^2} P^{\mu-1}_l(\zeta).
\label{formuladP/dx}
\end{eqnarray}
\end{comment}
%
We  suppose that the brane exists at $\zeta=\zeta_*$ and then, the Neumann boundary condition \eq{ddzPQ}
gives vanishing of $P^{-\mu+1}_l (\zeta)$ there, 
\begin{comment}
\begin{eqnarray}
P^{-\mu+1}_l (\zeta)=0.
\label{lmu1}
\end{eqnarray}
The latter two condition corresponds to the case  with
\begin{eqnarray}
\lambda_\rho = 0,
\label{lambda0}
\end{eqnarray}
which corresponds to massless spin-2 modes%
\footnote{
For the case with $\lambda_\rho = 0$, $R$ becomes constant.
Thus ,this massless spin-2 modes are non-normalizable modes in the braneworld context.
}.
a non-normalizable mode as we have shown in the previous section. 
For the case with $\lambda_\rho = 0$, we have 
\begin{eqnarray}
\mu= \frac{d}2, \qquad l=\mu -1.
\end{eqnarray}
With these value of $\mu$ and $l$, Eq.\eq{c1c2} gives $c_2=0$ and the associated Legendre function $P^\mu_{l}$ becomes
\begin{eqnarray}
P^\mu_{l}(\zeta) = P^\mu_{\mu-1}(\zeta) = P^\mu_{-\mu}(\zeta) \propto \left( 1 -\zeta^2\right)^{-\frac{d}{4}}.
\end{eqnarray}
Therefore, $R$ becomes constant, which is not normalizable as we have discussed in the previous section.

The first condition of Eq.~\eq{lmu1}
\end{comment}
which is written as
\begin{eqnarray}
0=P^{-\mu+1}_l (\zeta_*)
=\frac{\sqrt{2}  \left(\sin\theta\right)^{-(\mu-1)}}{\sqrt{\pi}\Ga{\mu-\frac12}}
\int_0^\theta \frac{\cos \left(\left(l+\frac12\right)\psi \right)}{\left( \cos \psi - \cos \theta\right)^{\frac32 - \mu}} d\psi ,
\label{inteq1}
\end{eqnarray}
where $\cos \theta := \zeta_*=\tanh\rho_*$.
%
%The above equation is satisfied iff the integration in the right hand side gives zero;
%\begin{eqnarray}
%\int_0^\theta \frac{\cos \left(\left(l+\frac12\right)\psi \right)}{\left( \cos \psi - \cos \theta\right)^{\frac32 - \mu}} d\psi =0. \label{inteq}
%\end{eqnarray}
This equation is generically difficult to solve for $l$, but in the case with $d=3$ it is solvable. 
We will analyse the case with $d=3$ in Sec.~\ref{subsecd=3}.
%
\begin{comment}
{\color{blue}
For $d=$ odd, we have $R=AQ_l^{-\m}(\z) (1-\z^2)^{\frac{d}{2}}$. As in the Legendre $P$ function case, the Neumann boundary condition gives
\begin{equation}
0 \, \hat{=} \, \frac{d}{d\zeta} \left(1-\zeta^2\right)^{\frac{d}{4}} Q^{-\mu}_l (\zeta)
\, = \, -\left(1-\zeta^2\right)^{\frac{d-2}{4}}  Q^{-\mu+1}_l (\zeta).
\end{equation}
Using the integral representation of the Legendre $Q$ function, this is written as
    \begin{align}
        0 \, = \, Q_l^{-\m+1}(\z) \, = \, \frac{\sqrt{2} e^{-(\m-1)\pi i} \left(\sinh\r_* \right)^{-(\mu-1)}}{\sqrt{\pi}\Ga{\mu-\frac12}}
        \int_\inf^{\r_*} dt \, \frac{e^{-\left(l+\frac12\right)t }}{\left( \cosh t - \cosh \r_* \right)^{\frac32 - \mu}} \, . 
    \end{align}
However, it seems difficult to solve this condition for $l$.
}
\end{comment}
%

Let us go back to the junction condition for the brane bending mode~\eq{BrBendJ}, which is written as
\begin{eqnarray}
2 \, \hat{=} \, {{\mathcal R}^{(\varphi)}}' = - B_1\left(1-\zeta^2\right)^{\frac{d-2}{4}}  P^{-\mu+1}_{\mu-2} (\zeta).
\label{JCRphi}
\end{eqnarray}
Therefore, if $P^{-\mu+1}_{\mu-2} (\zeta)$ is nonzero for any $\zeta$, 
we have a constant $B_1$ 
satisfying the above equation wherever the brane is.
Moreover, other modes with the same $\lambda_\rho$ as brane bending mode never exist, 
because then junction condition \eq{inteq1} is not satisfied. 
$P^{-\mu+1}_l (\zeta)$ is 
\begin{eqnarray}
P^{-\mu+1}_l (\zeta) = \frac{1}{\Gamma(\mu)}\left( \frac{1+\zeta}{1-\zeta} \right)^{\frac{1-\mu}{2}}
F\left( -l,l+1, \mu; \frac{1-\zeta}{2}\right),
\end{eqnarray}
where $F$ is the Gauss hypergeometric function. 
Euler's transformation gives 
\begin{eqnarray}
&&F\left( -l,l+1, \mu; \frac{1-\zeta}{2}\right)=\left( \frac{1+\z}{2} \right)^{\m-1}F\left( \mu+l,\mu-l-1, \mu; \frac{1-\zeta}{2}\right) \nonumber \\
&&\hspace{5mm}
= \left( \frac{1+\z}{2} \right)^{\m-1}\frac{\Gamma(\mu)}{\Gamma(\mu+l)\Gamma(\mu-l-1 )} \sum_{n=0}^{\infty} \frac{\Gamma(\mu+l+n)\Gamma(\mu-l-1+n )}{\Gamma(\mu+n)} \frac{(1-\zeta)^{n}}{ 2^n n!}.
\nonumber \\
\label{Feq}
\end{eqnarray}
For $-1<\zeta<1$, if $l$ satisfies $\mu-l-1>0$, each term in the sum is positive. Therefore, Eq.~\eq{Feq} does not have zero. 
This leads to the positivity (that is, nonzero) of $P^{-\mu+1}_l (\zeta)$. 
For $\mu-l-1=0$, $n=0$ term in the sum diverges but the divergence is canceled with the factor $1/\Gamma(\mu-l-1)$. 
Then, $n=0$ term becomes positive and other terms in the sum are also. 
Therefore, for $\mu-l-1=0$, $P^{-\mu+1}_l (\zeta)$ is positive. 
This means for modes satisfying the junction condition \eq{JCR}, 
$l$ is smaller larger than $\mu-1$, which gives
\begin{eqnarray}
\lambda_\rho>0, \ \ \ \ \nu> \frac{d-1}{2}.
\label{lowerboundl}
\end{eqnarray}
For the brane bending mode, $l$ is $\mu-2$, 
which satisfies $\mu-l-1\ge 0$. 
Thus, the brane bending mode exists wherever the brane is, and no other modes with the same $\lambda_\rho$ of the brane bending mode exist.

%%%%%%%%%%%%%%%%%%%%%%%%%%%%%%%%%%%%%%%%%%%%%%%%%%%%%%%%%%%%%%%%%
\subsection{Massive tensor on AdS boundary}
%%%%%%%%%%%%%%%%%%%%%%%%%%%%%%%%%%%%%%%%%%%%%%%%%%%%%%%%%%%%%%%%%
Here, we discuss the tensor $Y^{(i)}_{\mu\nu}$,
which is a symmetric tensor on the $d$-dimensional AdS spacetime
\begin{eqnarray}
d \tilde s ^2 &=& \gamma_{\mu\nu} dx^\mu dx^\nu 
\nonumber \\
&=&
\frac{1}{y^2} \left( dy^2 + \eta_{ij} dq^i dq^j\right),
\end{eqnarray}
and satisfies Eq.~\eq{eqy} and the transverse-traceless conditions
\begin{eqnarray}
\tilde \nabla^{\mu}Y^{(i)}_{\mu\nu}=0, \qquad Y^{(i)\mu}{}_{\mu}=0,
\label{TTconst}
\end{eqnarray}
which can be written as
\begin{eqnarray}
&&\pd^i Y_{iy}= -\pd_y Y_{yy} + \frac {d-2}{y} Y_{yy},\label{TV1}\\
&&\pd^j Y_{ji}= -\pd_y Y_{yi} + \frac {d-2}{y} Y_{yi},\label{TV2}\\
&&Y^i{}_{i}= - Y_{yy}.\label{TR}
\end{eqnarray}
With these conditions, Eq.~\eq{eqy} is written as 
\begin{eqnarray}
&&\Box_0Y_{yy} - \frac{d-2}{y} \pd_y Y_{yy} - \frac{\lambda_\rho}{y^2} Y_{yy} =0, \label{Beq1}\\
&&\Box_0Y_{yi} - \frac{d-4}{y} \pd_y Y_{yi} - \frac{d-2+\lambda_\rho}{y^2} Y_{yi} - \frac{2}{y}\pd_i Y_{yy} =0, \label{Beq2}\\
&&\Box_0Y_{ij} - \frac{d-6}{y} \pd_y Y_{ij} - \frac{2d-6+\lambda_\rho}{y^2} Y_{ij} 
- \frac{2}{y}\left( \pd_i Y_{yj}+\pd_j Y_{yi}\right) + \frac{2}{y^2}Y_{yy}\eta_{ij} =0,\label{Beq3}
\end{eqnarray}
where $\Box_0$ is the D'Alembertian of the flat metric, that is, 
\begin{eqnarray}
\Box_0:=\partial_y^2+\partial_i\partial^i.
\label{Box0}
\end{eqnarray}

\begin{comment}
Let us count the number of degrees of freedom with fixing $\lambda_\rho$. 
The number of components of symmetric 2-tensor in $d$-dimensional space is $d(d+1)/2$. 
We have transverse-traceless condition \eq{TTconst}, which number is $(d+1)$. 
Therefore, the number of dynamical degrees of freedom is $d(d+1)/2-(d+1)= (d+1)(d-2)/2$. 
\end{comment}
By virtue of the symmetry in the $(d-1)$-dimensional spacetime labeled by $(i,j,\cdots)$, 
these dynamical degrees of freedom are decomposed by helicity. 
$Y_{yy}$ is expressed only by helicity-0. 
$Y_{yi}$ and $Y_{ij}$ are decomposed as 
\begin{eqnarray}
&& Y_{yi}= \partial_i \phi + V_i, \\
&& Y_{ij}= \partial_i \partial_j \psi + \eta_{ij} \chi + \partial_i W_j + \partial_j W_i + M_{ij}
\end{eqnarray}
with 
\begin{eqnarray}
\partial^i  V_i=\partial^i  W_i=0 , \qquad \partial^i M_{ij} =0, \qquad M^i{}_i=0.
\end{eqnarray}
Here, $\phi$, $\psi$ and $\chi$ are helicity-0 modes, $V_i$ and $W_i$ are helicity-1 modes, and $M_{ij}$ is a helicity-2 mode.
Equations \eq{Beq1}-\eq{Beq3} can be solved separately for each helicity. 
The transverse-traceless conditions \eq{TV1}-\eq{TR} give the relations among $Y_{yy}$, $\phi$, $\psi$ and $\chi$ and between $V_i$ and $W_i$, 
and we will see that each number of dynamical helicity-0, 1 1nd 2 modes is one. 
Let us derive the solution for each helicity.

\paragraph{Helicity-0}
Since a helicity-0 mode includes $Y_{yy}$ component, we first solve Eq.~\eq{Beq1}. 
The solution is obtained as 
\begin{eqnarray}
Y_{yy}^{(0)}= y^{\frac{d-1}{2}} \left[ b_0 J_{\nu} (\sqrt{-\bs{k}^2} y) +c_0 Y_{\nu} (\sqrt{-\bs{k}^2} y) \right]\left[ e^{i \bs{k \cdot x}}+\mbox{c.c.}\right],
\label{Yyy0}
\end{eqnarray}
where $b_0$ and $c_0$ are constants and $\nu$ is defined in Eq.~\eq{lmunu}. 
The decaying solution would be chosen, that is, we set $c_0$ to be zero. 
The helicity-0 component of $Y_{yi}$ is expressed only by $\phi$ and it is obtained by Eq.~\eq{TV1} as
\begin{eqnarray}
Y_{yi}^{(0)} = \frac{\partial_i}{\bs{k}^2} \left( \partial_y Y_{yy}^{(0)} - \frac{d-2}{y}Y_{yy}^{(0)} \right).
\label{Yyi0}
\end{eqnarray}
Similarly, $Y_{ij}$ components are obtained from Eqs.~\eq{TV2} and \eq{TR} as
\begin{eqnarray}
&&Y_{ij}^{(0)} = \frac{1}{\bs{k}^2} \partial_i\partial_j Y_{yy}^{(0)} +
\frac{1}{(\bs{k}^2)^2}\left((d-1) \partial_i\partial_j + \bs{k}^2 \eta_{ij}  \right)\nonumber \\
&&\hspace{30mm}\times
\left[-\frac{1}{y}\partial_y Y_{yy}^{(0)} + \left((d-1)+\frac{\lambda_\rho}{d-2}\right)\frac{1 }{y^2}Y_{yy}^{(0)} \right].
\label{Yij0}
\end{eqnarray}
These solution \eq{Yyi0} and \eq{Yij0} are consistent with Eqs.~\eq{Beq2} and \eq{Beq3}.

The asymptotic behavior of $Y_{yy}$ in the boundary limit $y\to 0$ becomes 
\begin{eqnarray}
Y_{yy}^{(0)}= b_0 y^{\frac{d-1}{2}} J_{\nu} (\sqrt{-\bs{k}^2} y) \left[ e^{i \bs{k \cdot x}}+\mbox{c.c.}\right] \sim y^{\frac{d-1}{2} + \nu}
.
\label{estimate0yy}
\end{eqnarray}
Then, the asymptotic behavior of other components are estimated as%
\footnote{
In this estimate, nontrivial cancellation is not taken into account. 
We will see soon later that, with a special value of $\lambda_\rho$, the cancellation occurs.
}
\begin{eqnarray}
Y_{yi}^{(0)}\sim y^{\frac{d-3}{2} + \nu}, \qquad Y_{ij(T)}^{(0)}\sim y^{\frac{d-1}{2} + \nu}, \qquad Y_{(ij)}^{(0)}\sim y^{\frac{d-5}{2} + \nu},\label{estimate0yi}
\end{eqnarray}
where $Y_{ij(T)}^{(0)}$ and $Y_{(ij)}^{(0)}$ are the trace and traceless parts of $Y_{ij}^{(0)}$, respectively.

\paragraph{Helicity-1}
The helicity-1 mode has no $Y_{yy}$ component and thus Eq.~\eq{Beq1} is trivially satisfied. 
Equation \eq{Beq2} gives a solution
\begin{eqnarray}
Y_{yi}^{(1)}= y^{\frac{d-3}{2}} \left[ b_1 J_{\nu} (\sqrt{-\bs{k}^2} y) +c_1 Y_{\nu} (\sqrt{-\bs{k}^2} y) \right]\left[ e^{i \bs{k \cdot x}}+\mbox{c.c.}\right] E_i(\bs{k},s),
\end{eqnarray}
where $b_1$ and $c_1$ are constants
and $E_i(\bs{k},s)$ is basis vectors satisfying
\begin{eqnarray}
\bs{k}^i E_i(\bs{k},s) =0, \qquad E^i(\bs{k},s) E_i(\bs{k},s') = \delta^{ss'}. 
\label{basis1}
\end{eqnarray}
Here, $s$ is the label representing the different basis vectors. 
The number of its degrees of freedom is $(d-2)$ because the index $i$ is for $(d-1)$-dimensional space and we have the transverse constraint, the first equation of Eq.~\eq{basis1}.  
It may be required to choose the decaying mode, that is, $c_1=0$.
The helicity-1 mode of $Y_{ij}$ components is obtained from Eq.~\eq{TV2} as
\begin{eqnarray}
Y_{ij}^{(1)} = \frac{1}{\bs{k}^2} \left(\partial_y - \frac{d-2}{y} \right) \left(\partial_i Y_{yj}^{(1)}+ \partial_j Y_{yi}^{(1)} \right).  
\label{Yij1}
\end{eqnarray}
This solution satisfies Eq.~\eq{Beq3}.

The asymptotic behavior of each component in the limit $y\to 0$ is estimated as
\begin{eqnarray}
Y_{yi}^{(1)}\sim y^{\frac{d-3}{2} + \nu},  \qquad Y_{ij}^{(1)}\sim y^{\frac{d-5}{2} + \nu}.
\end{eqnarray}

\paragraph{Helicity-2}
For helicity-2 modes,%
\footnote{
Note that, for $d=3$, helicity-2 modes does not exist.
}
$Y_{yy}$ and $Y_{yi}$ components vanish. 
Only Eq.~\eq{Beq3} is nontrivial and it gives 
\begin{eqnarray}
Y_{ij}^{(2)}= y^{\frac{d-5}{2}} \left[ b_2 J_{\nu} (\sqrt{-\bs{k}^2} y) +c_2 Y_{\nu} (\sqrt{-\bs{k}^2} y) \right]\left[ e^{i \bs{k \cdot x}}+\mbox{c.c.}\right] E_{ij}(\bs{k},s),
\end{eqnarray}
where $b_1$ and $c_1$ are constants, 
and $E_i(\bs{k},s)$ is basis of symmetric tensor satisfying
\begin{eqnarray}
\bs{k}^i E_{ij}(\bs{k},s) =0, \qquad  E^i{}_i(\bs{k},s) =0, \qquad E^{ij}(\bs{k},s) E_{ij}(\bs{k},s') = \delta^{ss'}. 
\label{basis2}
\end{eqnarray}
Here, $s$ is the label for different basis of helicity-2 modes.
The number of its degrees of freedom can be counted as follows;
$E_{ij}$ is a symmetric tensor in the $(d-1)$ dimensional spacetime. 
We have the transverse and traceless conditions, which are the first and the second equations of 
Eq.~\eq{basis2}, respectively.
Then the dynamical number of degrees of freedom for a helicity-2 mode is $d(d-1)/2-(d-1)-1= d(d-3)/2$. 
The decaying modes would be chosen, that is, $c_2=0$, and its asymptotic behavior is 
\begin{eqnarray}
Y_{ij}^{(2)}\sim y^{\frac{d-5}{2} + \nu}.
\end{eqnarray}

\vspace{5mm}

The degrees of freedom that we obtained above is 1 for helicity-0, $(d-2)$ for helicity-1 and $d(d-3)/2$ for helicity-2. 
The total is $(d+1)(d-2)/2$, which is consistent with the number of degrees of freedom for $d$-dimensional transverse-traceless tensor, that is, $(d+1)d/2-d-1$ where $(d+1)d/2$ is the number of components of symmetric 2-tensor and $d$ and 1 correspond to the transverse and traceless conditions. 

From the estimates  \eq{estimate0yy} and \eq{estimate0yi}, 
if $Y_{yy}$ component exists, the asymptotic behavior of $Y_{yi}$ seems to be one order lower than that of $Y_{yy}$. 
However, there is an exception; we can see from Eq.~\eq{TV1} that, if $Y_{yy}$ asymptotically behave as $Y_{yy}\sim y^{d-2}$, the corresponding contribution in the right hand side of Eq.~\eq{TV1} is cancelled and the asymptotic behavior of $Y_{yi}$ becomes lower than the naive estimate through Eq.~\eq{estimate0yi}. 
Indeed, it occurs in the case of 
\begin{eqnarray}
\frac{d-1}{2} + \nu = d-2  \quad \Leftrightarrow \quad \lambda_\rho = -(d-2).
\end{eqnarray}
This value of $\lambda_\rho$ corresponds to that of the brane bending mode \eq{BBmodemass}.
As explained in the previous section, the brane bending mode produces a transeverse-traceless mode
even though it is a scalar mode. 
It appears as a helicity-0 mode of spin-2 and it does not excite $Y_{yi}$ and $Y_{yy}$ much, because it is originally scalar. 

Let us see that the brane bending mode corresponds to the helicity-0 mode of spin-2 with $\lambda_\rho = -(d-2)$. 
The brane bending mode $\varphi$ satisfies Eq.~\eq{varphieq}, a solution of which is written as
\begin{eqnarray}
\varphi= y^{\frac{d-1}{2}} \left[ b_\varphi J_{\frac{d+1}{2}} (\sqrt{-\bs{k}^2} y) +c_\varphi Y_{\frac{d+1}{2}} (\sqrt{-\bs{k}^2} y) \right]\left[ e^{i \bs{k \cdot x}}+\mbox{c.c.}\right],
\end{eqnarray}
where $b_\varphi$ and $c_\varphi$ are constants. 
The transverse-traceless tensor generating with $\varphi$ is written as 
\begin{eqnarray}
Y^{(\varphi)}_{\mu\nu}&=& \left( \tilde{\nabla}_\mu\tilde{\nabla}_\nu - \frac{1}{d}\tilde{\Box}\gamma_{\mu\nu} \right)\varphi
\nonumber \\
&=& \left( \tilde{\nabla}_\mu\tilde{\nabla}_\nu - \gamma_{\mu\nu} \right)\varphi ,
\end{eqnarray}
where we used Eq.~\eq{varphieq}. 
Its $(yy)$-component becomes
\begin{eqnarray}
Y^{(\varphi)}_{yy}
&=& \left( \tilde{\nabla}_y \tilde{\nabla}_y - \gamma_{yy} \right)\varphi
\nonumber \\
&=&\left( 
%\partial_y \partial_y
\partial_y^2
+ \frac{1}{y}\partial_y - \frac{1}{y^2} \right)
y^{\frac{d-1}{2}} \left[ b_\varphi J_{\frac{d+1}{2}} (\sqrt{-\bs{k}^2} y) +c_\varphi Y_{\frac{d+1}{2}} (\sqrt{-\bs{k}^2} y) \right]\left[ e^{i \bs{k \cdot x}}+\mbox{c.c.}\right]
\nonumber \\
&=&-\bs{k}^2 y^{\frac{d-1}{2}} \left[ b_\varphi J_{\frac{d-3}{2}} (\sqrt{-\bs{k}^2} y) +c_\varphi Y_{\frac{d-3}{2}} (\sqrt{-\bs{k}^2} y) \right]\left[ e^{i \bs{k \cdot x}}+\mbox{c.c.}\right].
\end{eqnarray}
One can see that this corresponds to Eq.~\eq{Yyy0} with $\lambda_\rho=-(d-2)$. 
Due to the transverse-traceless conditions, other components are uniquely fixed, 
and thus, the brane bending mode is confirmed to be the helicity-0 mode with $\lambda_\rho=-(d-2)$. 
This behaves $y^{d-2}$ in the limit $y\to0$.

To see the asymptotic behavior of other components, 
we derive their exact forms.
$(yi)$-component is obtained as 
\begin{eqnarray}
Y^{(\varphi)}_{yi}
&=& \tilde{\nabla}_y \tilde{\nabla}_i \varphi
\nonumber \\
&=&\partial_i \left( \partial_y+ \frac{1}{y}\right)
y^{\frac{d-1}{2}} \left[ b_\varphi J_{\frac{d+1}{2}} (\sqrt{-\bs{k}^2} y) +c_\varphi Y_{\frac{d+1}{2}} (\sqrt{-\bs{k}^2} y) \right]\left[ e^{i \bs{k \cdot x}}+\mbox{c.c.}\right]
\nonumber \\
&=& \sqrt{-\bs{k}^2} y^{\frac{d-1}{2}} \left[ b_\varphi J_{\frac{d-1}{2}} (\sqrt{-\bs{k}^2} y) +c_\varphi Y_{\frac{d-1}{2}} (\sqrt{-\bs{k}^2} y) \right] \partial_i \left[ e^{i \bs{k \cdot x}}+\mbox{c.c.}\right].
\end{eqnarray}
This behaves $y^{d-1}$ in the limit $y\to0$, which is one order higher than that of $Y^{(\varphi)}_{yy}$, that is, $Y^{(\varphi)}_{yi}$ decays faster than $Y^{(\varphi)}_{yy}$.
The trace part of $(ij)$-component can be easily obtained from the traceless condition~\eq{TR}, 
which asymptotic order is the same as that of $Y^{(\varphi)}_{yy}$. 
The traceless part of $(ij)$-component is calculated as 
\begin{eqnarray}
Y^{(\varphi)}_{(ij)}
&=& \left(\tilde{\nabla}_i \tilde{\nabla}_j- \frac1{d-1} \eta_{ij} \eta^{kl} \tilde{\nabla}_k \tilde{\nabla}_l  \right) \varphi
\nonumber \\
&=&
y^{\frac{d-1}{2}} \left[ b_\varphi J_{\frac{d+1}{2}} (\sqrt{-\bs{k}^2} y) +c_\varphi Y_{\frac{d+1}{2}} (\sqrt{-\bs{k}^2} y) \right]
\nonumber \\ 
&& \hspace{20mm} \times
\left( \partial_i \partial_j - \frac{1}{d-1} \eta_{ij} \eta^{kl}\partial_k\partial_l \right)
\left[ e^{i \bs{k \cdot x}}+\mbox{c.c.}\right].
\end{eqnarray}
This behaves $y^d$ in the limit $y\to0$.

\subsection{Holographic stress energy tensor in AdS/BCFT}
%%%%%%%%%%%%%%%%%%%%%%%%%%%%%%%%%%%%%%%%%%%%%%%%%%%%%%%%%%%%%%%%%

Now we would like to study the behavior of holographic stress tensor in the light of the analysis of metric perturbations in the previous subsections. 
The holographic stress energy tensor in AdS$_{d+1}/$BCFT$_d$ can be computed as 
\ba
{\cal T}_{ab}=\frac{d}{16\pi G_N}\lim_{z\to 0}g_{ab}(x,w,z)z^{2-d},
\ea
where $g_{ab}$ is the metric perturbation in the asymptotically Poincare AdS space (\ref{holEM}) and $x$ denotes $d-1$ dimensional coordinate $(x_0,x_1,\ddd,x_{d-2})$,
such that $(\xi_0,\ddd,\xi_{d-2},\xi_{d-1})=(x_0,\ddd,x_{d-2},w)$. 
Note that
this holographic stress tensor satisfies the traceless condition and momentum conservation (\ref{bcbfta}) owing to the Einstein equation \cite{Balasubramanian:1999re,deHaro:2000vlm}.

Then the holographic stress tensor can also be evaluated from the hyperbolic slice metric by using the coordinate transformation (\ref{cortra}) as follows
\ba
{\cal T}_{ab}=\frac{d}{16\pi G_N}\lim_{\rho\to \infty}\left[\left(\frac{e^{\rho}}{2y}\right)^{d-2}h_{ab}(x,y)\right]\Biggr|_{y=w},
\ea
where we identify $\xi_{d-1}(=w)$ with $y$ because we take $\rho\to\infty$ limit.  As we performed in the previous subsections,  
the metric perturbation 
$h_{\mu\nu}$ can be computed by imposing  the Dirichlet boundary condition $h_{\mu\nu}=O(e^{-\rho})$ in the boundary limit $\rho\to\infty$ and the Neumann boundary condition (\ref{beina}) on the EOW brane.

For $d=2$, i.e.\ AdS$_3/$BCFT$_2$, the only possible metric perturbation is the helicity $0$ mode with $\lambda_\rho=0$, 
which is the brane bending mode. This leads to the $O(1)$ metric perturbation in the AdS boundary limit $z\to 0$. This corresponds to the perturbations proportional to ${\cal T}_{++}$ and ${\cal T}_{--}$ in the asymptotically AdS$_3$ metric (\ref{metads}). Thus, the present metric perturbation analysis simply reproduces the result in section \ref{sec:ads3}.

In higher dimensions $d\geq 3$, there are helicity 0, 1 and 2 modes as we have seen.
(Helicity 2 mode does not appear for $d=3$.)
In the BCFT boundary limit $y\to 0$, these modes in general behave as (note $i,j=0,1,\ddd,d-2$)
\ba
&& \mbox{helicity 0}:\ \  Y_{yy}=O(y^{\frac{d-1}{2}+\nu}),\ \ \ Y_{yi}=O(y^{\frac{d-3}{2}+\nu}), \ \ \ Y_{(ij)}=O(y^{\frac{d-5}{2}+\nu}),  \label{scaleyy}  \no
&& \mbox{helicity 1}:\ \  Y_{yy}=0,\ \ \ Y_{yi}=O(y^{\frac{d-3}{2}+\nu}), \ \ \ Y_{(ij)}=O(y^{\frac{d-5}{2}+\nu}),  \label{scaleyya}  \no
&&\mbox{helicity 2}:\ \  Y_{yy}=0,\ \ \ Y_{yi}=0, \ \ \ Y_{(ij)}=O(y^{\frac{d-5}{2}+\nu}),  \label{scaleyyb} 
\ea
where $\nu=\s{\frac{(d-1)^2}{4}+\lambda_\rho}$. The indices $(ij)$ denotes the traceless symmetric part. By imposing the boundary conditions, the value of $\lambda_\rho$ gets quantized. 

As we have noted before, we only allow the values of $\nu$ which satisfy the previous constraint (\ref{lowerboundl}) with one exception, namely the brane bending mode.
In the helicity 0 perturbation, the lowest mode $\lambda_\rho=-(d-2)$, which is the brane bending mode, anomalously appears and this shows the following peculiar scaling different from (\ref{scaleyy}):
\ba
Y_{yy}=O(y^{d-2}),\ \ \  Y_{yi}=O(y^{d-1}),\ \ \ Y_{(ij)}=O(y^d).
\ea
Thus in this lowest mode, both ${\cal T}_{yy}$ and the trace part of ${\cal T}_{ij}$ take finite values, while 
we have ${\cal T}_{wi}={\cal T}_{(ij)}=0$ at the BCFT boundary $w=0$. Notice that this is the only mode which gives non-vanishing 
${\cal T}_{yy}$ and the trace part of ${\cal T}_{ij}$.

For higher modes of the helicity 0
perturbation follows the standard scaling (\ref{scaleyy}). This leads to the 
following behavior of stress tensor near the boundary $w\to 0$:
\ba
{\cal T}_{ww}=O(w^{-\frac{d-3}{2}+\nu}),\ \ \ 
{\cal T}_{wi}=O(w^{-\frac{d-1}{2}+\nu}),\ \ \ 
{\cal T}_{ij}=O(w^{-\frac{d+1}{2}+\nu}).
\ea
Owing to the constraint (\ref{lowerboundl}), the boundary condition $T_{wi}=0$ at $w=0$ is always satisfied as expected for BCFTs.

Similarly, we can read off the stress energy tensor for the helicity 1 and 2 perturbations. Again we find the behavior $T_{wi}=0$ at $w=0$ due to the constraint (\ref{lowerboundl}). In this way, we can confirm the expected boundary condition in BCFTs (\ref{bcbfta}), which basically argues the complete reflection at the boundary of a given BCFT on a half plane $w>0$. Notice also that in the bulk of BCFT, there are $\frac{(d-2)(d+1)}{2}$ independent components of stress tensor after we imposed the conservation and traceless condition. This agrees with the number of independent metric perturbations.  

\subsection{Case with $d=3$}\label{subsecd=3}
As an explicit example, below we focus on $d=3$ and study the perturbation modes in detail. 
The helicity 2 mode is absent at $d=3$.

For $d=3$, $\mu$ is $3/2$ and thus the denominator of the integrand of Eq.~\eq{inteq1} becomes unity and it becomes
%
\begin{comment}
Moreover, the gamma functions are simplified as
\begin{eqnarray}
\frac{ \Ga{l+\mu}}{\Ga{l-\mu+2}}
=\frac{ \Ga{l+\frac{3}{2}}}{\Ga{l+\frac{1}{2}}}
=l + \frac{1}{2}.
\end{eqnarray}
Then, Eq.\eq{inteq1} becomes
\end{comment}
%
\begin{eqnarray}
0&=&\frac{\sqrt{2}  \left(\sin\theta\right)^{-\frac{1}{2} }}{\sqrt{\pi}\Ga{1}}
\int_0^\theta \cos \left(\left(l+\frac12\right)\psi \right) d\psi 
\nonumber \\
&=&
\frac{\sqrt{2}  \left(\sin\theta\right)^{-\frac{1}{2}} \sin \left(\left(l+\frac12\right)\theta \right)}
{\sqrt{\pi}\left(l+\frac12\right)}
\end{eqnarray}
and the solution for $l$ is 
\begin{eqnarray}
l= \frac{\pi}{\theta} a -\frac12 \qquad (a \in \N).
\label{l_d=3}
\end{eqnarray}
(Here we use the definition that $\N$ does not include zero.)
%
\begin{comment}
For normalizability, $l$ should satisfies
\begin{eqnarray}
l> \mu-1 = \frac{1}{2}.
\end{eqnarray}
Since $0<\theta<\pi$, $l$ satisfies the above inequality for any $a(\in \Z)$.
\end{comment}
%
From Eq.~\eq{lmunu}, $\lambda_\rho$ is written as
\begin{eqnarray}
\left(m_{KK}^2=\right)\lambda_\rho = \left( \frac{2l+1}{2}\right)^2-1 = \left( \frac{\pi}{\theta} a \right)^2-1.
\label{KKmass}
\end{eqnarray}
Since $\theta$ takes a value in the range $0<\theta<\pi$, $\lambda_\rho$ is positive, 
which is consistent with Eq.~\eq{lowerboundl}.

 The previous analysis leads to the expression of perturbative solutions:
\ba
h_{ab}=\s{\cosh\rho}\cdot P^{-\frac{3}{2}}_{\s{1+\lambda_\rho}-\frac{1}{2}}(\tanh\rho)\cdot Y_{ab}(\tau,x,y),
\ea
where we set $(x^0,x^1,x^2)=(\tau,x,y)$. Note that the Legendre function 
%$Q$ 
behaves like 
\ba
P^{-\frac{3}{2}}_{\s{1+\lambda_\rho}-\frac{1}{2}}(\tanh\rho)=  O(e^{-\frac{3\rho}{2}}),\ \ \ (\rho\to\infty),
\ea
so that it fits nicely with the Dirichlet boundary condition. 
The Neumann boundary condition leads to the quantization (\ref{KKmass}) in the $\rho$ direction, which leads to  
\ba
\nu=\s{1+\lambda_\rho}=\frac{\pi a}{\theta},\ \ \ (a=1,2,\ddd),
\label{1+lambda}
\ea
where we also impose the condition (\ref{lowerboundl}) i.e.\ $\nu>1$.
Here $\theta$, which takes values in the range $0<\theta<\pi$, is the angle of the EOW brane. 
$\sigma>0\ (\rho_*<0)$, 
$\sigma=0\ (\rho_*=0)$ and $\sigma<0\ (\rho_*>0)$ correspond to 
$\theta>\frac{\pi}{2}$, $\theta=\frac{\pi}{2}$, and $\theta<\frac{\pi}{2}$, respectively.

We have the other extra mode, the brane bending mode. 
For the brane bending mode, we have $\s{1+\lambda_\rho}=0$, which corresponds to $a=0$ case of Eq.~\eq{1+lambda}. 
Note again that the brane bending mode generates only helicity 0 mode, and 
helicity 1 modes are absent for $a=0$.
Below we study the behavior of the corresponding stress tensor for each value of $a$.

Let us start with the anomalous mode of the helicity 0 perturbation at $a=0$ or equally $\lambda_\rho=-1$, which is the brane bending mode. 
For any values of tension or angle $\theta$, we find
\ba
&&\s{\cosh\rho}\cdot P^{-\frac{3}{2}}_{-\frac{1}{2}}(\tanh\rho)\no
&& \propto \left[\cosh\rho\left(1+2\sinh\rho\arctan\left[\tanh\frac{\rho}{2}\right]\right)
-\frac{\pi}{2}\sinh\rho\cosh\rho\right] \sim_{\rho\to\infty} e^{-\rho},\no
&& Y_{yy}\propto y,\ \ \ \ Y_{yi}\propto y^2,\ \ \ \ Y_{ij(T)}\propto y,\ \ \ \ Y_{(ij)}\propto y^3,
\ \ \ (y\to 0),
\label{azero}
\ea
where $Y_{ij(T)}$ shows the trace part of $Y_{ij}$.
For this mode, which corresponds to the brane bending mode, 
the stress tensor behaves in the boundary limit $w\to 0$ as follows 
\ba
\mbox{Brane bending mode\ ($a=0)$}: {\cal T}_{yy}=\mbox{finite},\ \ \ {\cal T}_{wi}=0,\ \ \  {\cal T}_{ij}=\mbox{finite}.  \label{bendiq}
\ea
In particular, when $Y_{yy}$ does not depend on $x^i$, we can set $Y_{yi}=Y_{(ij)}=0$.
In this case the solution describes the constant stress energy tensor 
which satisfies $T_{ww}=-2T_{\tau\tau}=-2T_{xx}$ with all other components vanishing. This corresponds to the $U_1$ and $U_2$ mode (\ref{pertsolb})
in the appendix, by turning on the $\vp(x)$ mode in (\ref{vpmode}).
As an illustration, we show the time evolution of a wave packet of the brane bending mode and corresponding ${\cal T}_{yy}$ in Fig.~\ref{Fig:wave_phi} (see also section~\ref{sec:reflection}).

For $a\geq 1$, the helicity 0 mode looks like
\ba
&&\s{\cosh\rho}\cdot P^{-\frac{3}{2}}_{\frac{\pi a}{\theta}-\frac{1}{2}}(\tanh\rho)\sim_{\rho\to\infty} e^{-\rho},\no
&& Y_{yy}\propto y^{\frac{\pi a}{\theta}+1},\ \ \ \ Y_{yi}\propto y^{\frac{\pi a}{\theta}},\ \ \ \ Y_{(ij)}\propto y^{\frac{\pi a}{\theta}-1}.\ \ \ (y\to 0).
\label{aoneone}
\ea
This leads to the following behavior of stress tensor in the boundary limit $w\to 0$:
\ba
\mbox{helicity 0} : {\cal T}_{ww}\propto w^{\frac{\pi a}{\theta}},\ \ \ 
{\cal T}_{wi}\propto w^{\frac{\pi a}{\theta}-1},\ \ \ 
{\cal T}_{ij}\propto w^{\frac{\pi a}{\theta}-2}.\ \ \  (a=1,2,3,\ddd).   \label{delzoe}
\ea
We can analyze the helicity 1 similarly. Their stress tensor behaves like
\ba
&& \mbox{helicity 1} :  {\cal T}_{ww}=0,\ \ \  {\cal T}_{wi}\propto w^{\frac{\pi a}{\theta}-1},\ \ \ 
{\cal T}_{ij}\propto w^{\frac{\pi a}{\theta}-2}.\ \ \    \label{delzoee}
\ea
Note that owing to the condition (\ref{lowerboundl}) i.e.\ $\frac{\pi a}{\theta}>1$, we find $T_{wi}=0$ at the boundary $w=0$ for both the helicity 0 result (\ref{delzoe}) and the helicity 1 one (\ref{delzoee}), as expected.

It is also intriguing to look into the tensionless case $\sigma=0$ i.e.\ $\theta=\frac{\pi}{2}$ specially. In this case, both the helicity 0 mode (\ref{delzoe}) and helicity 1 one  (\ref{delzoee})  at $a=1$ leads to finite values of stress tensor ${\cal T}_{ij}$ in addition to the brane bending mode (\ref{bendiq}). Moreover, for any $a$, the power of $w$ of the stress tensor in the limit $w\to 0$ is always integer at $\sigma=0$.
All of the above analysis at $d=3$ perfectly matches with the direct perturbation analysis in appendix A as in (\ref{compareeqa}).

%%%%%%%%%%%%%%%%%%%%%%%%%%%%%%%%%%%%%%%%%%%%%%%%
\section{Scalar field excitations in higher dimensional AdS/BCFT}
\label{sec:SF}
%%%%%%%%%%%%%%%%%%%%%%%%%%%%%%%%%%%%%%%%%%%%%%%

In the previous section, we have studied metric perturbations in the AdS/BCFT setup, where the EOW brane plays a very important role. As a next step, we would like to analyze the dynamical properties of matter fields in the presence of EOW brane. In this section, as such an example, we study the free massive scalar fields in AdS$_{d+1}$ with the EOW brane and their reflection at the brane.

%%%%%%%%%%%%%%%%%%%%%%%%%%%%%%%%%%%%%%%%%%%%%%%
\subsection{Massive scalar field}
%%%%%%%%%%%%%%%%%%%%%%%%%%%%%%%%%%%%%%%%%%%%%%%

%We consider a massive scalar field in the spacetime (\ref{metric}):
A massive scalar field in higher dimensional AdS/BCFT setting can be treated in a parallel manner to the gravitational wave discussed in the previous section.
The equation of motion of a massive scalar field
on the spacetime~\eq{metric ansatz}
is given by
\begin{equation}
0=\left(\square - m^2\right)\phi(\rho,x^\mu)
=
\frac{1}{\cosh^2\rho}
\left[
y^2
%\left( -\phi_{,tt}+\phi_{,yy}+\bm{\nabla}^2\phi\right)
\Box_0 
-(d-2)y \phi_{,y}
\right]
+\phi_{,\rho\rho} + d\, \tanh\rho \phi_{,\rho}
-m^2 \phi~,
\label{EoM}
\end{equation}
where $\Box_0$ is the flat D'Alembertian (\ref{Box0}).
This equation may be solved using separation of variable
by writing the general solution as 
$\phi={\mathcal R}(\rho)Y(x^\mu)$, where ${\mathcal R}(\rho)$ and $Y(x^\mu)$ obey
\begin{gather}
{\mathcal R}_{,\rho\rho} + d\, \tanh \rho {\mathcal R}_{,\rho}
-m^2 {\mathcal R} = -\frac{\lambda_\rho}{\cosh^2(\rho)}  {\mathcal R}~,
\label{rhoeq}
\\
%y^2 Y_{,yy}-(d-2)y Y_{,y}+\left[y^2(\omega^2 - \mathbf{k^2}) - \lambda_\rho \right]Y
\Box_0 Y - \frac{d-2}{y} \partial_y Y
- \frac{\lambda_\rho}{y^2} Y
=0~.
\label{yeq}
\end{gather}
The ${\mathcal R}$ equation is modified from Eq.~\eq{eqR} just by the mass term,
and the $Y$ equation is the same as Eq.~\eq{Beq1} for $Y_{yy}$.

Because the separated equations \eq{rhoeq}, \eq{yeq} are similar to those for the gravitational perturbations,
their general solutions can be immediately found from the results in the previous section.
Introducing $\zeta = \tanh\rho$ as we did in Eq.~\eq{zetadef},
we find that
the solution to Eq.~\eq{rhoeq} for the bulk direction $\rho$ 
satisfying the Dirichlet boundary condition at the AdS$_{d+1}$ boundary $\zeta \to 1$
is given by Eq.~\eq{P-}, where the index $\mu$ of the associated Legendre functions is promoted to
\begin{equation}
\mu = \frac12 \sqrt{d^2 + 4 m^2}~,
\end{equation}
while the other indices $l,\nu$ are the same as those in Eq.~\eq{lmunu}.
$\mu$ is guaranteed to be a real number if the Breitenlohner-Freedman bound $m^2 \geq -\frac14 d^2$ is satisfied.
Equation~\eq{yeq} for the AdS$_d$ direction is the same as Eq.~\eq{Beq1}, then its solution is given by Eq.~\eq{Yyy0}.
If we assume the Dirichlet condition at the AdS boundary $y=0$
on each AdS$_d$ slice $\rho=\text{const.}$, the decaying (normalizable) solution with $c_0=0$ will be taken as the solution.

\begin{figure}[htbp]
\centering
\includegraphics[width=8cm]{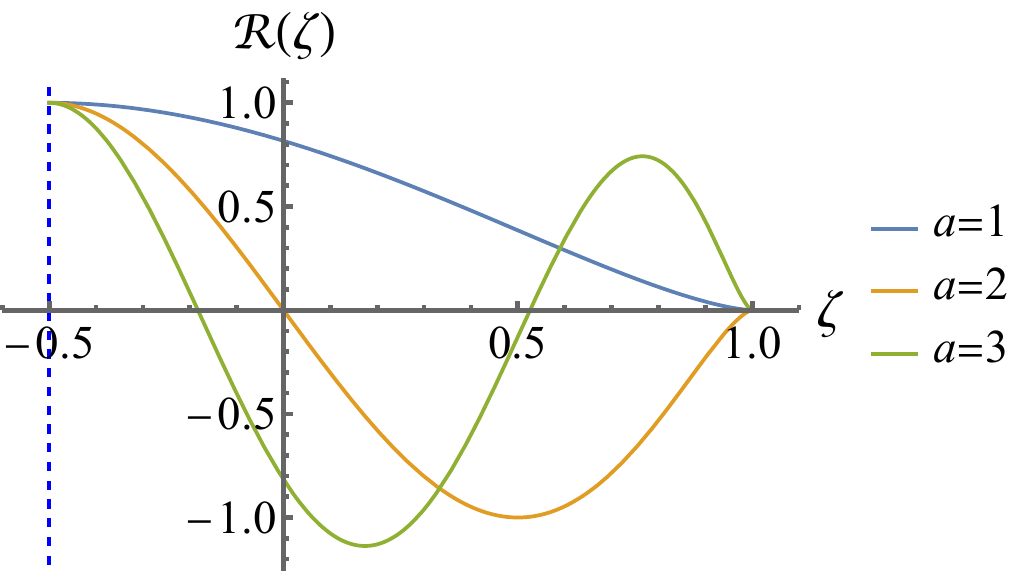}
\caption{The first three eigenfunctions ${\mathcal R}(\zeta)$ for $d=3, m^2=0$ and the brane located at $\zeta=\zeta_*=-1/2$, corresponding to $\theta=2\pi/3$. ${\mathcal R}(\zeta)$ is normalized so that ${\mathcal R}(\zeta_*)=1$.}
\label{Fig:R}
\end{figure}

We assume that the scalar field minimally couples with the EOW brane,\footnote{Later in section \ref{sec:one-point}, we will introduce a non-trivial coupling between the scalar field and the EOW brane so that the dual scalar operator has a non-zero expectation value. Here we assume that a scalar field whose dual operator has a vanishing one point function i.e.\ $a_0=0$.} and then the scalar field satisfies the Neumann boundary condition $\phi_{,\zeta}\hat =0$ at the brane located at $\zeta=\zeta_*$.
%This condition is satisfied by discrete values of $\lambda_\rho$.
%We also impose the Dirichlet boundary condition $\phi\to 0$ at the boundary of AdS$_{d+1}$.
The separation constant $\lambda_\rho$ is fixed by solving Eq.~\eq{rhoeq} as an eigenvalue equation imposing the Dirichlet condition at the AdS$_{d+1}$ boundary $\zeta=1$ and the Neumann condition at the brane $\zeta = \zeta_*$, as argued in Sec.~\ref{sec:JC}.
Using the solution~\eq{P-}, the condition to fix $\lambda_\rho$ in terms of $\zeta_*$ is given by
\begin{equation}
0\hat=\frac{d}{d\zeta} \left(1-\zeta^2\right)^{\frac{d}{4}}  P^{-\mu}_l (\zeta)
%\nonumber \\ &=&
 =
 -\frac{d}{2}\left(1-\zeta^2\right)^{\frac{d}{4}-1} \zeta P^{-\mu}_l (\zeta)  +\left(1-\zeta^2\right)^{\frac{d}{4}}  \frac{d}{d\zeta} P^{-\mu}_l (\zeta)~.
%\nonumber \\ 
%&=&
%-\left(1-\zeta^2\right)^{\frac{d-2}{4}}  P^{-\mu+1}_l (\zeta).
%\label{ddzPQ}
\label{ddzPQ-gen}
\end{equation}
This expression can be further simplified and reduces to Eq.~\eq{ddzPQ} for a massless scalar field $m^2=0$, 
while for general $m^2\neq 0$ we need to use Eq.~\eq{ddzPQ-gen}.
Once $\lambda_\rho$ is fixed, then $Y(x^\mu)$ is given by Eq.~\eq{Yyy0} with $c_0=0$.
As an example, in Fig.~\ref{Fig:R} we show the first three eigenfunctions for $d=3, m^2=0$ and the brane located at $\zeta=\zeta_*=-1/2$, corresponding to the cases $a=1,2,3$ of Eq.~\eq{l_d=3}.

Any solution 
$\phi(\rho,x^\mu)$
can be expressed in terms of the general solutions ${\mathcal R}(\rho), Y(x^\mu)$ constructed above. In principle, a given initial data of the scalar field distribution may be decomposed into mode solutions given above, and then the time dependence scalar field will be described analytically.

\subsection{Scalar field reflection at the boundary}
\label{sec:reflection}

As an example of nontrivial dynamics realized in our setup,
in this section
let us construct bulk dual of waves that propagate toward and reflect at the boundary in a BCFT.
We model such waves by normalizable modes of a bulk scalar field, which may be identified as the expectation value of a scalar operator in the BCFT.
A reflecting solution can be constructed by taking an appropriate superposition of the mode solutions constructed in the previous section, but instead of that we directly solve Eq.~(\ref{EoM}) as a time-evolution equation.

Expressing a solution as $\phi={\mathcal R}(\rho)Y(t,y)e^{i\mathbf{k}\cdot\mathbf{x}}$,
the scalar field equation of motion~(\ref{EoM}) is separated into the $\rho$ equation~(\ref{rhoeq}) and an equation for $Y(t,y)$ given by
\begin{equation}
y^2 
\left(
- Y_{,tt}
+Y_{,yy}
\right)
-(d-2)y Y_{,y}
-\left(
\lambda_\rho
+\mathbf{ k}^2 y^2
\right)
Y
=0~.
\label{yteq}
\end{equation}
Once an initial data for $Y(t,r)$ on a time slice is given, Eq.~(\ref{yteq}) fixes the time evolution of $\phi$ for given parameters $d, \mathbf{k}$ and $\lambda_\rho$.
%As explained in section~\ref{Sec:rho}, 
We consider solutions satisfying the Dirichlet condition at the boundary of AdS$_{d+1}$, ${\mathcal R}(\rho=\infty) = 0$, which corresponds to the normalizable mode. Then the eigenvalue $\lambda_\rho$ is fixed by specifying the brane position $\rho_*$ and the mode number for ${\mathcal R}(\rho)$.
Because (the real part of) $Y(t,r)e^{i\mathbf{k}\cdot\mathbf{x}}$ is proportional to the normalizable mode in the $\rho$ direction, its value may be interpreted as the expectation value $\langle\phi\rangle$ for a scalar operator in the BCFT realized on AdS$_d$.

We numerically solve Eq.~(\ref{yteq}) to construct time-dependent solutions $Y(t,y)$.
%, which will be proportional to the expectation value of the scalar operator when $R(\rho)$ is taken to be a normalizable mode in AdS$_{d+1}$.
In order to conduct a stable numerical calculation,
it is useful to introduce the ingoing Eddington-Finkelstein coordinate $v=t-y$, with which the metric~(\ref{metric ansatz}) is expressed as
\begin{equation}
ds^2 = d\rho^2 + \cosh^2\rho\left(
\frac{-dv^2 - 2 dv dy + d\mathbf{x}^2}{y^2}
\right)
\label{metricEF}
\end{equation}
and Eq.~(\ref{yteq}) becomes an equation for $Y(v,y)$ given by
%\footnote{Numerical calculation using \texttt{NDSolve} becomes more stable in the EF coordinate compared to solving Eq.~(\ref{yteq}) using $t$ as the time coordinate.}
\begin{equation}
y^2 
\left(
- 2 Y_{,vy}
+Y_{,yy}
\right)
-(d-2)y 
\left(
-Y_{,v} + Y_{,y}
\right)
-\left(
\lambda_\rho
+ \mathbf{k}^2 y^2
\right)
Y
=0~.
\label{yveq}
\end{equation}
In this coordinate, ingoing and outgoing null rays are expressed by $v=\text{const}$ and $v+2y=\text{const}$, respectively.

%To construct a time-dependent solution,
The setting for the numerical solution construction is as follows.
We take an initial time slice given by $0\leq y \leq y_\text{fin}$ at $v=0$, 
and solve Eq.~(\ref{yveq}) in the domain of dependence of the initial time slice, which is given by
$v\in[0,v_\text{fin}]$ and $y\in[0,y_\text{fin}-\frac12 v]$ with $v_\text{fin} = 2 y_\text{fin}$.
On the initial time slice, we specify the profile of $Y(v,y)$ as
\begin{equation}
Y(v=0,y) = \exp\left[
-\frac12 \left(\frac{y-y_0}{s}\right)^2
\right]~,
\end{equation}
which describes a gaussian wave packet in the bulk that propagates toward the boundary of AdS$_d$ at $y=0$.
We also need to specify the boundary condition at 
the AdS$_d$ boundary at $y=0$.
We assume that $Y(y)$ behaves as the normalized mode in AdS$_d$, hence we impose the Dirichlet condition $Y(y=0) = 0$. In the numerical calculation, for simplicity we instead impose this condition at $y=\epsilon>0$, where the cutoff $\epsilon$ is taken to be sufficiently small.

\begin{figure}[htbp]
\centering
\subfigure[Time evolution of $Y$]{\includegraphics[height=5cm]{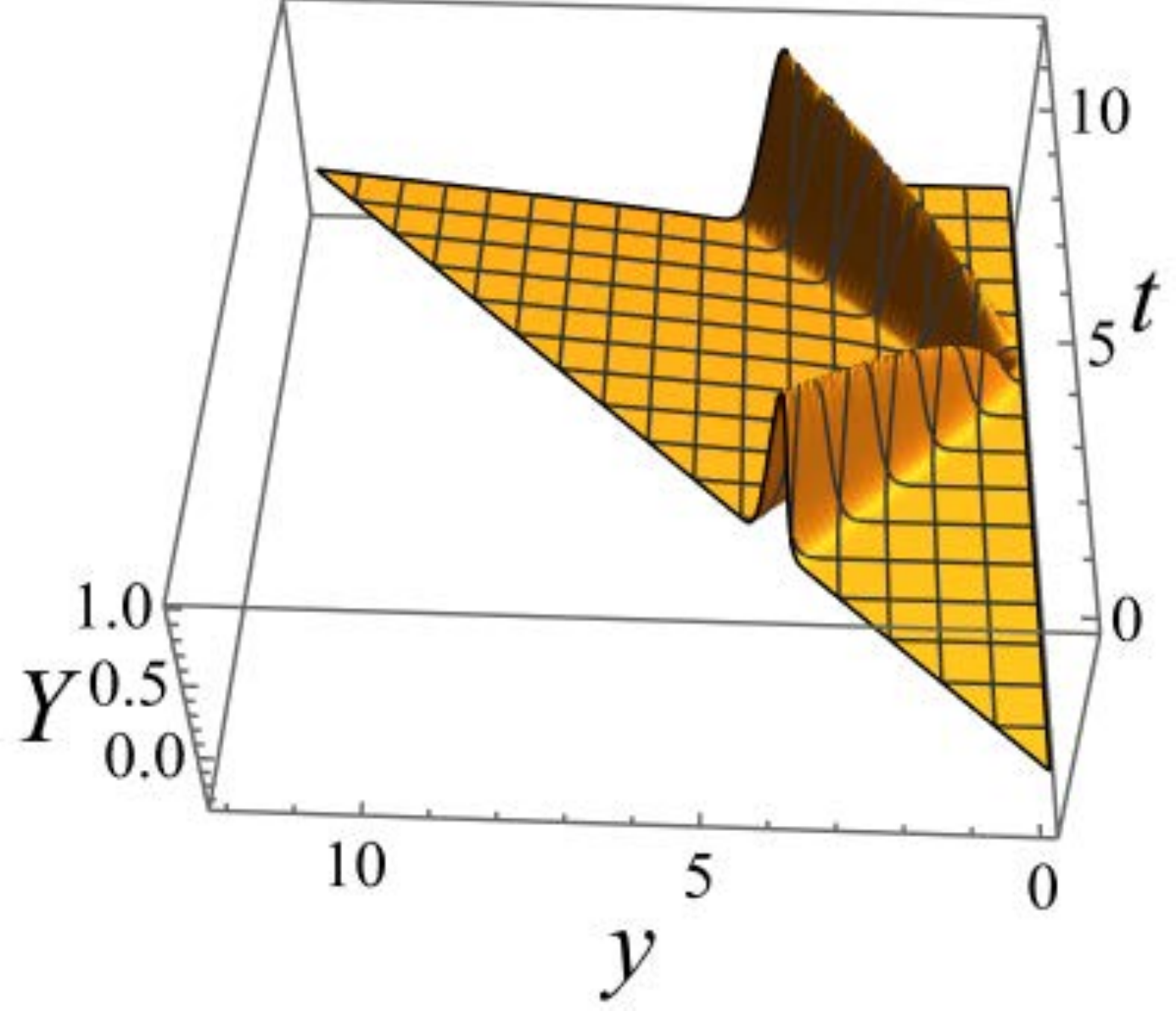}}
\hspace{0.5cm}
\subfigure[Near reflection point]{\includegraphics[height=5.5cm]{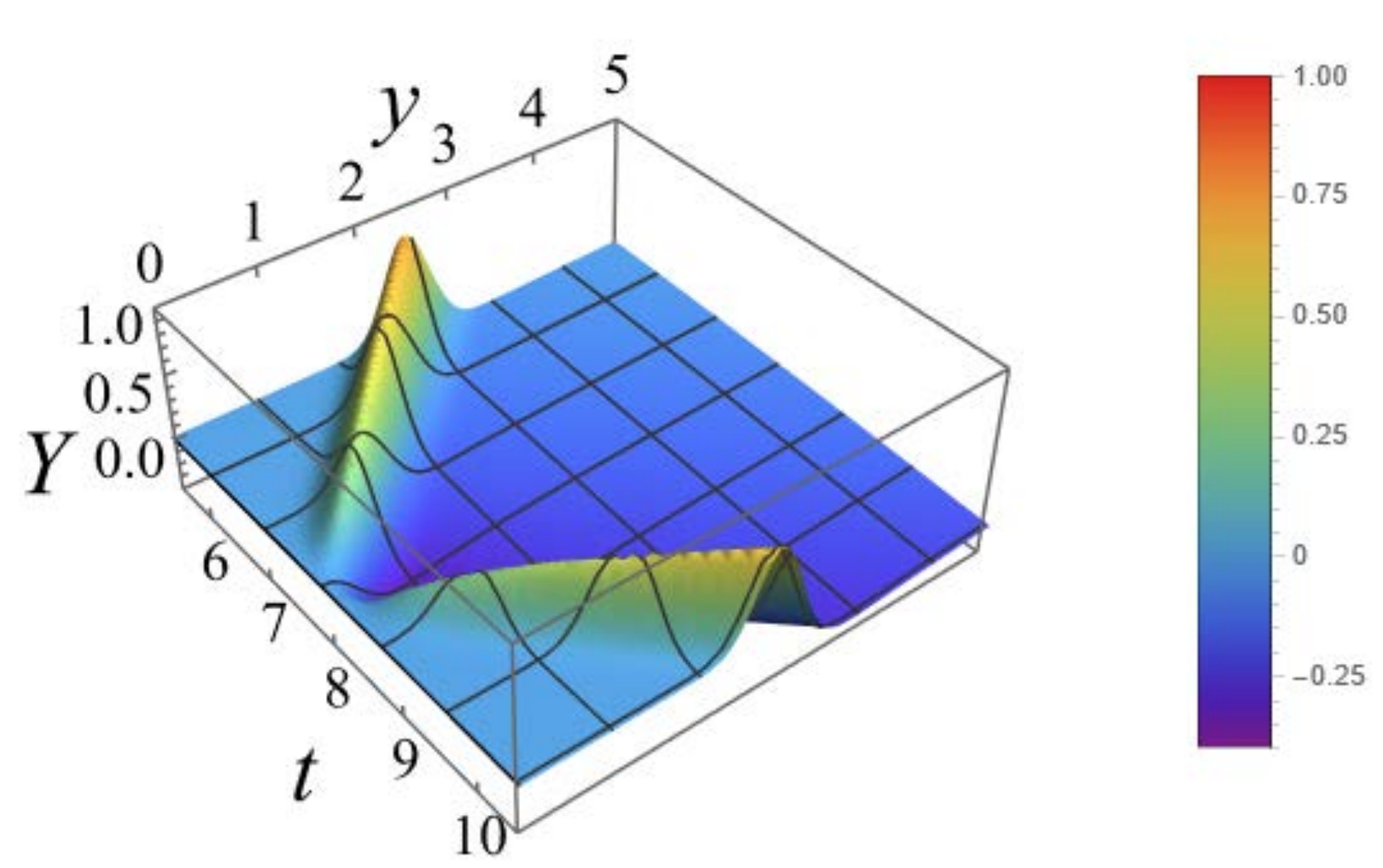}\label{fig:reflection}}
\caption{Numerical solution for $d=3, m^2=0, \mathbf{k}=0, \lambda_\rho= 5/4$ and initial wave packet with $y_0=4, s=1/8$.
In panel (a), only a part of the numerical domain for $0<t<12$ is shown.
Panel (b) shows the solution near the reflection point $(t,y)\simeq (8,0)$.}
\label{Fig:wave}
\end{figure}

In Fig.~\ref{Fig:wave}, we show an example of numerical solution in the case $d=3, m^2=0, \mathbf{k}=0$ for the initial wave packet with $y_0=4, s=1/8$. 
We have chosen ${\mathcal R}(\rho)$ to be the lowest mode ($a=1$) for the brane located at $\zeta_*=-1/2$, for which $\lambda_\rho = 5/4$ and $\theta=2\pi/3$.
For numerical calculation we used $y_\text{fin}=12$ and $\epsilon = 10^{-4}$.
We can observe that the scalar field wave is almost completely reflected by the AdS$_d$ boundary at $y=0$, and it approximately follows null lines given by $t\pm y = \text{const}$.

\begin{figure}[htbp]
\centering
\subfigure[${\mathcal R}^{(\varphi)}(\zeta)$]{\includegraphics[height=4.5cm]{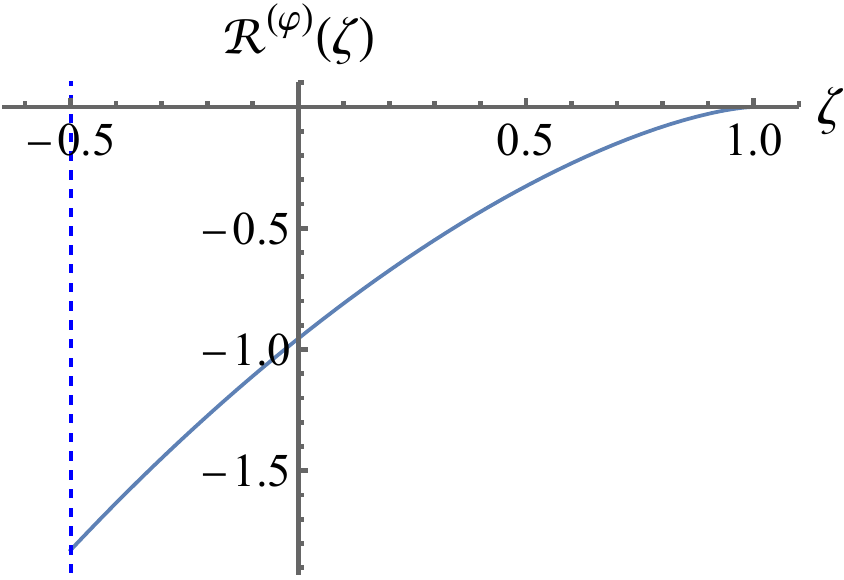}}
\hspace{0.5cm}
\subfigure[$Y_{yy}^{(\varphi)}$ near reflection point]{\includegraphics[height=5.5cm]{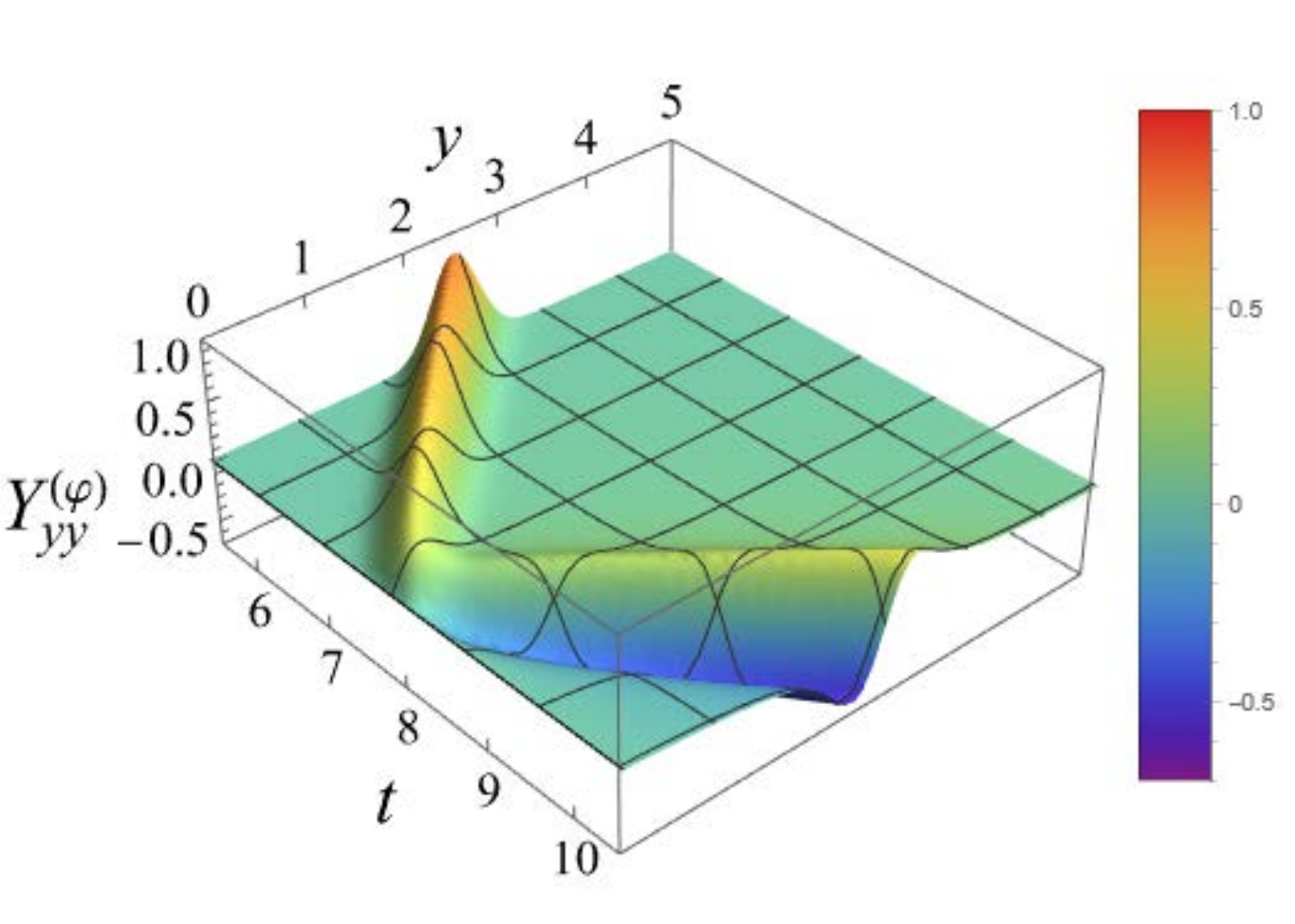}}
\caption{The dynamics of the brane bending mode in $d=3$ for the brane located at $\zeta=\zeta_* = -1/2$.
Panel (a) shows ${\mathcal R}^{(\varphi)}(\zeta)$ analyzed in Sec.~\ref{sec:JC}. Panel (b) shows the time evolution of $Y_{yy}^{(\varphi)}$ with $\mathbf{k}=0$, for which $m^2=0, \lambda_\rho= -1$.
The initial field configuration is the same as that for Fig.~\ref{Fig:wave}.}
\label{Fig:wave_phi}
\end{figure}

The equation governing the $Y_{yy}$ component of the gravitational perturbation, Eq.~\eq{Beq1}, is the same as that for the scalar field \eq{yeq}.
Hence the numerical method introduced above can be straightforwardly applied to the dynamics of $Y_{yy}$.
As an example, in Fig.~\ref{Fig:wave_phi} we show the time evolution of the $h_{yy} = \mathfrak{a}^2 {\mathcal R}^{(\varphi)}(\rho) Y^{(\varphi)}_{yy}$ of the brane bending mode analysed in Sec.~\ref{sec:GWa} in the case $d=3, m^2=0, \mathbf{k}=0$ and $\zeta_*=-1/2$. 
Panel (a) shows the profile of ${\mathcal R}^{(\varphi)}(\rho)$, which is given by Eq.~\eq{P-} with the junction condition \eq{JCRphi} imposed at $\zeta=\zeta_*$.
Panel (b) shows the time evolution of $Y_{yy}^{(\varphi)}$,
which obeys Eq.~\eq{yeq} with $\lambda_\rho = -1$,
for the initial wave packet with $y_0=4, s=1/8$.
As argued in Sec.~\ref{subsecd=3}, $Y_{yy}^{(\varphi)}$ is proportional to the holographic stress energy tensor component ${\cal T}_{yy}$.
The qualitative feature of its time evolution is similar to those for the massless scalar field shown in Fig.~\ref{fig:reflection}. The wave approximately follows a null line reflecting at the AdS$_d$ boundary $y=0$, while the shape of the wave packet after the reflection is slightly distorted compared to that of the scalar field due to the difference in the value of $\lambda_\rho$.

In this section, we showed time evolution of the scalar field and the scalar part of the gravitational perturbations assuming that the mode function in the bulk direction $\rho$ has only one mode given by Eq.~\eq{P-} for given $l$ and $\mu$.
This field profile corresponds to a plane wave
that has a $y=\text{const.}$ surface as the wavefront and propagates toward the $y$ direction keeping its profile in the $\rho$ direction invariant.
In the bulk gravity viewpoint, this structure is the origin of the almost complete reflection at the AdS$_d$ boundary observed in Figs.~\ref{Fig:wave} and \ref{Fig:wave_phi}.
The time evolution around and after the reflection may be qualitatively different for waves with more general shapes, and it would be interesting to study the dynamics of such waves with general profiles and their interpretations in the BCFT viewpoint.
We will address such issues in the future work.

%%%%%%%%%%%%%%%%%%%%%%%%%%%%%%%%%%%%%%%%%%%%%%%%%%%
\section{Island/BCFT correspondence in higher dimensions} 
\label{sec:island}
%%%%%%%%%%%%%%%%%%%%%%%%%%%%%%%%%%%%%%%%%%%%%%%%%%%%

Another important aspect of the AdS/BCFT correspondence is that it has the third description where the $d$ dimensional CFT on a half plane $w>0$ is coupled to $d$ dimensional gravity on AdS$_d$ which extends in $w<0$. This description is qualitatively expected from the brane-world holography \cite{Randall:1999ee,Randall:1999vf,Gubser:1999vj,Karch:2000gx,Giddings:2000mu,Shiromizu:2001jm,Shiromizu:2001ve,Nojiri:2000eb,Nojiri:2000gb,Hawking:2000kj,Koyama:2001rf,Kanno:2002iaa} and has been beautifully applied to the black hole information problem in the light of the Island formula \cite{Penington:2019npb,Almheiri:2019psf,Almheiri:2019hni}.
This triality includes a purely $d$ dimensional statement that a $d$ dimensional BCFT is equivalent to a $d$ dimensional CFT on a half plane, coupled to $d$ dimensional gravity on AdS$_d$, which is called Island/BCFT correspondence \cite{Suzuki:2022xwv}. Even though this duality can be formally obtained by combining the AdS/BCFT and brane-world holography, it is fair to say that the precise and quantitative justification of the Island/BCFT correspondence is still poorly understood. In the recent paper \cite{Suzuki:2022xwv}, we examined this problem and presented quantitative evidences in the lowest dimension $d=2$. Below we would like to test the Island/BCFT correspondence via computations of entanglement entropy in higher dimensions $d>2$. Since we do not have powerful controls of entanglement entropy, as opposed to the $d=2$ case studied in \cite{Suzuki:2022xwv}, we will focus on the area law term contributions.

%%%%%%%%%%%%%%%%%%%%%%%%%%%%%%%%%%%%%%%%%%%%%%%%%%%%%%%%
\subsection{Entanglement entropy from island formula}
\label{sec:Islands in AdS_{d+1}}
%%%%%%%%%%%%%%%%%%%%%%%%%%%%%%%%%%%%%%%%%%%%%%%%%%%%%%%%

Consider a $d$ dimensional BCFT on a half plane $w>0$ in the flat Lorentzian spacetime
	\begin{align}
		ds_d^2 \, = \, - dt^2 + dw^2 + \sum_{i=1}^{d-2} dx_i^2 \, ,
	\end{align}
and take the subregion $B$ as a half ball 
	\begin{align}
		w^2 \, + \, \sum_{i=1}^{d-2} x_i^2 \, \le \, l^2 \, , \qquad w > 0 \, .
		\label{halfBa}
	\end{align}
We call its complement the subsystem $A$ for which we consider the entanglement entropy, denoted by $S_A$.
We couple this BCFT$_d$ to the AdS$_d$ gravity at $w=0$, so that we have AdS$_d$ in $w < 0$ and BCFT$_d$ in $w > 0$.
Then the corresponding metric of AdS$_d$ is given by
	\begin{align}
		ds_d^2 \, = \, \frac{\d^2}{(\d - w)^2} \left( - dt^2 + dw^2 + \sum_{i=1}^{d-2} dx_i^2 \right) \, , \qquad (w < 0),
	\end{align}
where in this AdS$_d$ spacetime, $\d$ can be understood as an UV cutoff.
By the following coordinate transformation
	\begin{align}
		\delta - w \, &= \, r \cos \th \, , \\[-10pt]
		x_i \, &= \, r \sin \th \cos \phi_i \prod_{n=1}^{i-1} \sin \phi_n \, , \qquad (i = 1, \cdots, d-3) \\[-10pt]
		x_{d-2} \, &= \, r \sin \th \prod_{n=1}^{d-3} \sin \phi_n \, ,
	\label{coordchange}
	\end{align}
we can also bring the metric into the spherical coordinates
	\begin{align}
		ds_d^2 \, = \, \frac{\d^2}{r^2 \cos\th^2} \left( - dt^2 + dr^2 + r^2 d \O_{d-2}^2 \right) \, .
	\end{align}
	
\begin{figure}
  \centering
  \includegraphics[width=10cm]{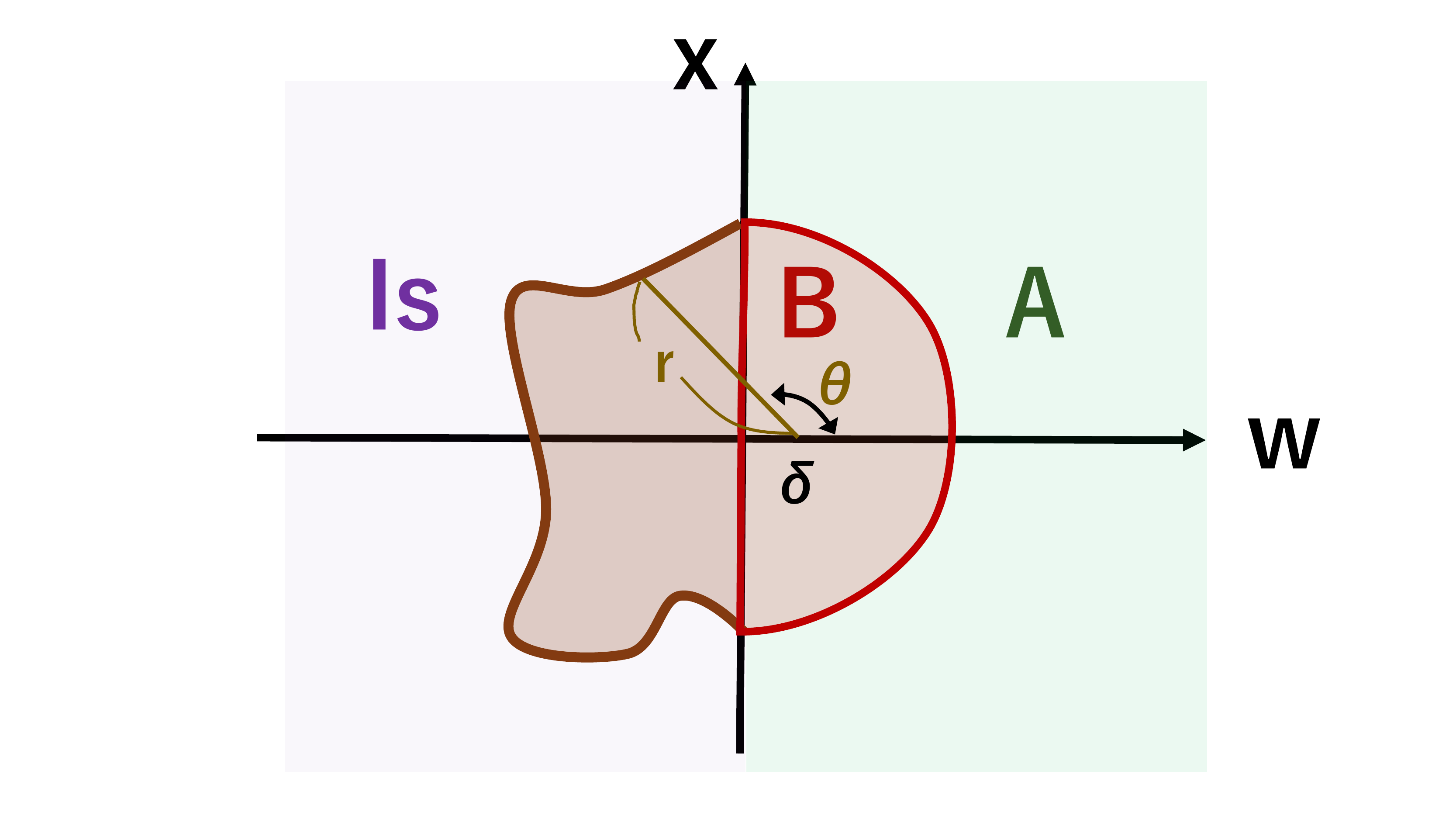}
  \caption{Island setup in BCFT$_3$}
\label{fig:IS4}
\end{figure}

In such a system of a CFT coupled to gravity, the entanglement entropy $S_A$ can be computed by applying the Island prescription \cite{Penington:2019npb,Almheiri:2019psf,Almheiri:2019hni}. Namely, the entanglement entropy is obtained by the formula (refer to Fig.~\ref{fig:IS4})
\ba
S_A=\mbox{Ext}_{Is}\left[S^{(G)}_{\de Is}+S^{(Q)}_{A\cup Is}\right].
\ea
Here $Is$ denotes the Island which is a region in the AdS$_d$ which is originally taken to be arbitrary under the condition that it connects with the region $A$ at the boundary $w=0$ as in Fig.~\ref{fig:IS4}. The entropy  $S^{(G)}$ denotes the classical gravity contribution, which is the area of the boundary of Island $\de Is$ divided by $4G_N$ in Einstein gravity. In our case, we expect that the brane-world gravity is purely induced from quantum corrections of matter fields and thus we simply set $S^{(G)}=0$. 
In other words, we can set $\frac{1}{G^{(d)}_N}=0$ before we integrate over the matter fields, where $G^{(d)}_N$ is the $d$ dimensional Newton constant.
On the other hand, $S^{(Q)}$ describes all contributions from the quantum fields to entanglement entropy, where the subsystem is taken to be an union of $A$ and the Island. Finally we extremize the total entropy by allowing to change the shape of the Island. This is the Island prescription in our setup. In our higher dimensional case $d>2$,  we expect that $S^{(Q)}$ follows the area law \cite{Bombelli:1986rw,Srednicki:1993im} and thus the leading divergent contribution looks like
\ba
S^{(Q)}_{A\cup Is}\simeq \mbox{Ext}_{Is}\left[\gamma\cdot \frac{\mbox{Area}(\de Is)}{\ep^{d-2}}\right]+S_0+O(\ep^{-(d-4)}),
\ea
where $S_0$ represents the area law contribution from the subsystem $A$, being proportional to $\frac{l^{d-2}}{\ep^{d-2}}$. The extremization procedure is equivalent to finding an extremal surface as $\de Is$. For static backgrounds, which we focus in this section, this becomes a minimal surface on a constant time slice.

%%%%%%%%%%%%%%%%%%%%%%%%%%%%%%%%%%%%%%%%%%%%%%%%%%%%%%%%
%%%%%%%%%%%%%%%%%%%%%%%%%%%%%%%%%%%%%%%%%%%%%%%%%%%%%%%%

We parameterize a co-dimension two extremal surface on a constant time slice in the AdS$_d$, we parameterize the surface by
	\begin{align}
		r \, = \, r(\th, \phi_1, \cdots, \phi_{d-3}) \, ,
	\end{align}
with a boundary condition $r = l$ at $w=0$.
Then, the induced metric on the surface is given by
	\begin{align}
		ds_{d-2}^2=\frac{\d^2}{\cos^2\th} \left[ \left( \frac{\pa \log r}{\pa \th} d\th
		 + \sum_{i=1}^{d-3} \, \frac{\pa \log r}{\pa \phi_i} d\phi_i \right)^2 \, + \, d \O_{d-2}^2 \right] \, .
	\end{align}
For the above induced metric, we denote the contribution from $d\O_{d-2}^2$ by $(h_{d-2}^{(0)})_{\m\n}$ and the contribution from $dr^2$ by $(\tilde{h}_{d-2})_{\m\n}$ as
	\begin{align}
		\big( h_{d-2} \big)_{\m\n}
		\, = \, \frac{\d^2}{\cos^2\th} \left[ \big( h_{d-2}^{(0)} \big)_{\m\n} \, + \, \big( \tilde{h}_{d-2} \big)_{\m\n}\right] \, ,
	\end{align}
where
	\begin{align}
    	\big( h_{d-2}^{(0)} \big)_{\m\n} \, &= \, 
        \begin{pmatrix}
        1 & 0 & \cdots & 0\\
        0 & \sin^2\th & \cdots & 0 \\
        \vdots & \vdots & & \vdots  \\
        0 & 0 & \cdots & \sin^2\th \prod_{i=1}^{d-4} (\sin\p_i)^2
        \end{pmatrix} \, , \\[4pt]
    	\big( \tilde{h}_{d-2} \big)_{\m\n} \, &= \, v_\m \, (v^T)_\n \, , \\[4pt]
        v^T_\m \, &= \, \left( \frac{\pa \log r}{\pa \th} \, , \, \frac{\pa \log r}{\pa \p_1} \, , \,  \cdots \, , \, \frac{\pa \log r}{\pa \p_{d-3}} \right) \, .
	\end{align}
The areal law term of the entanglement entropy for $A\cup Is$ reads 
	\begin{align}
		S^{(Q)}_{A \cup Is} \, \approx \, \frac{\g}{\e^{d-2}} \int_{- \frac{\pi}{2} + \d'}^{\frac{\pi}{2} - \d'} d\th
		\int_0^{\pi} \prod_{i=1}^{d-3} d\phi_i \sqrt{\det(h_{d-2})} \, + \, S_0 \, ,
	\end{align}
where the regularization parameter $\delta'$ is defined by $l \cos\delta'=\delta$.
	
Using the decomposition of the determinant, the determinant of the induced metric is given by
	\begin{align}
		\det(h_{d-2}) \, &= \, \left( \frac{\d}{\cos\th} \right)^{2(d-2)} \, \det\Big(h_{d-2}^{(0)} + v v^T \Big) \nn\\
		\, &= \, \left( \frac{\d}{\cos\th} \right)^{2(d-2)} \, \det\big(h_{d-2}^{(0)} \big) \det\big(I_{d-2} + v^T (h_{d-2}^{(0)})^{-1} v \big) \, ,
	\end{align}
where $I_{d-2}$ is the $d-2$ dimensional identity matrix and we used Sylvester's determinant theorem (or the matrix determinant lemma) for the second line.
Since the first determinant $\det\big(h_{d-2}^{(0)} \big)$ does not depend on the surface profile $r=r(\th, \p_i)$, we need to consider to minimize the second determinant $\det\big(I_{d-2} + v^T (h_{d-2}^{(0)})^{-1} v \big)$.
This is archived by the vanishing vector $v=0$, so that $\det\big(I_{d-2} + v^T (h_{d-2}^{(0)})^{-1} v \big) = 1$.
Therefore, the minimal surface is semi-circle with the constant radius $r=l$.
Hence in this case, the entanglement entropy is given by
	\begin{align}
		S_A=S^{(Q)}_{A \cup Is} \, &= \, \frac{\g \d^{d-2}}{\e^{d-2}} \int_{- \frac{\pi}{2} + \d'}^{\frac{\pi}{2} - \d'} d\th \, \frac{(\sin\th)^{d-3}}{(\cos\th)^{d-2}}
		\int_0^{\pi} \prod_{i=1}^{d-3} d\phi_i \big( \sin \phi_i \big)^{d-3-i} \, + \, S_0 \nn\\
		&\approx \, \frac{\g \d S_{d-3} l^{d-3}}{(d-3) \e^{d-2}} \, + \, S_0 \, ,
	\label{SAI_AdS_d}
	\end{align}
where $S_{d-3} = 2 \pi^{(d-2)/2}/\G((d-2)/2)$.
The above result of the $\th$ integral assumes $d>3$ and the case of $d=3$ must be treated separately.

For $d=3$, we actually obtain 
\ba
S_A\simeq 2\gamma\cdot \frac{\delta}{\ep}\log \frac{2l}{\delta}+S_0,  \label{logIS}
\ea
where we have $S_0=\gamma\frac{\pi l}{\ep}$.

We also would like to note an important fact. We initially assumed the subsystem $A$ to be the outside of a round half ball $B$ specified by (\ref{halfBa}). However, the above calculation based on the Island formula only uses the location of the boundary $\de A$ at the interface between the CFT and the gravity i.e.\ we have $x=\pm l$ at $w=0$. Thus we obtain the same result (\ref{logIS}) for any shape of the subsystem $A$ as long as it ends on $|x|=\pm l$ at $w=0$.

%%%%%%%%%%%%%%%%%%%%%%%%%%%%%%%%%%%%%%%%%%%%%%%%%%%%%%%%
\subsection{Entanglement entropy in BCFT from AdS/BCFT}
%%%%%%%%%%%%%%%%%%%%%%%%%%%%%%%%%%%%%%%%%%%%%%%%%%%%%%%%

\begin{figure}
  \centering
  \includegraphics[width=10cm]{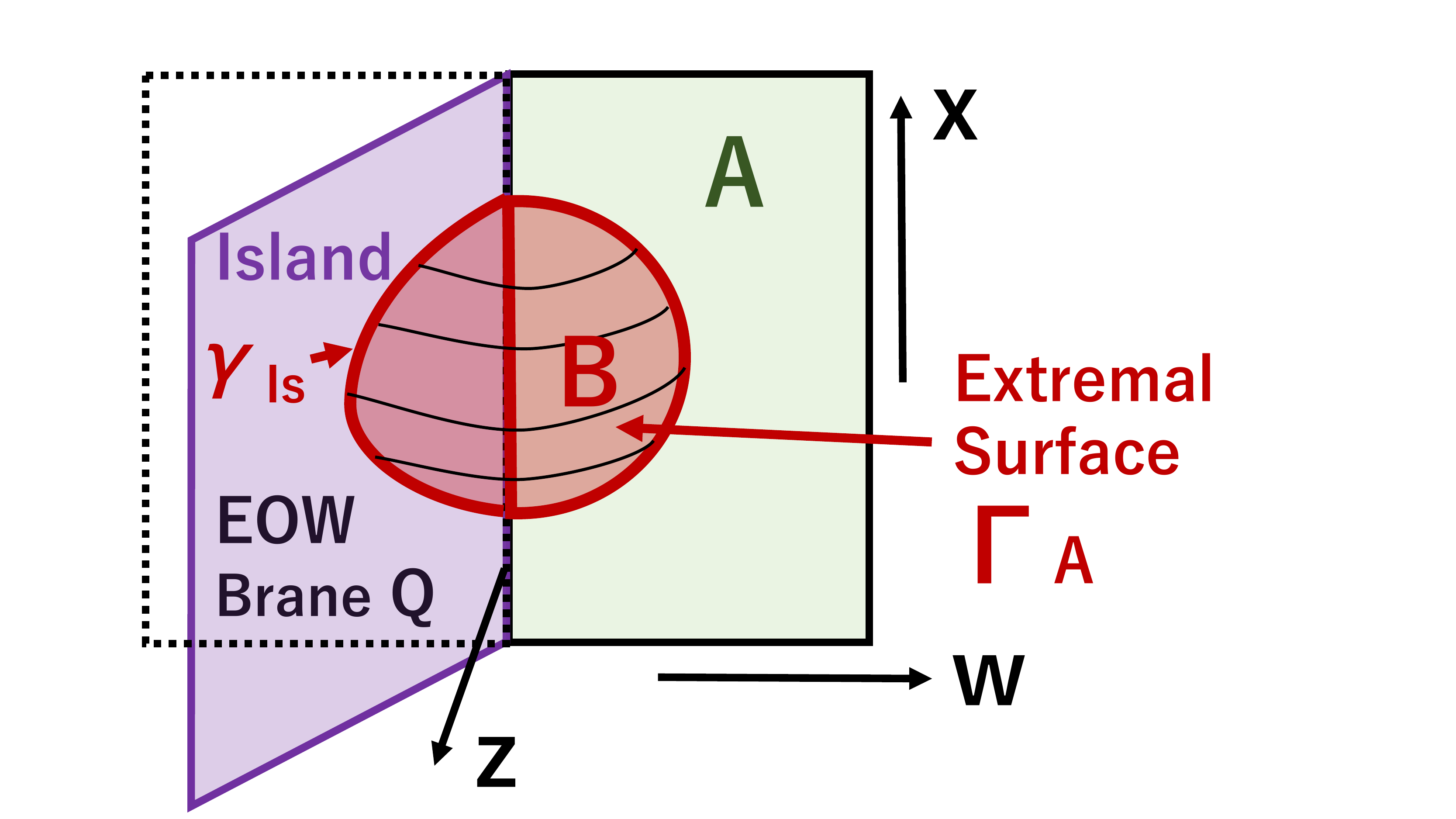}
  \caption{Calculation of holographic entanglement entropy in AdS/BCFT}
\label{fig:IS5}
\end{figure}

Next we would like to compare the previous result based on the Island formula with the result obtained from a holographic BCFT, using the AdS/BCFT.
In the AdS/BCFT, the holographic entanglement entropy $S_A$ \cite{Ryu:2006ef,Ryu:2006bv,Hubeny:2007xt}, is given by 
\ba
S_A=\mbox{Ext}_{\Gamma_A}\left[\frac{\mbox{Area}(\Gamma_A)}{4G_N}\right].
\ea
with an extra rule \cite{Takayanagi:2011zk,Fujita:2011fp} that 
the surface $\gamma_A$ is a $d-1$ dimensional surface whose boundary is given by  $\de A\cup \gamma_{Is}$, where $\gamma_{Is}$ is a $d-2$ dimensional surface on the EOW brane $Q$. In the end, $\gamma_{Is}$ will be identified with the boundary of the Island i.e.\ $\de Is$. Refer to Fig.~\ref{fig:IS5} for a sketch. Note that for static backgrounds, which we focus in this section, $\Gamma_A$ becomes a minimal surface on a constant time slice.

We choose the coordinates of the AdS$_{d+1}$ to be 
	\begin{align}
		ds_{d+1}^2 \, = \, d\r^2 + \frac{\cosh^2 \r}{y^2} \left( - dt^2 + dy^2 + \sum_{i=1}^{d-2} dx_i^2 \right) \, .
	\label{ds_d+1}
	\end{align}
	
When $A$ is a semi-disk with radius $l$ as before,  the minimal surface is given by a half sphere even in the presence of EOW brane as we explain in appendix~\ref{app:minimal surface}. We employ the spherical coordinates by $y=l \cos \th$ and (\ref{coordchange}) for $x_i$'s such that the minimal surface is given by $r=l$. Then, the induced metric on the two minimal surface in this AdS$_{d+1}$ is given by
	\begin{align}
		ds_{d-1}^2 \, = \, d\r^2 + \frac{\cosh^2 \r}{\cos^2 \th} \, d \O_{d-2}^2 \, .
	\end{align}
In the previous sections, we placed the EOW brane at $\r=\r_*$ with $\r_*<0$,
but for this section, it's more convenient to denote the brane location by $\r=- \r_*$ with $\r_*>0$,
so the sign of $\r_*$ here is different from the one used in the previous sections.
Then, the holographic EE is computed as 
	\begin{align}
		S_A \, &= \, \frac{1}{4G_N} \int_{-\r_*}^{\r_{\inf}} d\r \, (\cosh \r)^{d-2}
		\int_{- \frac{\pi}{2} + \frac{\e}{l} \cosh\r}^{\frac{\pi}{2} - \frac{\e}{l} \cosh\r} d\th \, \frac{(\sin\th)^{d-3}}{(\cos\th)^{d-2}}
		\int_0^{\pi} \prod_{i=1}^{d-3} d\phi_i \big( \sin \phi_i \big)^{d-3-i} \nn\\
		&\approx \, S^{half}_A \, + \, \frac{S_{d-3} l^{d-3} L^{d-1}}{4(d-3) G_N \e^{d-3}} \, \sinh \r_* \, .
	\label{SA_ads_d+1}
	\end{align}
Again the $\th$ integral assumes $d>3$ and the case of $d=3$ must be treated separately.

For $d=3$, we actually obtain 
\ba
S_A&=&\frac{1}{2G_N}\int^{\rho_\infty}_{-\rho_*} d\rho \cosh\rho \int^l_{\ep\cosh\rho}dy\frac{l}{y\s{l^2-y^2}}\no
&\simeq & \frac{L^2}{2G^{(4)}_N}\sinh\rho_*\log\frac{2l}{\ep}+S_0.
\label{heeis}
\ea

Comparing (\ref{SA_ads_d+1}) with (\ref{SAI_AdS_d}), the corresponding dictionary for AdS$_{d+1}$/BCFT$_d$ is given by
	\begin{align}
		\frac{\g \d}{\e} \, = \, \frac{L^{d-1}}{4G_N^{(d+1)}} \, \sinh \r_* \, .
	\end{align}
Note that since we expect $\gamma\sim \frac{L^{d-1}}{2G^{(d+1)}_N}$, which is a sort of the central charge, we expect $\sinh\rho_*\sim \frac{\delta}{\ep}$. This is indeed what is expected in the AdS/BCFT as we have 
\ba
\frac{w}{z}\left(=\frac{\delta}{\ep}\right)=\sinh\rho_*,
\ea
on the EOW brane. In this way, we find a nice agreement of entanglement entropy between the Island and the holographic BCFT calculation.

Notice that this analysis of entanglement entropy in a holographic BCFT is correct for any value of tension $\sigma$ or angle $\r_*$ of the EOW brane. However, when the tension takes a generic value, we need to take into account the subleading terms other than the area law term in the Island calculation of the dual system, via the Island/BCFT correspondence. This is clear in our AdS/BCFT analysis presented above because the BCFT contribution to the holographic entanglement entropy comes from the region deep in the bulk. When we deform the shape of the subsystem $A$ with keeping the location at the interface $w=0$, the extremal surface $\Gamma_A$ will clearly be deformed accordingly. Nevertheless, if we assume the tension of EOW brane is very large ($\sigma\gg 1$ or $\theta\simeq \pi$) such that it sits very close to the asymptotically AdS boundary, then it is obvious that the leading contribution to the holographic entanglement entropy is still given by (\ref{heeis}) even if we deform the shape of subsystem $A$. This agrees with our observation in the Island calculation mentioned in the final part of the previous subsection.

%%%%%%%%%%%%%%%%%%%%%%%%%%%%%%%%%%%%%%%%%%%%%%%%%%%%%%%%
%%%%%%%%%%%%%%%%%%%%%%%%%%%%%%%%%%%%%%%%%%%%%%%%%%%%%%%%
\section{One point functions in higher dimensional AdS/BCFT}
\label{sec:one-point}
%%%%%%%%%%%%%%%%%%%%%%%%%%%%%%%%%%%%%%%%%%%%%%%%%%%%%%%%
%%%%%%%%%%%%%%%%%%%%%%%%%%%%%%%%%%%%%%%%%%%%%%%%%%%%%%%%
In general, one-point functions in BCFT are non-vanishing \cite{Cardy:1984bb}.
For a scalar primary operator $\mathcal{O}$ with a conformal dimension $\D_{\mathcal{O}}$, the one-point function has a form
	\begin{align}
		\big\la \mathcal{O}(w) \big\ra \, = \, \frac{\mathcal{N}_\mathcal{O}}{|w|^{\D_{\mathcal{O}}}} \, ,
	\end{align}
where $|w|$ is the distance from the boundary with a numerical normalization factor $\mathcal{N}_\mathcal{O}$.
In this section, we will see that in order to reproduce this non-vanishing one-point function,
the dual bulk scalar field must couple to the bulk gravity with a non-trivial expectation value.
This was pointed out in \cite{Suzuki:2022xwv} in the case of AdS$_3/$BCFT$_2$ (see also earlier work \cite{Fujita:2011fp, Kastikainen:2021ybu}). Below we will extend this to higher dimensional cases.

%%%%%%%%%%%%%%%%%%%%%%%%%%%%%%%%%%%%%%%%%%%%%%%%%%%%%%%%
\subsection{The model}
\label{sec:model}
%%%%%%%%%%%%%%%%%%%%%%%%%%%%%%%%%%%%%%%%%%%%%%%%%%%%%%%%
We study the Einstein-dilaton theory defined by the action
	\begin{align}
		I \, = \, - \frac{1}{16\pi G} \int d^{d+1}x \sqrt{g} \Big[ R - g^{\m\n} \pa_\m \p \pa_\n \p - U(\p) \Big] \, ,
	\end{align}
where $R$ is the $d+1$-dimensional Ricci scalar and $U(\p)$ is a generic potential of the dilaton $\p$.
This model and the solutions we are looking for can be thought as a generalization of the Janus solutions \cite{Bak:2003jk, Bak:2007jm}.
(See also \cite{Suzuki:2022xwv}).

We can obtain a class of setups of AdS$_{d+1}$/BCFT$_d$ by placing the EOW brane $Q$ at $\r=\r_*$, such that the bulk gravity extends in the region $\r_* < \r < \inf$.
For the scalar field $\p$, we assume the linear interaction on the brane
	\begin{align}
		I_\textrm{bdy} \, = \, \frac{a_0}{8\pi G} \int_Q d^dx \sqrt{h} \, \p \, ,
	\end{align}
where $a_0$ is a coupling constant and $h_{ij}$ is the induced metric on $Q$.
Combined with the bulk action, the variation of $\p$ leads to the Neumann-like boundary condition at $Q$:
	\begin{align}
		\pa_\r \p \big|_{\r = \r_*} \, =  \, a_0 \, .
	\label{ModNeumannBC}
	\end{align}
The Neumann boundary condition of the gravity is now written as
	\begin{align}
		K_{ij} - h_{ij} K \, = \, - \left(\frac{\sigma}{2} + a_0 \p \right) h_{ij} \, .
	\end{align}

From the bulk action, the Einstein equation and the Klein-Gordon equation are given by
	\begin{gather}
		R_{\m\n} \, - \, \frac{U(\p)}{d-1} \, g_{\m\n} \, = \, \pa_\m \p \pa_\n \p \, , \\
		2\pa_\m \big( \sqrt{g} g^{\m\n} \pa_\n \p \big) \, = \, \sqrt{g} \pa_{\p} U(\p) \, .
	\end{gather}
For the background solution, we consider the Janus type ansatz
\footnote{When there is no dilaton field, the warp factor $f(\r)$ coincides with $\mathfrak{a}^2(\r)=\cosh^2(\r)$ defined in (\ref{defa}).
However in this section, we consider non-trivial backreaction of the dilaton $\phi$,
so it differs from $\mathfrak{a}^2(\r)$ and we instead use $f(\r)$ which is the standard notation for the Janus solutions.}
	\begin{align}
		ds_{d+1}^2 \, &= \, d\r^2 + f(\r) ds_{AdS_d}^2 \, , \\
		\p \, &= \, \p(\r) \, .
	\end{align}
Given this ansatz, the Einstein equation and the Klein-Gordon equation are reduced to 
	\begin{gather}
		2(d-1) f f'' \, + \, (d-1)(d-2)(f')^2 \, + \, 4 (d-1)^2 f \, + \, 4U f^2 \, = \, 0 \, , \label{Eeq1} \\
		d(d-1)\big( 2f f'' - (f')^2 \big) \, + \, 4U f^2 \, + \, 4(d-1) f^2 (\p')^2 \, = \, 0 \, , \label{Eeq2} \\
		2f \p'' \, + \, d \, f' \p' - f \pa_{\p} U \, = \, 0 \, , \label{KGeq}
	\end{gather}
where the prime denotes a derivative with respect to $\r$.
In fact, the two equations (\ref{Eeq1}), (\ref{Eeq2}) coming from the Einstein equation are not independent to each other provided the Klein-Gordon equation (\ref{KGeq}) and they can be combined into 
	\begin{align}
		d(d-1) (f')^2 \, + \, 4d(d-1) f \, + \, 4 U f^2 \, - \, 4 f^2 (\p')^2 \, = \, 0 . \label{Eeq3}
	\end{align}
Therefore, we need to solve the Klein-Gordon equation (\ref{KGeq}) and the Einstein equation (\ref{Eeq3}) simultaneously.

%%%%%%%%%%%%%%%%%%%%%%%%%%%%%%%%%%%%%%%%%%%%%%%%%%%%%%%%
\subsection{Free massless case}
\label{sec:massless}
\begin{figure}[t!]
	\begin{center}
		\scalebox{0.62}{\includegraphics{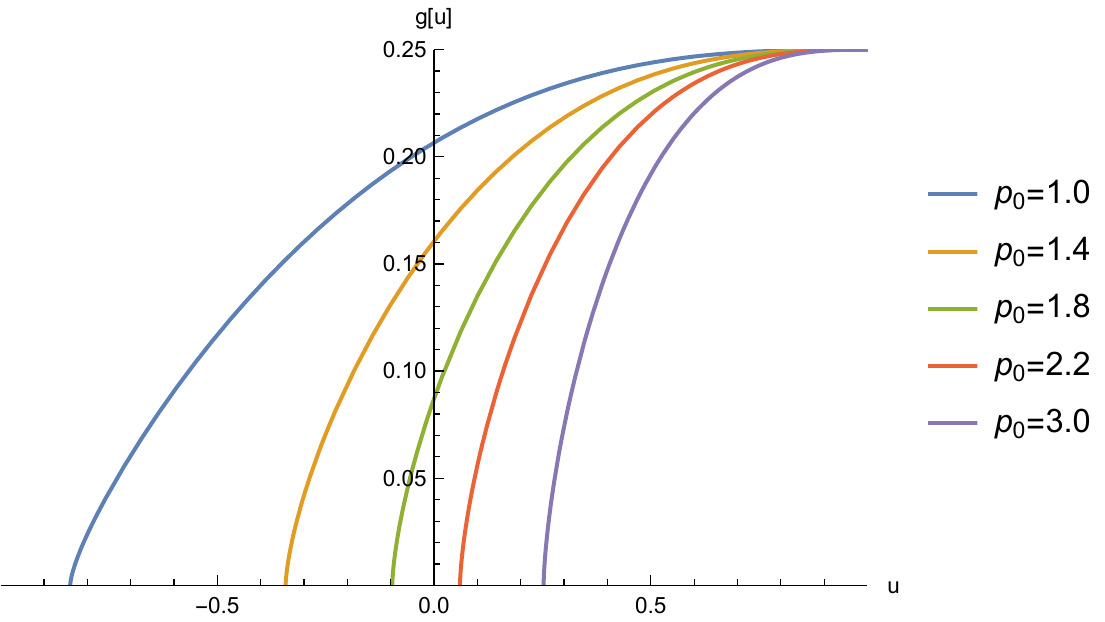}} \ \ \scalebox{0.62}{\includegraphics{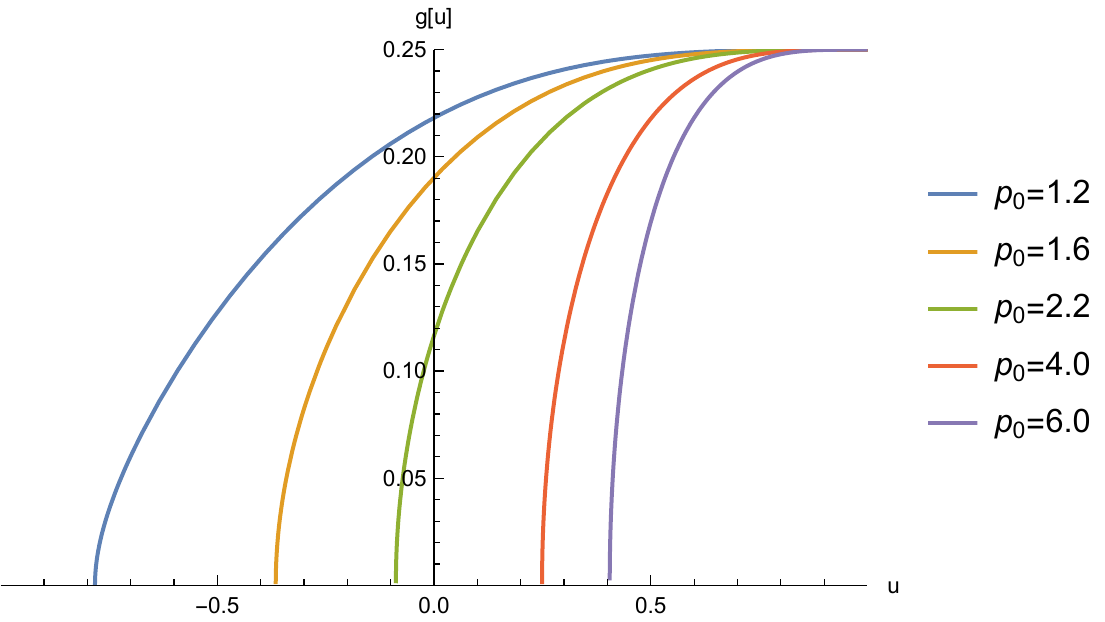}} \\[8pt]
		\scalebox{0.62}{\includegraphics{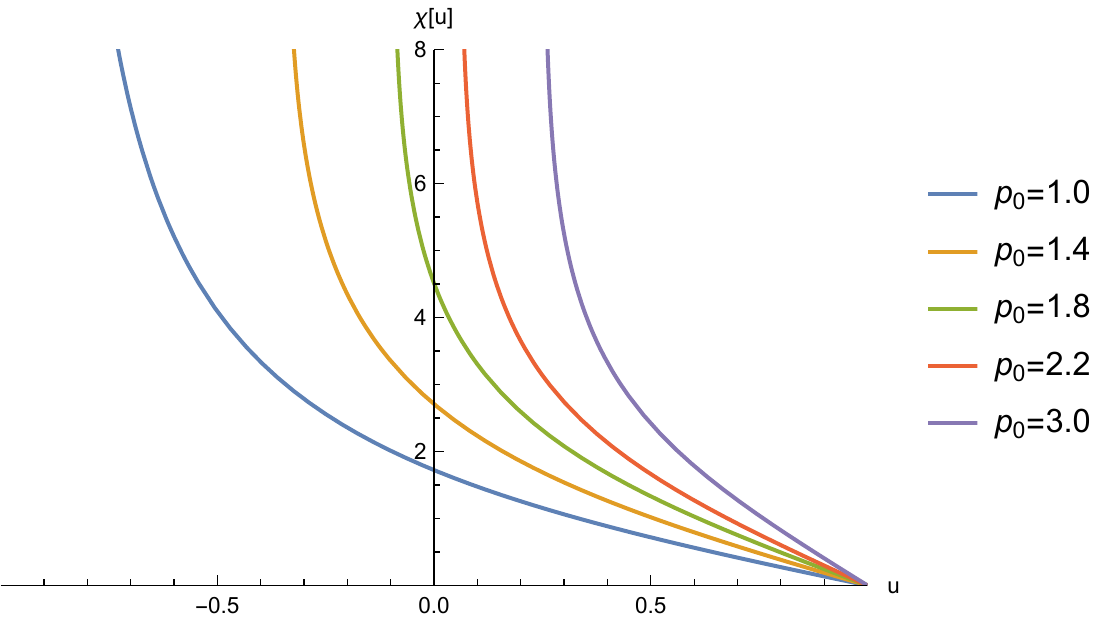}} \ \ \scalebox{0.62}{\includegraphics{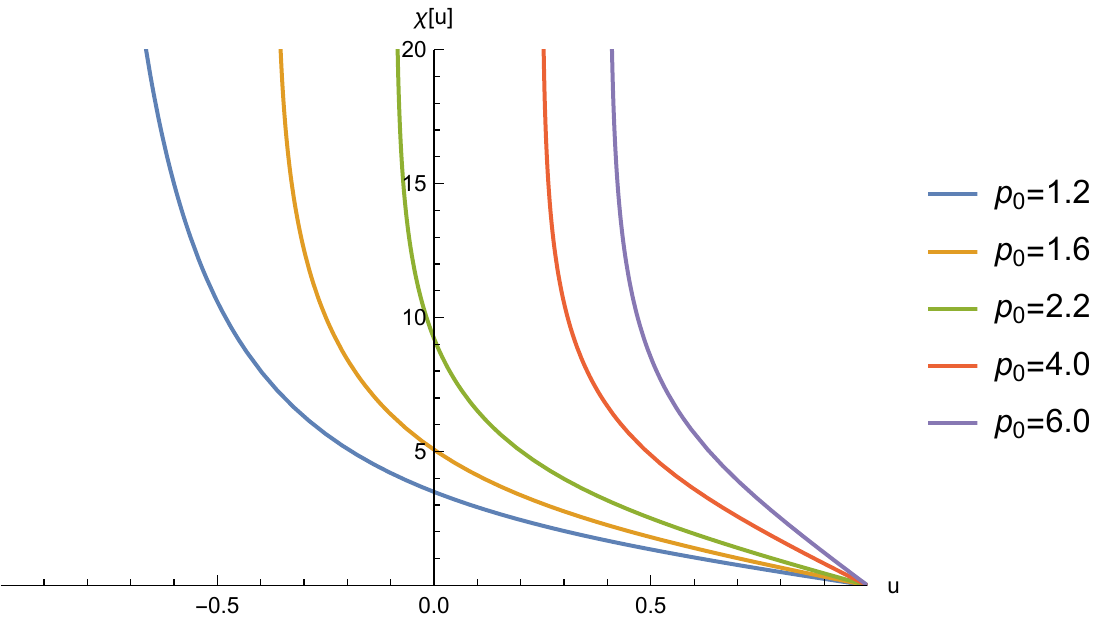}} 
	\caption{Massless solutions for $d=3$ (left) and $d=4$ (right).}
	\label{fig:massless}
	\end{center}
\end{figure}

%%%%%%%%%%%%%%%%%%%%%%%%%%%%%%%%%%%%%%%%%%%%%%%%%%%%%%%%
For free massless case, we have $U(\p) = 2\L = -d(d-1)$.
For this case, the Einstein equation and the Klein-Gordon equation are
	\begin{gather}
		d(d-1) (f')^2 \, + \, 4d(d-1) f \, - \, 4 d(d-1) f^2 \, - \, 4 f^2 (\p')^2 \, = \, 0 , \\
		2f \p'' \, + \, d \, f' \p' \, = \, 0 \, .
	\end{gather}
Since the Klein-Gordon equation is a total derivative, it is solved by
	\begin{align}
		\p ' \, = \, \frac{p_0}{f^{d/2}} \, ,
	\label{phiprime}
	\end{align}
where $p_0$ is an integration constant and this constant is fixed by the parameter $a_0$ in (\ref{ModNeumannBC}) by $p_0 = a_0 f^{d/2}(\r_*)$.
Substituting this into the Einstein equation, we find
	\begin{align}
		d(d-1) (f')^2 \, + \, 4d(d-1) f \, - \, 4 d(d-1) f^2 \, - \, 4 c^2 f^{2-d} \, = \, 0 .
	\end{align}
Since this is a first order differential equation, we need only one boundary condition which we choose
	\begin{align}
		f(\r) \, \to \, \frac{1}{4} \, e^{2\r} \, + \, \cdots  \qquad \textrm{for} \ \ \r \to \inf \, .
	\label{masslessfBC}
	\end{align}	
This equation is still difficult to find analytical solutions, except for $d=2$ case \cite{Bak:2003jk}.
However, numerically we can easily find solutions and we plotted some numerical solutions for $d=3$ and $d=4$ in Fig.~\ref{fig:massless}.
As in \cite{Suzuki:2022xwv}, we defined 
	\begin{align}
		u \, = \, \tanh \r \, ,
	\end{align}
which map the coordinate into $-1 \le u \le 1$ and
	\begin{align}
		f(u) \, = \, \frac{4 g(u)}{1-u^2} \, , \qquad \p(u) \, = \, \left(\frac{1-u^2}{4} \right)^{\frac{\D}{2}-1} \c(u) \, .
	\label{g&chi}
	\end{align}
These solutions are singular solutions in the sense that we see a naked singularity at $u=u_s$ where $f(u_s)=0$ and $\p(u_s)$ diverges.
Such a naked singularity must be prohibited for the usual AdS/CFT without an EOW brane.
However, in the context of the AdS/BCFT setup,
this is not a problem \cite{Suzuki:2022xwv} but just means that we have to place the EOW brane before this singularity ($u_s < u_*=\tanh\r_*$).

%%%%%%%%%%%%%%%%%%%%%%%%%%%%%%%%%%%%%%%%%%%%%%%%
%%%%%%%%%%%%%%%%%%%%%%%%%%%%%%%%%%%%%%%%%%%%%%%%

%%%%%%%%%%%%%%%%%%%%%%%%%%%%%%%%%%%%%%%%%%%%%%%%%%%%%%%%
\subsection{Free massive case}
\label{sec:massive}
\begin{figure}[t!]
	\begin{center}
		\scalebox{0.5}{\includegraphics{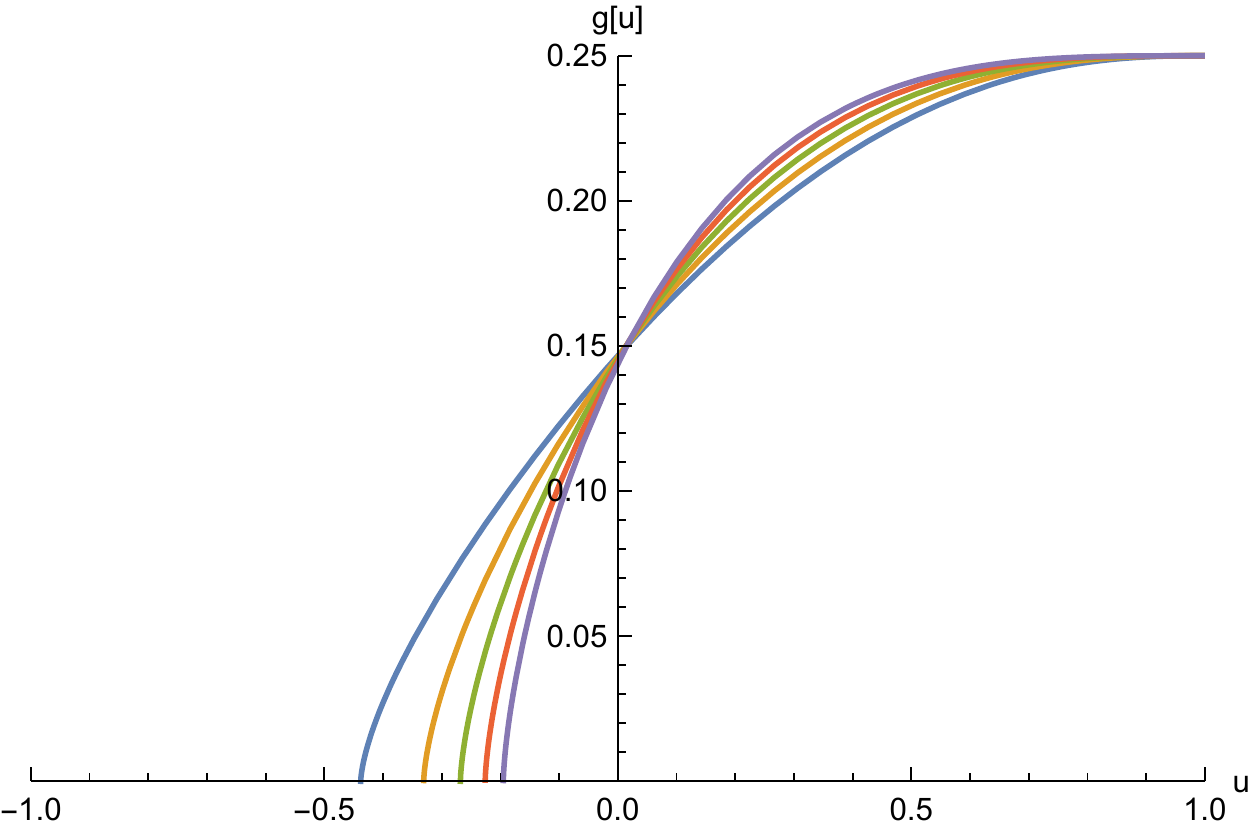}} \ \ \scalebox{0.5}{\includegraphics{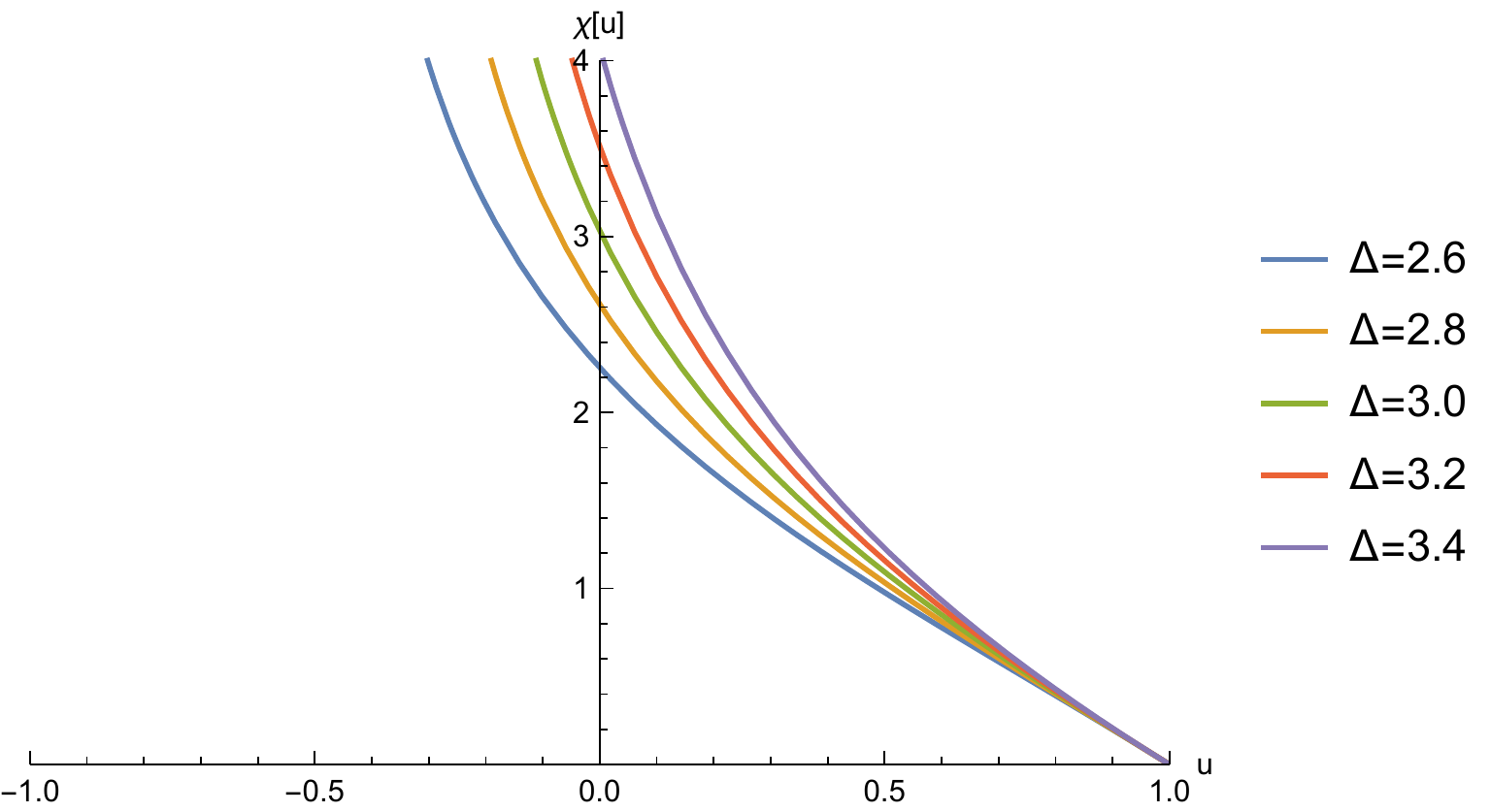}} \\
		\scalebox{0.5}{\includegraphics{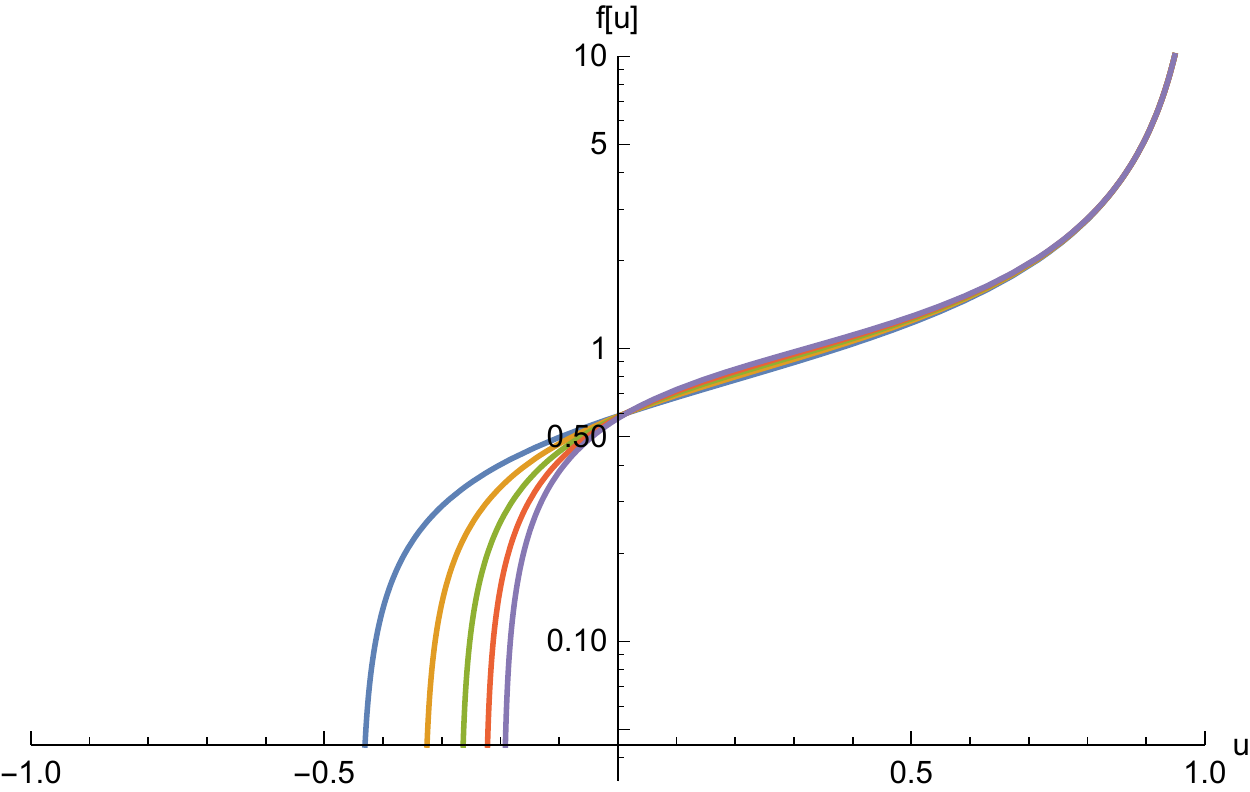}} \ \ \scalebox{0.5}{\includegraphics{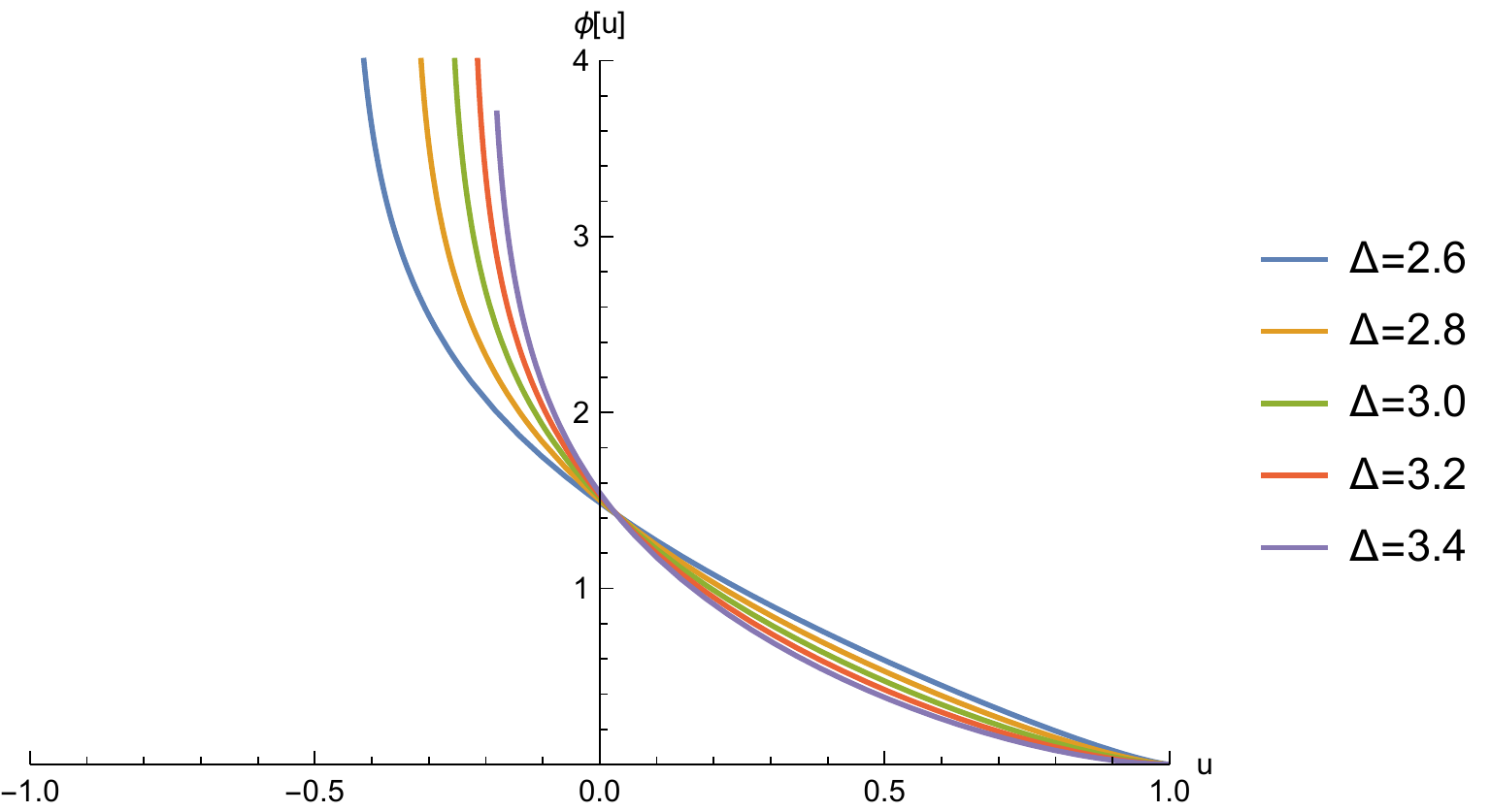}} 	
	\caption{Massive singular solutions for $d=3$ with $\a=4.0$.}
	\label{fig:massive(d=3)}
	\end{center}
\end{figure}

\begin{figure}[h!]
	\begin{center}
		\scalebox{0.5}{\includegraphics{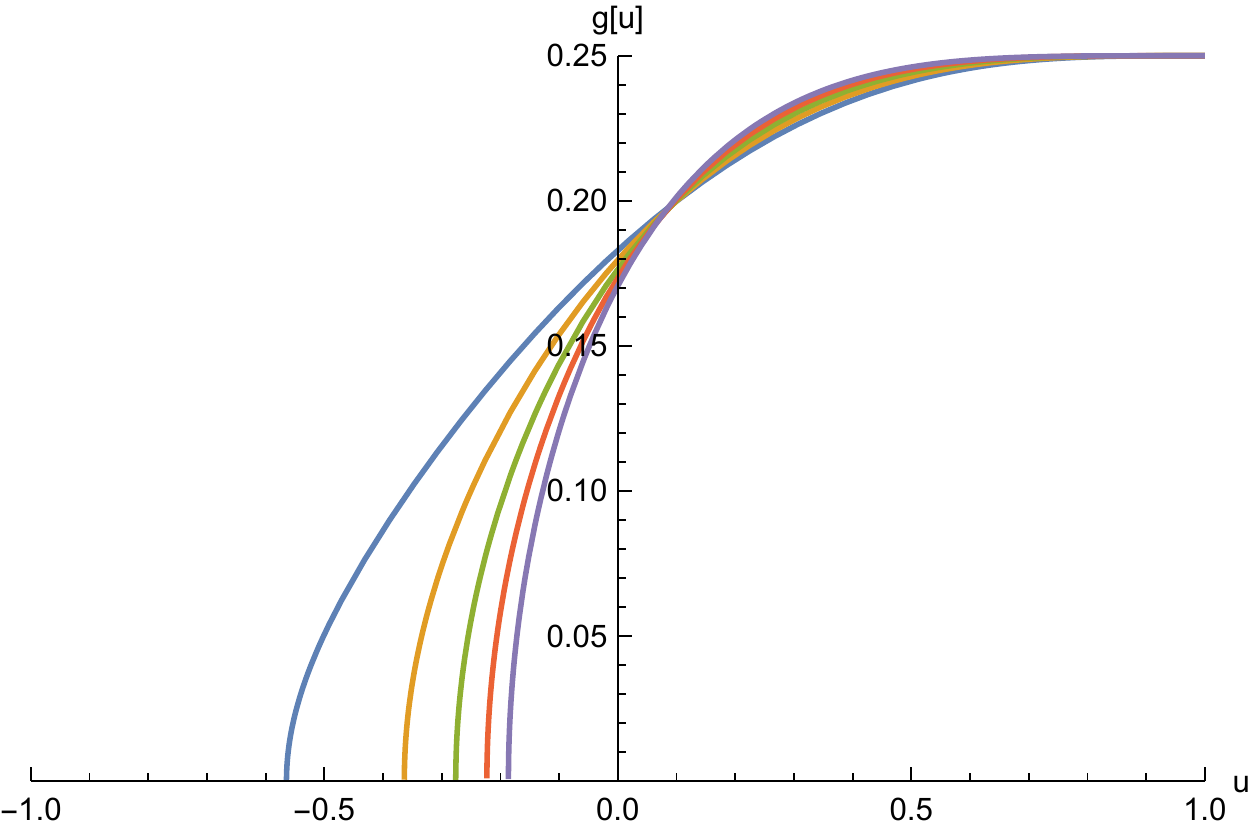}} \ \ \scalebox{0.5}{\includegraphics{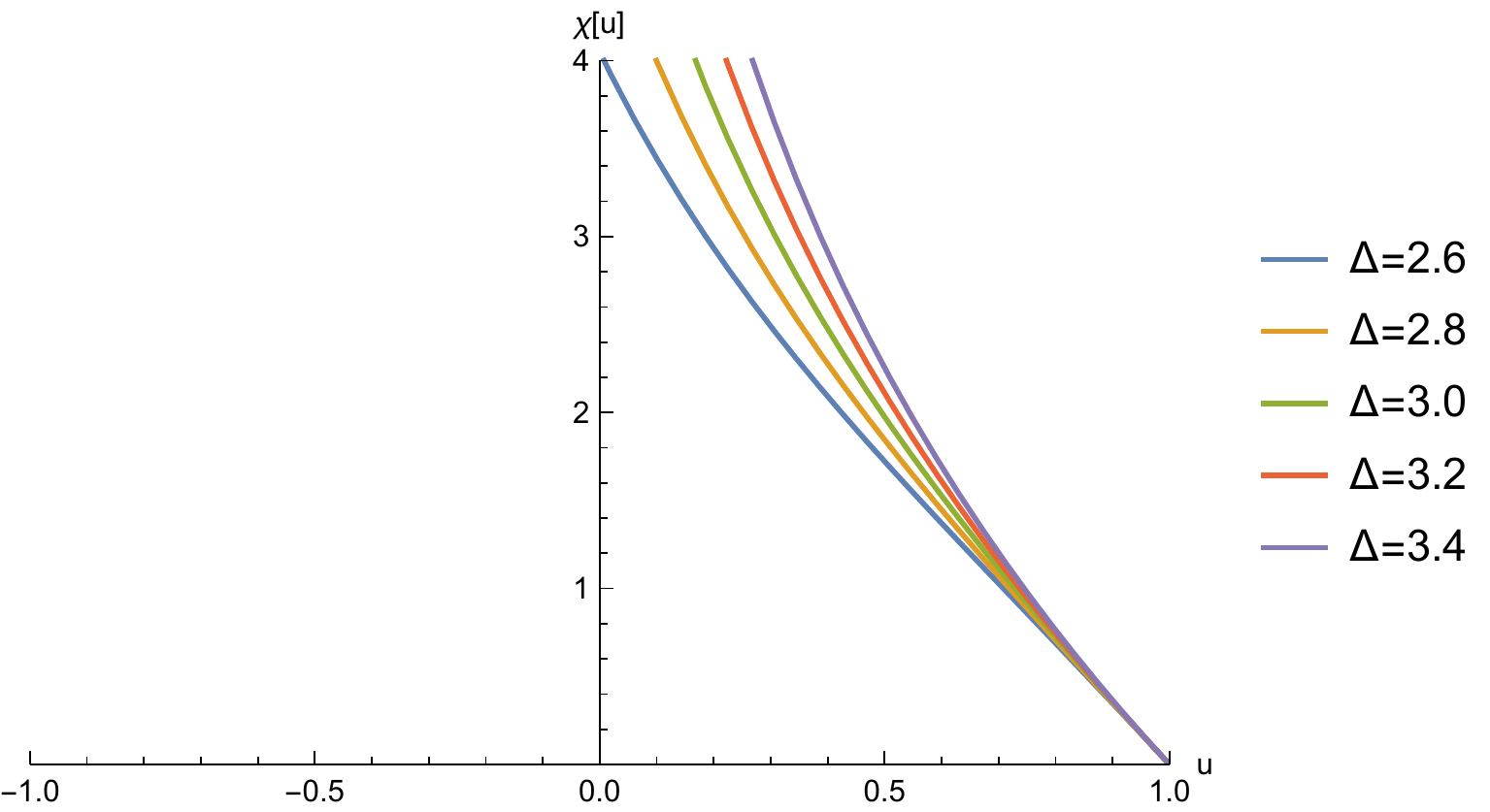}} \\
		\scalebox{0.5}{\includegraphics{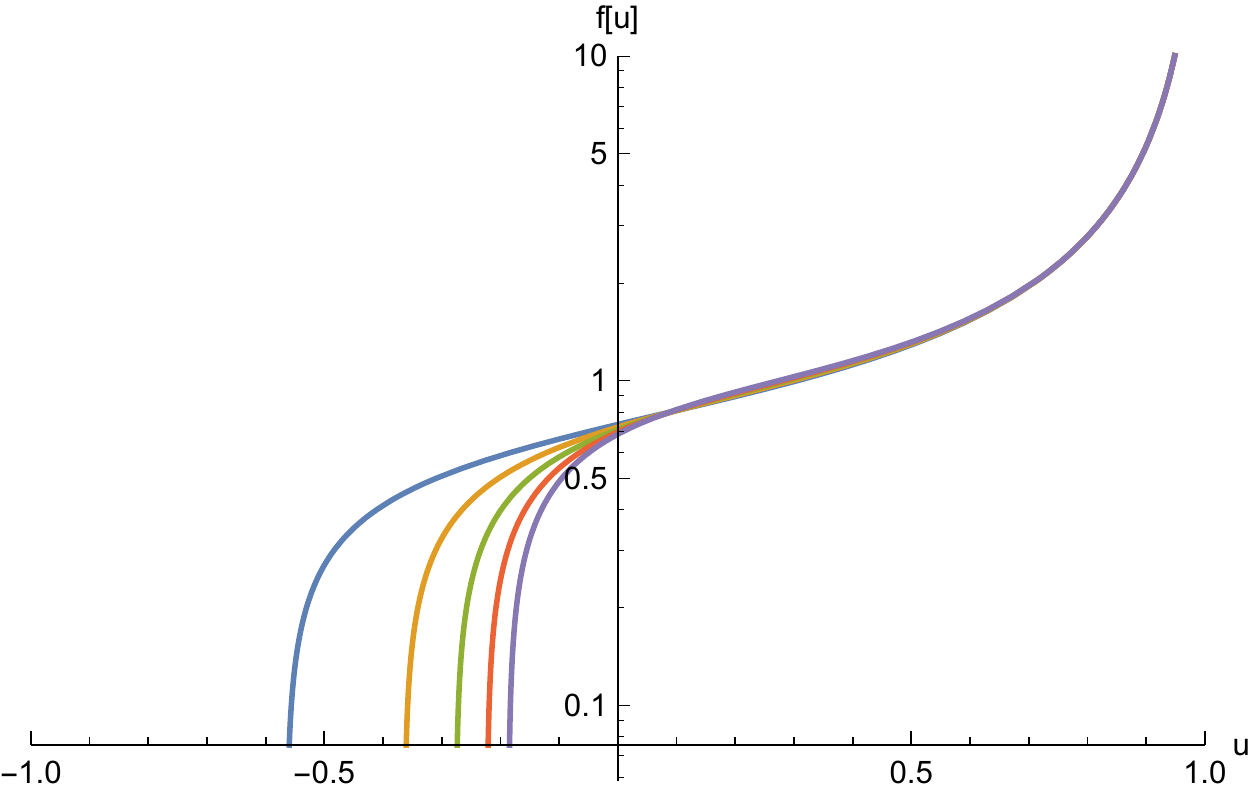}} \ \ \scalebox{0.5}{\includegraphics{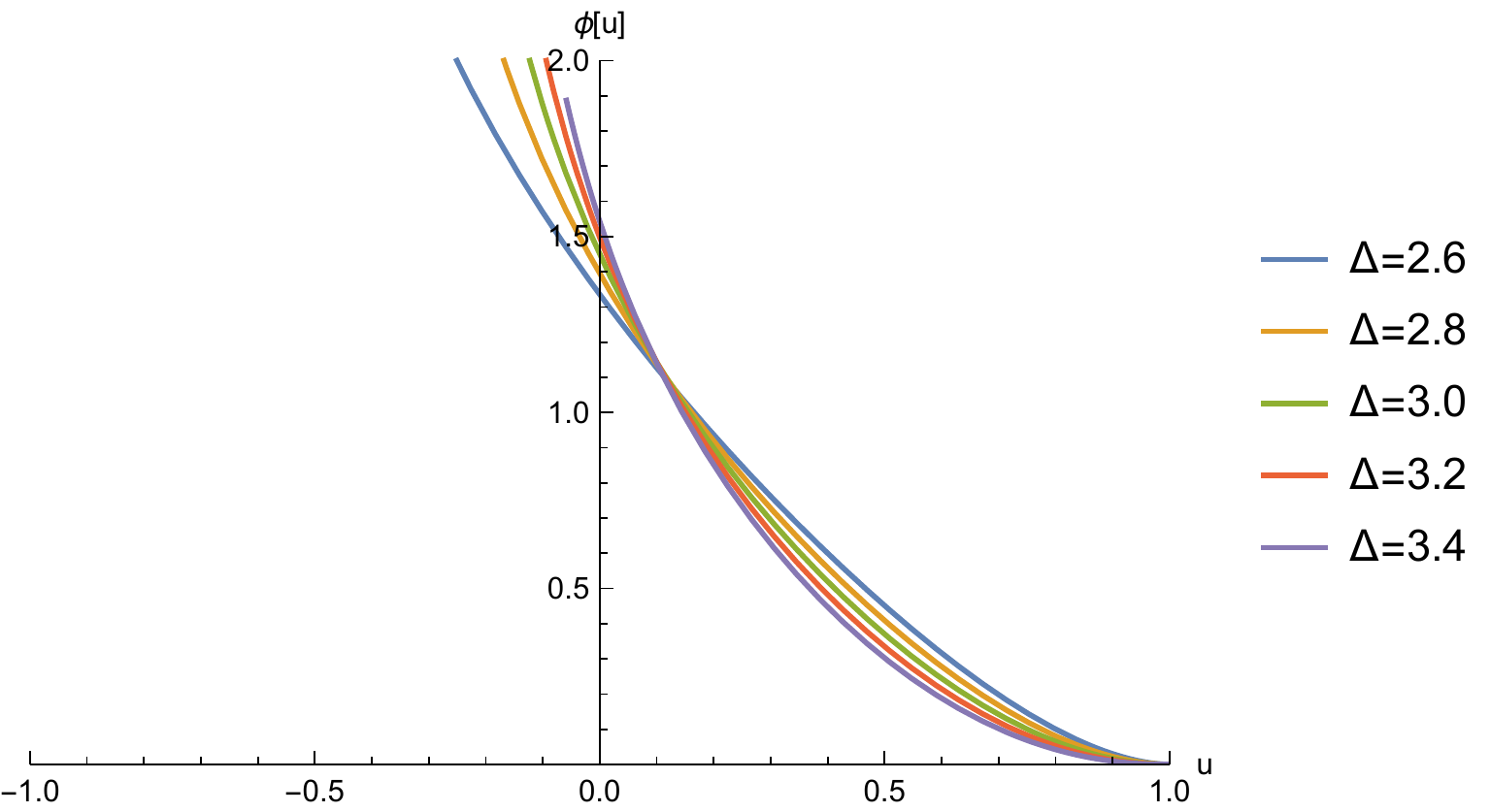}} 
	\caption{Massive singular solutions for $d=4$ with $\a=7.0$.}
	\label{fig:massive(d=4)}
	\end{center}
\end{figure}

\begin{figure}[h!]
	\begin{center}
		\scalebox{0.7}{\includegraphics{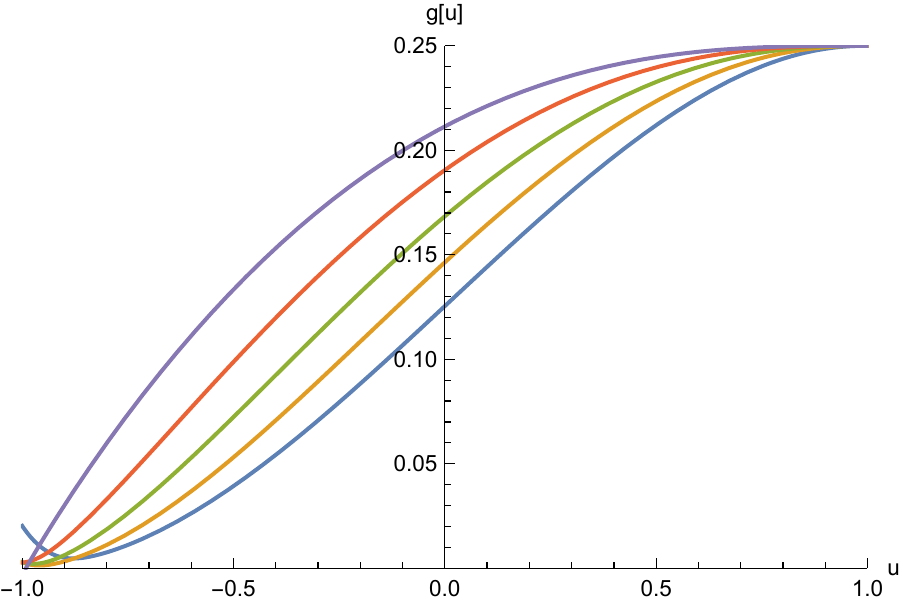}} \ \ \scalebox{0.7}{\includegraphics{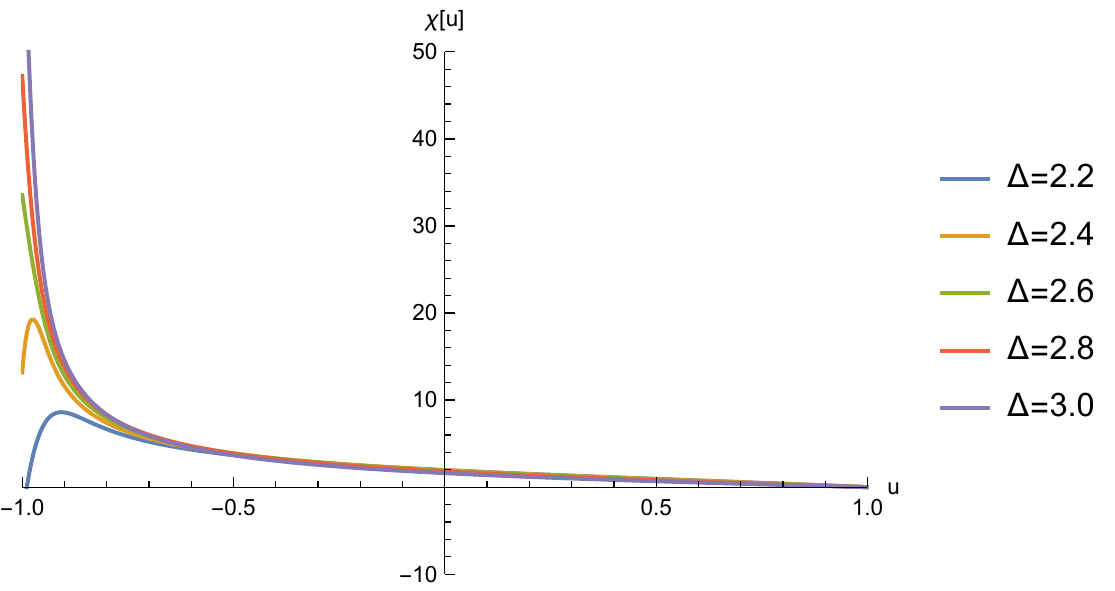}} \\
		\scalebox{0.7}{\includegraphics{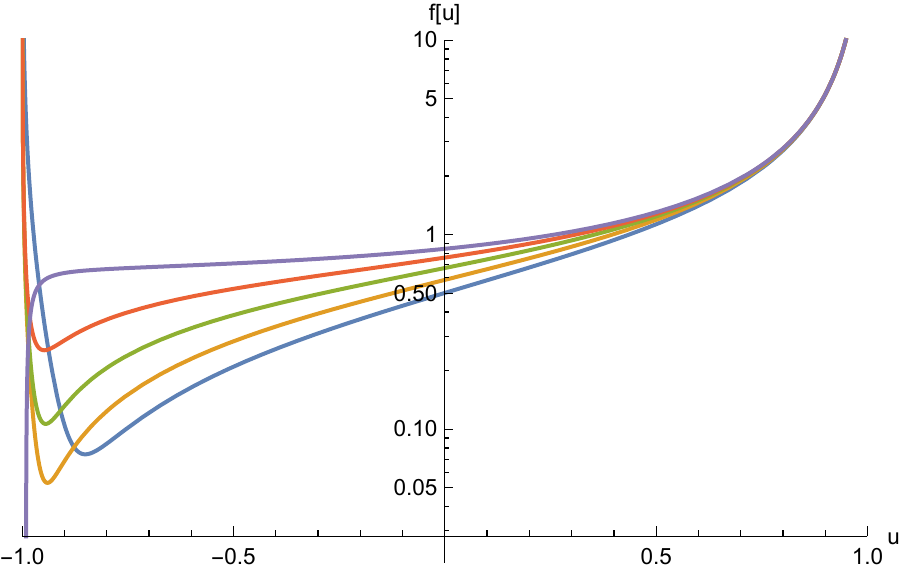}} \ \ \scalebox{0.7}{\includegraphics{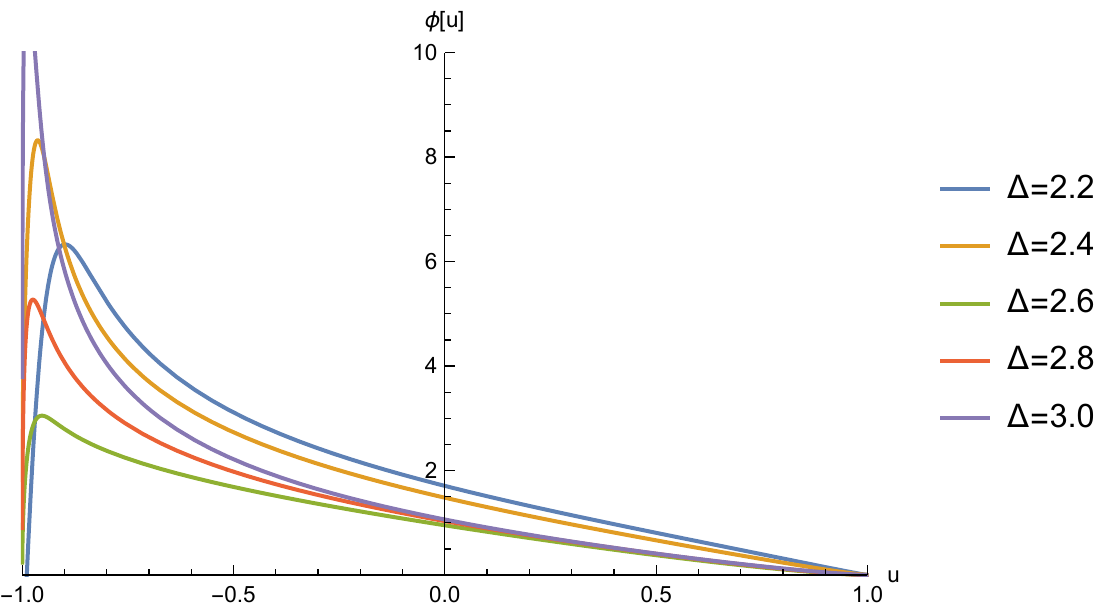}} 	
	\caption{Massive non-singular solutions for $d=3$ with $\a=$4.2756, 4.0026, 3.6125, 3.1150, 2.5210 for $\D=$2.2, 2.4, 2.6, 2.8, 3.0, respectively. }
	\label{fig:massive(d=3nonsingular)}
	\end{center}
\end{figure}

%%%%%%%%%%%%%%%%%%%%%%%%%%%%%%%%%%%%%%%%%%%%%%%%%%%%%%%%
For free massive case, we have $U(\p) = -d(d-1) + m^2 \p^2$.
For this case, the Einstein equation and the Klein-Gordon equation are
	\begin{gather}
		d(d-1) (f')^2 \, + \, 4d(d-1) f \, - \, 4 d(d-1) f^2 \, + \, 4m^2 f^2 \p^2 \, - \, 4 f^2 (\p')^2 \, = \, 0 , \\
		2f \p'' \, + \, d \, f' \p' \, - \, 2m^2 f \p \, = \, 0 \, .
	\end{gather}
The boundary conditions we impose are
	\begin{align}
		f(\r) \, &\to \, \frac{1}{4} \, e^{2\r} \, + \, \frac{1}{2} \, + \, \cdots \, , \\
		\p(\r) \, &\to \, \a \, e^{-\D \r} \, + \, \cdots \, , 
	\label{massiveBC}
	\end{align}	
where we parametrized the mass by $m^2 = \D(\D -d)$.
In this note, we focus on the masses above the BF bound $m^2 \ge - (d/2)^2$, this implies for the dimension we have $\D\ge d/2$.
In terms of $g$ and $\c$ defined in (\ref{g&chi}), the boundary conditions are written as
	\begin{align}
		g(u) \, \to \, \frac{1}{4} \, + \, \cdots \, , \qquad 
		\c(u) \, \to \, \a \left(\frac{1-u^2}{4} \right) \, + \, \cdots \, .
	\end{align}
We plotted numerical singular solutions for $d=3$ in Fig.~\ref{fig:massive(d=3)} and $d=4$ in Fig.~\ref{fig:massive(d=4)}.
In Fig.~\ref{fig:massive(d=3nonsingular)}, we also showed some non-singular solutions for $d=3$.

%%%%%%%%%%%%%%%%%%%%%%%%%%%%%%%%%%%%%%%%%%%%%%%%
%%%%%%%%%%%%%%%%%%%%%%%%%%%%%%%%%%%%%%%%%%%%%%%%

%%%%%%%%%%%%%%%%%%%%%%%%%%%%%%%%%%%%%%%%%%%%%%%%
%%%%%%%%%%%%%%%%%%%%%%%%%%%%%%%%%%%%%%%%%%%%%%%%

%%%%%%%%%%%%%%%%%%%%%%%%%%%%%%%%%%%%%%%%%%%%%%%%
%%%%%%%%%%%%%%%%%%%%%%%%%%%%%%%%%%%%%%%%%%%%%%%%

%%%%%%%%%%%%%%%%%%%%%%%%%%%%%%%%%%%%%%%%%%%%%%%%%%%%%%%%
\subsection{Boundary one-point function}
%\label{sec:massive}
%%%%%%%%%%%%%%%%%%%%%%%%%%%%%%%%%%%%%%%%%%%%%%%%%%%%%%%%
If we take the Poincare coordinates for AdS$_{d+1}$
\ba
ds^2 \, = \, \frac{d\t^2 + dz^2 + dw^2 + \sum_{i=0}^{d-2}dx_i^2}{z^2} \, , 
\ea
the bulk scalar behaves near the AdS boundary $z \to 0$ as
	\begin{align}
		\p(\t, z, w, x_i) \, = \, J(\t, w, x_i) \, z^{d-\D} \, + \, \a(\t, w, x_i) z^\D \, + \, \cdots \, ,
	\end{align}
where $J$ is the source for the dual operator $\mathcal{O}$ and $\a$ corresponds to the expectation value as
	\begin{align}
		\big\la \mathcal{O}(\t, w, x_i) \big\ra \, = \, \a(\t, w, x_i) \, .
	\end{align}
Using the coordinate change $z = w/\sinh\r$, we can move from the hyperbolic slicing coordinates to the Poincare coordinates.
In particular, near the boundary we have
	\begin{align}
		e^{- \r} \, \simeq \, \frac{z}{2w} \, .
	\end{align}
Therefore, given the boundary condition of the scalar field as in (\ref{massiveBC}),
	\begin{align}
		\big\la \mathcal{O}(\t, w, x_i) \big\ra \, = \, \frac{\a}{(2w)^\D} \, .
	\end{align}
This agrees with the form of the expected one-point function from the BCFT side.

For the free massless case, with the solution (\ref{phiprime}) and the boundary behaviour of $f$ (\ref{masslessfBC}), 
the scalar field behaves near the boundary as
	\begin{align}
		\p \, \simeq \, - \frac{p_0}{d} \, (2e^{-\r})^d \, + \, \cdots \, .
	\end{align}
Since for the massless case we have $\D=d$, the one-point function is obtained as
	\begin{align}
		\big\la \mathcal{O}(\t, w, x_i) \big\ra \, = \, - \frac{p_0}{d} \frac{1}{(w)^d} \, .
	\end{align}
This again agrees with the form of the expected one-point function from the BCFT side.

%%%%%%%%%%%%%%%%%%%%%%%%%%%%%%%%%%%%%%%%%%%%%%
\section{Conclusions and Discussions}
\label{sec:conclusions}
%%%%%%%%%%%%%%%%%%%%%%%%%%%%%%%%%%%%%%%%%%%%%%

In this paper, we have studied several dynamical aspects of holographic boundary conformal field theories (BCFTs), by employing a gravity dual construction known as the AdS/BCFT correspondence. A key feature of AdS/BCFT is the presence of end of the world-brane (EOW brane), which stretches from the boundary of the BCFT towards the AdS bulk. When we excite a holographic BCFT, the EOW brane starts to vibrate  in addition to bulk metric excitations in the gravity dual. The EOW brane is characterized by the value of its tension $\sigma$. In this way, we can analyze the dynamical property of holographic BCFT via the gravity dual in a way similar to the one employed in earlier studies of brane-world models.

First we noted that the gravity dual of a two dimensional BCFT, i.e.\ AdS$_3/$BCFT$_2$ is special in that there are no propagating modes in the bulk of the pure AdS$_3$ gravity. Owing to this property, we can find analytical solutions to the gravity duals of an arbitrary excited states in a holographic BCFT, whose stress energy tensor satisfies the boundary condition of complete reflection. However, the main aim of this paper is to investigate dynamical properties of AdS/BCFT in higher dimensions.

In higher dimensional AdS$_{d+1}/$BCFT$_d$ ($d\geq 3$), the bulk graviton mode is propagating and its dynamics becomes much richer. In this paper we performed a complete analysis of metric perturbations in the presence of EOW brane. We found that they are organized into the helicity 0, 1 and 2 mode. In particular, the lightest mode of helicity 0 can be identified with the brane bending mode. We can calculate the holographic stress energy tensor from these solutions to the metric perturbations. The 
brane bending mode provides a diagonal component of stress energy tensor which shows an universal behavior such that it approaches a constant value at the boundary of BCFT, being independent from the value of brane tension. On the other hand, all other modes of helicity 0, 1 and 2 are sensitive to the value of tension and the corresponding stress energy tensors are proportional to characteristic powers of the distance $w$ between the point and the boundary. These powers depend on the tension and interestingly, they become integer valued when the tension vanishes. We confirmed that the boundary condition of stress energy tensor $T_{iw}|_{w=0}=0$, which is expected for any BCFTs, is indeed satisfied for any values of tension. It will be an important future problem to understand the above behaviors of stress energy tensor, predicted by the gravity dual analysis, from field theoretic calculations in BCFTs.

We also examined solutions of the scalar field perturbations in the AdS/BCFT setup. This is dual to the scalar operator excitation in a holographic BCFT.
This analysis again confirms that the bulk excitation propagates to the EOW brane and is completely reflected back, as expected from the dual BCFT. We construct an explicit numerical solution which shows this behavior. It would be an interesting future problem to study both scalar field and metric perturbations at the same time and construct a gravity dual of realistic black hole evaporation in higher dimensions, where we expect that the dynamics of brane world plays a crucial role.

Then we analyzed another important aspect of AdS/BCFT, namely this has the third description in terms of a CFT coupled to gravity along the boundary. This is expected from the brane-world holography, so called double holography. If we focus on the duality in lower dimensional (i.e.\ $d$ dimensional) descriptions, this leads to the equivalence that a $d$ dimensional BCFT is equivalent to a $d$ dimensional CFT on a half plane, coupled to $d$ dimensional gravity on AdS$_d$, which is called Island/BCFT correspondence. In this paper, we provided an evidence which supports this duality by calculating the entanglement entropy for a semi-disk subsystem. The upshot is that the entanglement entropy in a $d$ dimensional holographic BCFT agrees with that in the theory defined by a $d$ dimensional CFT coupled to a $d$ dimensional gravity, where the latter gravitational theory is assumed to be an induced gravity. It will be intriguing to generalize this analysis to more general subsystems. 

Finally, we present a gravity dual computation of one point functions in holographic BCFTs, where the scalar field solution was given numerically in the presence of a non-trivial boundary condition on EOW brane. This reproduces the expected form of one-point functions in BCFTs. An interesting future problem is to consider a string theory embedding of the AdS/BCFT and calculate the one point functions both from string theory and the dual BCFT.

%%%%%%%%%%%%%%%%%%%%%%%%%%%%%%%%%%%%%%%%%%%%%%%%%%%%%%%%
%%%%%%%%%%%%%%%%%%%%%%%%%%%%%%%%%%%%%%%%%%%%%%%%%%%%%%%%
\section*{Acknowledgements}
%%%%%%%%%%%%%%%%%%%%%%%%%%%%%%%%%%%%%%%%%%%%%%%%%%%%%%%%
%%%%%%%%%%%%%%%%%%%%%%%%%%%%%%%%%%%%%%%%%%%%%%%%%%%%%%%%
We are grateful to Shan-Ming Ruan and Zixia Wei for useful comments on the draft of this paper. We also thank Yu-ki Suzuki and Seiji Terashima for informing us of their independent work \cite{Suzuki:2022yru} when we were writing up this draft. This work is supported by MEXT KAKENHI Grant-in-Aid for Transformative Research Areas (A) through the ``Extreme Universe'' collaboration: Grant Number 21H05182, 21H05187 and 21H05189. 
KI and TS is supported by JSPS Grants-in-Aid for Scientific Research (A) 17H01091. KI is also supported by JSPS Grants-in-Aid for Scientific Research (B) JP20H01902. TS is also supported by JSPS Grants-in-Aid for Scientific Research (C) JP21K03551.
KS and TT are supported by the Simons Foundation through the ``It from Qubit'' collaboration. TT is also supported by Inamori Research Institute for Science, World Premier International Research Center Initiative (WPI Initiative) from the Japan Ministry of Education, Culture, Sports, Science and Technology (MEXT), and JSPS Grant-in-Aid for Scientific Research (A) No.~21H04469.
NT is supported in part by JSPS Grants-in-Aid for Scientific Research (C) 18K03623.

\appendix
%%%%%%%%%%%%%%%%%%%%%%%%%%%%%%%%%%%%%%%%%%%%%%%%%%%%%%%%
%%%%%%%%%%%%%%%%%%%%%%%%%%%%%%%%%%%%%%%%%%%%%%%%%%%%%%%%
\section{A class of explicit metric perturbations at $d=3$}
\label{app:d=3}
%%%%%%%%%%%%%%%%%%%%%%%%%%%%%%%%%%%%%%%%%%%%%%%%%%%%%%%%
%%%%%%%%%%%%%%%%%%%%%%%%%%%%%%%%%%%%%%%%%%%%%%%%%%%%%%%%
Here, we show a class of explicit solutions for the perturbative Einstein equation under the following 
metric ansatz for $d=3$:
\ba
ds^2=d\rho^2+\hat{H}_{\mu\nu}(\rho,\tau,x,y)dx^\mu dx^\mu,
\ea
where we assume the following form
\ba
\hat{H}_{\mu\nu}(\rho,\tau,x,y)=\delta_{\mu\nu}\frac{\cosh^2\rho}{y^2}+y\cdot \hat{h}_{\mu\nu}(\rho,\tau,x)+O(y^2).
\ea
This form is motivated by picking up perturbations which lead to a constant value of holographic stress energy tensor. We work with the Euclidean signature. As opposed to the section \ref{sec:GWa}, we will not impose the TT-gauge condition. Instead we require that the location of EOW brane is at $\rho=\rho_*$ even after we take into account the gravitational backreaction.

%%%%%%%%%%%%%%%%%%%%%%%%%%%%%%%%%%%%%%%%%%%%%%%%%%%%%%%%
\subsection{Perturbative solutions to Einstein equation}
%%%%%%%%%%%%%%%%%%%%%%%%%%%%%%%%%%%%%%%%%%%%%%%%%%%%%%%%

By solving the Einstein equation perturbatively, we find the solutions as follows
\ba
&& \hat{h}_{\tau\tau}=-\frac{1}{2}B\cosh^2\rho+\frac{U_1}{2}\sinh\rho\cosh\rho+\frac{U_2}{2}\cosh\rho\left(1+2\sinh\rho\cdot\arctan\left[\tanh\frac{\rho}{2}\right]
\right)\no
&& \ \ \ \ \ \ -V_1\tanh\rho(\sinh^2\rho+3)-\frac{V_2}{\cosh\rho},\no
&& \hat{h}_{xx}=-\frac{1}{2}B\cosh^2\rho+\frac{U_1}{2}\sinh\rho\cosh\rho+\frac{U_2}{2}
\cosh\rho\left(1+2\sinh\rho\cdot\arctan\left[\tanh\frac{\rho}{2}\right]
\right)\no &&\ \ \ \ \ \ +V_1\tanh\rho(\sinh^2\rho+3)+\frac{V_2}{\cosh\rho},\no
&& \hat{h}_{yy}=\frac{3}{2}B\cosh^2\rho-U_1\sinh\rho\cosh\rho-U_2
\cosh\rho\left(1+2\sinh\rho\cdot\arctan\left[\tanh\frac{\rho}{2}\right]
\right),\no
&& \hat{h}_{\tau x}=A_1\tanh\rho(\sinh^2\rho+3)+\frac{A_2}{\cosh\rho},\no
&& \hat{h}_{\tau y}=Q_1\cosh^2\rho +Q_2\cosh^2\rho\left(2\arctan\left[\tanh\frac{\rho}{2}\right]+
\frac{\sinh\rho}{\cosh\rho^2}\right),\no
&& \hat{h}_{xy}=R_1\cosh^2\rho +R_2\cosh^2\rho\left(2\arctan\left[\tanh\frac{\rho}{2}\right]+
\frac{\sinh\rho}{\cosh\rho^2}\right),\label{pertsolb}
\ea
where $A_1,A_2,B,U_1,U_2,V_1,V_2,P_1,P_2,Q_1,Q_2$ are arbitrary functions of $\tau$ and $x$. Here we imposed the Einstein equation up to $O(y^3)$, $O(y^2)$ and $O(y)$ for $(\rho,\rho)$, $(\rho,\mu)$ and $(\mu,\nu)$ components, respectively. It is also useful to note that we can rewrite 
\ba
\arctan\left[\tanh\frac{\rho}{2}\right]=\frac{i}{2}\cdot\log\left[\frac{e^{\rho/2}+ie^{-\rho/2}}{e^{\rho/2}-ie^{-\rho/2}}\right]+\frac{\pi}{4}.
\ea

%%%%%%%%%%%%%%%%%%%%%%%%%%%%%%%%%%%%%%%%%%%%%%%%%%%%%%%%
\subsection{Comparison with AdS$_4$ black brane solution}
\label{comparison}
%%%%%%%%%%%%%%%%%%%%%%%%%%%%%%%%%%%%%%%%%%%%%%%%%%%%%%%%

Consider the static black brane solution (AdS$_4$ Schwarzschild)
\ba
ds^2=\frac{f(\hat{z})}{\hat{z}^2}dw^2+\frac{d\hat{z}^2}{f(\hat{z})\hat{z}^2}
+\frac{d\tau^2+dx^2}{\hat{z}^2},
\ea
where $f(\hat{z})=1-3\ap \hat{z}^3$. Via the coordinate transformation
\ba
z=\hat{z}+\frac{\ap}{2}\hat{z}^4+\ddd,
\ea
we can write the metric in the Graham-Fefferman form
\ba
ds^2\simeq \frac{dz^2+(1-2\ap z^3)dw^2+(1+\ap z^3)d\tau^2+(1+\ap z^3)dx^2}{z^2}.
\ea
More generally, at the first order of perturbation, the solution to the Einstein equation looks like
\ba
&& ds^2\simeq  \frac{dz^2+(1+\ap z^3)d\tau^2+(1+\beta z^3)dx^2+(1+\gamma z^3)dw^2}{z^2},\no
&& \mbox{with the condition:}\ \ap+\beta+\gamma=0.
\ea
Note that the above condition is equivalent to the traceless condition of holographic stress energy tensor.

Now by applying the coordinate transformation (\ref{cortra}) or equally
\ba
y=\s{z^2+w^2},\ \ \ \cosh\rho=\frac{\s{w^2+z^2}}{z}, \label{cortrb}
\ea
we obtain
\ba
ds^2&\simeq& d\rho^2+\cosh^2\rho\left[\frac{d\tau^2+dx^2+dy^2}{y^2}\right]
+\frac{\ap y}{\cosh\rho}d\tau^2+\frac{\beta y}{\cosh\rho}dx^2\no
&&\ +\frac{\gamma y}{\cosh\rho}\left(\tanh\rho dy+\frac{yd\rho}{\cosh^2\rho}\right)^2\no
&=&\left(1+\frac{\gamma y^3}{\cosh^5\rho}\right)d\rho^2
+2\frac{\gamma y^2\sinh\rho}
{\cosh^4\rho}dyd\rho+\left(\frac{\cosh^2\rho}{y^2}
+\gamma y\frac{\sinh^2\rho}{\cosh^3\rho }\right)dy^2\no
&& \  +\left(\frac{\cosh^2\rho}{y^2}+\frac{\ap y}{\cosh\rho}\right) d\tau^2
+\left(\frac{\cosh^2\rho}{y^2}+\frac{\beta y}{\cosh\rho}\right) dx^2.
\ea
To make this metric into that of a Gaussian normal coordinate, we further perform
 the coordinate transformation into the new coordinate $\ti{\rho}$ and $\ti{y}$:
\ba
\rho=\ti{\rho}+\eta(\ti{\rho},\ti{y}), \ \ \ y=\ti{y}+\xi(\ti{\rho},\ti{y}).  \label{shiftf}
\ea
We require 
\ba
&& g_{\ti{\rho}\ti{\rho}}=1+\frac{\gamma \ti{y}^3}{\cosh^5\ti{\rho}}+2\de_{\ti{\rho}}\eta\equiv 1,\no
&& g_{\ti{\rho}\ti{y}}=\gamma \ti{y}^2\frac{\sinh\ti{\rho}}{\cosh^4\ti{\rho}}
+\de_{\ti{y}}\eta+\frac{\cosh^2\ti{\rho}}{\ti{y}^2}\de_{\ti{\rho}}\xi\equiv 0. \label{dififv}
\ea
These are solved as follows:
\ba
&& \eta=-\frac{\gamma}{2}\ti{y}^3\left(
\frac{3}{4}\arctan\left[\tanh\frac{\ti{\rho}}{2}\right]+\frac{3}{8}
\frac{\sinh\ti{\rho}}{\cosh^2\ti{\rho}}+\frac{1}{4}\frac{\sinh\ti{\rho}}{\cosh^4\ti{\rho}}\right),\no
&& \xi=\gamma \ti{y}^4\left(\frac{9}{16}\frac{1}{\cosh\ti{\rho}}-\frac{3}{16}\frac{1}{\cosh^3\ti{\rho}}+\frac{1}{8}\frac{1}{\cosh^5\ti{\rho}}-\frac{9}{8}\tanh\rho\cdot \arctan\left[\tanh\frac{\ti{\rho}}{2}\right]\right).\no
\ea
This leads to the solution in the Gaussian normal coordinate:
\ba
ds^2&\simeq& d\ti{\rho}^2+\cosh^2\ti{\rho}\left[\frac{d\tau^2+dx^2+d\ti{y}^2}{\ti{y}^2}\right]+3\gamma \ti{y}\cosh\ti{\rho}\left(1+2\arctan\left[\tanh\frac{\ti{\rho}}{2}\right]\sinh\ti{\rho}\right)d\ti{y}^2\no
&&+\ti{y}\left(-\frac{3}{2}\gamma\cosh\ti{\rho}+\left(\ap+\frac{\gamma}{2}\right)\frac{1}{\cosh\ti{\rho}}-\frac{3\gamma}{2}\arctan\left[\tanh\frac{\ti{\rho}}{2}\right]\sinh 2\ti{\rho}\right)d\tau^2\no
&&+\ti{y}\left(-\frac{3}{2}\gamma\cosh\ti{\rho}+\left(\beta+\frac{\gamma}{2}\right)\frac{1}{\cosh\ti{\rho}}-\frac{3\gamma}{2}\arctan\left[\tanh\frac{\ti{\rho}}{2}\right]\sinh 2\ti{\rho}\right)dx^2.  \label{modePoin}
\ea

Due to the constraint $\ap+\beta+\gamma=0$, there are two independent solutions:
one is $(\ap,\beta,\gamma)=(-2V_2,2V_2,0)$ and the other is 
$(\ap,\beta,\gamma)=(U_2/6,U_2/6,-U_2/3)$. Each of them coincide with those in the perturbative solutions in (\ref{pertsolb}). Moreover, the solutions to 
(\ref{dififv}) have two ambiguities which allow us to shift $\eta\to \eta+C_1\ti{y}^3$
and $\xi\to\xi+C_2 \ti{y}^4$. They correspond to $U_1$ and $B$ modes in (\ref{pertsolb}), respectively.
In the same way, we can add a perturbation
\ba
ds^2\to ds^2+\mbox{(const.)}\cdot  z dtdx.
\ea
This corresponds to the mode $A_2$ in (\ref{pertsolb}),
In this way, these static AdS deformations can explain the modes $(U_2,V_2,A_2)$ and 
the redundant modes $(U_1,B)$. 

\subsection{Full solutions with boundary condition imposed and relation to the TT gauge result}

Now we impose the Neumann boundary condition (\ref{bcondq}) to find perturbative solutions in the AdS/BCFT.
We can asssume that the EOW brane is situated at $\rho=\rho_*$. Then the Neumann boundary condition leads to:
\ba
\frac{1}{2}\frac{\de}{\de\rho}\hat{h}_{\mu\nu}=\tanh\rho_*\cdot \hat{h}_{\mu\nu}.  \label{bcondq}
\ea
First note that the modes $B$, $Q_1$ and $R_1$ automatically satisfy this condition (\ref{bcondq}), though they are non-normalizble in the $\rho\to\infty$ region. This means that we cannot turn on either $B$, $Q_1$, $R_2$, $Q_2$ or $R_2$. Then we can also find that for any $\rho_*$, a suitable linear combination of $U_1$ and $U_2$ can satisfy the normalizability and the boundary condition.  Indeed, as we have seen in (\ref{modePoin}), it is clear that the modes $(U_1,U_2)$ are normalizable, which corresponds to the $a=0$ mode (\ref{azero}) in the TT gauge analysis.

On the other hand, since $A_1$ and $V_1$ are non-normalizable and thus are not available in our setup of AdS/BCFT, $V_2$
and $A_2$ can satisfy the boundary condition (\ref{bcondq}) only when $\rho_*=0$ i.e.\ the vanishing tension case. This is consistent with the speciality of the vanishing tension case we found in subsection 
\ref{subsecd=3}. 

If we also take into account non-normalizable modes (i,e, which gets larger than $O(e^{-\rho})$ in $\rho\to\infty$ limit), which are not dynamical in AdS/BCFT, we have the complete matching between our analysis here and that in the TT gauge result in section 3 as follows (note that Legendre $P$ function gives non-normalizable mode near the AdS boundary): 
\ba
 && (i)\ \lambda_\rho=-1\ (a=0) :\no
 &&  \s{\cosh\rho}\ P^{3/2}_{-1/2}(\tanh\rho)\propto \sinh\rho\cosh\rho,\ \to U_1\no
&& \s{\cosh\rho}\ P^{-3/2}_{-1/2}(\tanh\rho) \propto \left[\cosh\rho(1+2\sinh\rho \arctan\left(\tanh\frac{\rho}{2}\right)
-\frac{\pi}{2}\sinh\rho\cosh\rho\right],\ \to (U_1,U_2)\no
 && (ii)\ \lambda_\rho=0\ (a=1): \no
&& \s{\cosh\rho}\ P^{3/2}_{1/2}(\tanh\rho)\propto \cosh^2\rho,\ \to Q_1,R_1 \no
&&\s{\cosh\rho}\ P^{-3/2}_{1/2}(\tanh\rho)\propto \cosh^2\rho\left[2\arctan\left(\tanh\frac{\rho}{2}\right)+\frac{\sinh\rho}{\cosh^2\rho}-\frac{\pi}{2}\right],\ \to (Q_1,Q_2),(R_1,R_2) \no
&&  (iii)\ \lambda_\rho=3\ (a=2):\no
&& \s{\cosh\rho}\ P^{3/2}_{3/2}(\tanh\rho)\propto \tanh\rho(\sinh^2\rho+3),\ \to A_1,V_1\no
&& \s{\cosh\rho}\ P^{-3/2}_{3/2}(\tanh\rho)\propto \frac{1}{\cosh\rho}.\ \to A_2,V_2.
\label{compareeqa}
\ea
In the above, physical modes are $(U_1,U_2)$, $A_2$ and $V_2$.
Other modes are either non-normalizable or violating the boundary condition. The mode (ii) $\lambda_\rho=0$ 
is not allowed due to the condition (\ref{lowerboundl}).
Note also that $B$ mode does not satisfy the traceless gauge condition $h^\mu_\mu=0$. These perfectly match with the physical modes in the TT gauge, discussed in subsection \ref{subsecd=3}.

%%%%%%%%%%%%%%%%%%%%%%%%%%%%%%%%%%%%%%%%%%%%%%%%%%%%%%%%
%%%%%%%%%%%%%%%%%%%%%%%%%%%%%%%%%%%%%%%%%%%%%%%%%%%%%%%%
\section{Minimal surface in AdS$_{d+1}$}
\label{app:minimal surface}
%%%%%%%%%%%%%%%%%%%%%%%%%%%%%%%%%%%%%%%%%%%%%%%%%%%%%%%%
%%%%%%%%%%%%%%%%%%%%%%%%%%%%%%%%%%%%%%%%%%%%%%%%%%%%%%%%
In this appendix, we study the condimension-2 minimal surface in AdS$_{d+1}$ with the metric (\ref{ds_d+1})
	\begin{align}
		ds_{d+1}^2 \, = \, d\r^2 + \frac{\cosh^2 \r}{y^2} \left( - dt^2 + dy^2 + \sum_{i=1}^{d-2} dx_i^2 \right) \, ,
	\end{align}
with a condition that the boundary of the surface in the $\r \to \inf$ limit is given by
	\begin{align}
		y^2 \, + \, \sum_{i=1}^{d-2} x_i^2 \, = \, l^2 \, , \qquad y \, \ge \, 0 \, .
    \label{BC_AdS_d+1}
	\end{align}
The determination of the minimal surface in this case is very much similar to the one we discussed in section~\ref{sec:Islands in AdS_{d+1}} for AdS$_d$ space.
 
First, we again move to the spherical coordinates by 
	\begin{align}
		y \, &= \, r \cos \th \, , \\[-10pt]
		x_i \, &= \, r \sin \th \cos \phi_i \prod_{n=1}^{i-1} \sin \phi_n \, , \qquad (i = 1, \cdots, d-3) \\[-10pt]
		x_{d-2} \, &= \, r \sin \th \prod_{n=1}^{d-3} \sin \phi_n \, ,
	\end{align}
which brings the metric into 
	\begin{align}
		ds_{d+1}^2 \, = \, d\r^2 + \frac{\cosh^2 \r}{r^2\cos^2\th} \big( - dt^2 + dr^2 + r^2 d\O_{d-2}^2 \big) \, .
	\end{align}
We specify the surface by 
	\begin{align}
		r \, = \, r(\r, \th, \phi_1, \cdots, \phi_{d-3}) \, ,
	\end{align}
so that the induced metric is found as
	\begin{align}
		ds_{d-1}^2 \, &= \, d\r^2 + \frac{\cosh^2 \r}{\cos^2\th} \, d\O_{d-2}^2 \nn\\
		&\qquad \, + \, \frac{\cosh^2\r}{\cos^2\th} \left( \frac{\pa \log r}{\pa \r} d\r + \frac{\pa \log r}{\pa \th} d\th + \sum_{i=1}^{d-3} \frac{\pa \log r}{\pa \p_i} d\p_i \right)^2 \, .
    \label{induced_metric_AdS_d+1}
	\end{align}
As before, we decompose this induced metric into two contributions as one from the first line and the other from the second line of (\ref{induced_metric_AdS_d+1})
	\begin{align}
		\big( h_{d-1} \big)_{\m\n}
		\, = \, \big( h_{d-1}^{(0)} \big)_{\m\n} \, + \, \big( \tilde{h}_{d-1} \big)_{\m\n} \, .
	\end{align}
Again the $\big( \tilde{h}_{d-1} \big)_{\m\n}$ metric is given by a tensor product of a vector
	\begin{align}
	    &\big( \tilde{h}_{d-1} \big)_{\m\n} \, = \, v_\m \, (v^T)_\n \, , \\
	    &v^T_\m \, = \, \frac{\cosh\r}{\cos\th} \left( \frac{\pa \log r}{\pa \r} \, , \, \frac{\pa \log r}{\pa \th} \, , \, \frac{\pa \log r}{\pa \p_1} \, , \,  \cdots \, , \, \frac{\pa \log r}{\pa \p_{d-3}} \right) \, ,
	\end{align}
so that the determinant is decomposed as 
	\begin{align}
		\det(h_{d-1}) \, &= \, \det\Big(h_{d-1}^{(0)} + v v^T \Big) \nn\\
		\, &= \, \det\big(h_{d-1}^{(0)} \big) \det\big(I_{d-1} + v^T (h_{d-1}^{(0)})^{-1} v \big) \, ,
	\end{align}
Minimization of this determinant requires $v=0$, therefore, the minimal surface is given by $y=l$.

%%%%%%%%%%%%%%%%%%%%%%%%%%%%%%%%%%%%%%%%%%%%%%%%%%%%%%%%
%%%%%%%%%%%%%%%%%%%%%%%%%%%%%%%%%%%%%%%%%%%%%%%%%%%%%%%%
% bibliography via bibtex
\bibliographystyle{JHEP}
\bibliography{AdSBCFT}

%%%%%%%%%%%%%%%%%%%%%%%%%%%%%%%%%%%%%%%%%%%%%%%%%%%%%%%%
%%%%%%%%%%%%%%%%%%%%%%%%%%%%%%%%%%%%%%%%%%%%%%%%%%%%%%%%

\end{document}